\documentclass[aps,prb,twocolumn,showpacs]{revtex4}
\usepackage{graphicx} 
\usepackage{amsmath}
\usepackage{color}
\newcommand{\be}{\begin{equation}}
\newcommand{\ee}{\end{equation}}
\newcommand{\bea}{\begin{eqnarray}}
\newcommand{\eea}{\end{eqnarray}}
\newcommand{\bse}{\begin{subequations}}
\newcommand{\ese}{\end{subequations}}

\begin{document}

\title{Thermodynamic Properties of the van der Waals Fluid}

\author{David C. Johnston}

\affiliation{Department of Physics and Astronomy, Iowa State University, Ames, Iowa 50011, USA}

\date{\today}

\begin{abstract}

The van der Waals (vdW) theory of fluids is the first and simplest theory that takes into account interactions between the particles of a system that result in a phase transition versus temperature.  Combined with Maxwell's construction, this mean-field theory predicts the conditions for equilibrium coexistence between the gas and liquid phases and the first-order transition between them.  However, important properties of the vdW fluid have not been systematically investigated.  Here we report a comprehensive study of these properties.  Ambiguities about the physical interpretation of the Boyle temperature and the influence of the vdW molecular interactions on the pressure of the vdW gas are resolved.  Thermodynamic variables and properties are formulated in reduced units that allow all properties to be expressed as laws of corresponding states that apply to all vdW fluids.  Lekner's parametric solution for the vdW gas-liquid coexistence curve in the pressure-temperature plane and related thermodynamic properties [Am.\ J. Phys.\ {\bf 50}, 161 (1982)] is explained and significantly extended.  Hysteresis in the first-order transition temperature on heating and cooling is examined and the maximum degrees of superheating and supercooling determined.  The latent heat of vaporization and the entropy change on crossing the coexistence curve are investigated.  The temperature dependences of the isothermal compressibility, thermal expansion coefficient and heat capacity at constant pressure for a range of pressures above, at and below the critical pressure are systematically studied from numerical calculations including their critical behaviors and their discontinuities on crossing the coexistence curve.  Joule-Thomson expansion of the vdW gas is investigated in detail and the pressure and temperature conditions for liquifying a vdW gas on passing through the throttle are determined.

\end{abstract}


\maketitle

\tableofcontents

\section{Introduction}

The van der Waals (vdW) fluid is the first, simplest and most widely known example of an interacting system of particles that exhibits a phase transition, in this case a first-order transition between liquid and gas (vapor) phases.\cite{vanderWaals1873, vanderWaals1910}  For these reasons the vdW fluid and associated phase transition are presented in most thermodynamics and statistical mechanics courses and textbooks (see, e.g., Refs.~\onlinecite{Reif1965, Kittel1980, Schroeder2000}), where, however, the treatment is often limited to a discussion of the pressure~$p$ versus volume~$V$ isotherms and their interpretation in terms of the Maxwell construction\cite{Maxwell1875} to define the regions of coexistence of gas and liquid.  In addition,  critical exponents of several thermodynamic properties on approaching the critical point termininating the $p$ versus temperature~$T$ liquid-gas coexistence curve are well known in the context of critical phenomena.\cite{Stanley1971, Heller1967}  On the other hand, for example, to our knowledge there have been no systematic studies of the temperature dependences of thermodynamic properties of the vdW fluid such as the heat capacity at constant pressure~$C_{\rm p}$, the isothermal compressibility $\kappa_{\rm T}$ or the volume thermal expansion coefficient $\alpha$, and how those properties are influenced by proximity to the critical point or by crossing the liquid-gas coexistence curve in the $p$-$T$ plane.  Therefore the landscape of thermodynamic properties of the vdW fluid is unclear.

Here a comprehensive analytical and numerical study of the van der Waals fluid and its thermodynamic properties is presented.  All thermodynamic properties are formulated in terms of reduced parameters that result in many laws of corresponding states, which by definition are the same for any fluid satisfying the assumptions of the vdW theory.  These formulations allow the discussed thermodynamic properties to describe all vdW fluids.

The first few Secs.~\ref{Eq:BckgrndIG}--\ref{Sec:EOSRedCorr}  are short introductory sections.  In Sec.~\ref{Eq:BckgrndIG} the nomenclature and definitions of thermodynamics functions and properties used here are briefly discussed along with the well-known properties of the ideal gas for reference.  The vdW molecular interaction parameters~$a$ and~$b$ are discussed in Sec.~\ref{Sec:vdWInteractions} in terms of the Lennard-Jones potential where the ratio $a/b$ is shown to be a fixed value for a particular vdW fluid which is determined by the depth of the Lennard-Jones potential well for that fluid.  We only consider here molecules without internal degrees of freedom.  The Helmholtz free energy~$F$, the critical pressure~$p_{\rm c}$, temperature $T_{\rm c}$ and volume~$V_{\rm c}$ and critical compressibility factor~$Z_{\rm c}$ are defined in terms of $a$ and~$b$ Sec.~\ref{Sec:Fab}, which then allows the values of $a$ and~$b$ for a particular fluid to be determined from the measured values of $p_{\rm c}$ and $T_{\rm c}$ for the fluid.  The  entropy~$S$, internal energy~$U$ and heat capacity at constant volume~$C_{\rm V}$ for the vdW fluid are written in terms of $a$, $b$, the volume $V$ occupied by the fluid and the number~$N$ of molecules in Sec.~\ref{Sec:SUCV} and the pressure and enthalpy~$H$ in Sec.~\ref{Sec:PH}.  The vdW equation of state is written in terms of dimensionless reduced variables in Sec.~\ref{Sec:EOSRedCorr} and the definition of laws of corresponding states reviewed.

There has been much discussion and disagreement over the past century about the influence of the vdW molecular interaction parameters $a$ and/or~$b$ on the pressure of a vdW gas compared to that of an ideal gas at the same volume and temperature.  This topic is quantitatively discussed in Sec.~\ref{Sec:Pvsab} where it is shown that the pressure of a vdW gas can either increase or decrease compared to that of an ideal gas depending on the volume and temperature of the gas.  A related topic is the Boyle temperature $T_{\rm B}$ at which the ``compression factor'' $Z \equiv pV/(Nk_{\rm B}T_{\rm B})$ is the same as for the ideal gas as discussed in Sec.~\ref{Sec:Boyle}, where $k_{\rm B}$ is Boltzmann's constant.  It is sometimes stated that at the Boyle temperature the properties of a gas are the same as for the ideal gas; we show that this inference is incorrect for the vdW gas because even at this temperature other thermodynamic properties are not the same as those of an ideal gas.  In Secs.~\ref{Eq:PropsRedVars} and~\ref{Sec:ChemPot} the thermodynamic variables, functions and chemical potential~$\mu$ are written in terms of dimensionless reduced variables that are used in the remainder of the paper.  Representative $p$-$V$, $T$-$V$ and $\mu$-$T$ vdW isotherms are presented in terms of the reduced parameters, where unstable or metastable regions are present that are removed when the equilibrium properties are obtained from them.

The equilibrium values of the pressure and coexisting liquid and gas volumes are calculated in Sec.~\ref{Sec:EquilpVTBpT} using traditional methods such as the Maxwell construction and the equilibrium $p$-$V$, $T$-$V$ and $p$-$T$ isotherms and phase diagrams including gas-liquid coexistence regions are presented.  An important advance in calculating the gas-liquid coexistence curve in the $p$-$T$ plane and associated properties was presented by Lekner in 1982, who formulated a parametric solution in terms of the entropy difference between the gas and liquid phases.\cite{Lekner1982} Lekner's results were extended by Berberan-Santos~et al.\ in 2008.\cite{Berberan-Santos2008} In Sec.~\ref{Sec:Lekner} Lekner's results are explained and further extended for the the full temperature region up to the critical temperature and the limiting behaviors of properties associated with the coexistence curve for $T\to0$ and $T\to T_{\rm c}$ are also calculated.  In the topics and regions of overlap these results agree with the previous ones.\cite{Lekner1982, Berberan-Santos2008} In Secs.~\ref{Sec:Lekner2} and~\ref{Sec:LatentHeat}  the coexisting liquid and gas densities, the difference between them which is the order parameter for the gas-liquid phase transition, the temperature-density phase diagram, and the latent heat and entropy of vaporization utilizing Lekner's parametrization are calculated and plotted.  Tables of calculated values of parameters and properties obtained using both the conventional and Lekner parametrizations are given in Appendix~\ref{App:Tables}.  Some qualitatively similar numerical calculations of thermodynamic properties of the vdW fluid were recently reported in 2013 by Swendsen.\cite{Swendsen2013}

Critical exponents and amplitudes for the vdW fluid are calculated in Sec.~\ref{Sec:CritExps}, where their values can depend on the path of approach to the critical point.  We express the critical amplitudes in terms of the universal reduced parameters used throughout this paper.  The asymptotic critical behavior for the order parameter is found to be accurately followed from~$T_{\rm c}$ down to about $0.97T_{\rm c}$.  In Sec.~\ref{Sec:Hysteresis} hysteresis in the transition temperature on heating and cooling through the first-order equilibrium liquid-gas phase transition temperature at constant pressure is evaluated.  Numerical calculations of  $\kappa_{\rm T}$, $\alpha$ and $C_{\rm p}$ versus temperature at constant pressure for $p > p_{\rm c}$, $p = p_{\rm c}$ and $p < p_{\rm c}$ are presented in Sec.~\ref{Sec:ConstPcalcs}, where the fitted critical exponents and amplitudes for $p = p_{\rm c}$ are found to agree with the corresponding behaviors predicted analytically in Sec.~\ref{Sec:CritExps}.  The discontinuities in the calculated $\kappa_{\rm T}$, $\alpha$ and $C_{\rm p}$ on crossing the coexistence curve at constant pressure with $p < p_{\rm c}$ are also shown to be in agreement with the analytic predictions in Appendix~\ref{Sec:Discont} that were derived based on Lekner's parametric solution to the coexistence curve.

Cooling the vdW gas by adiabatic free expansion and cooling and/or liquifying the vdW gas by Joule-Thomson expansion are discussed in Sec.~\ref{Sec:Expansions}, where the conditions for liquification of a vdW gas on passing through a throttle are presented.  An analytical equation for the inversion curve associated with the Joule-Thomson expansion of a vdW fluid is derived and found to be consistent with that previously reported by Le~Vent in 2001.\cite{LeVent2001}  A brief summary of the paper is given in Sec.~\ref{Eq:Summary}.

\section{\label{Eq:BckgrndIG} Background and Nomenclature: The Ideal Gas}

An ideal gas is defined as a gas of noninteracting particles in the classical regime where the number density of the gas is small.\cite{Kittel1980}  In this case one obtains an equation of state called the ideal gas law
\bse
\be
pV = Nk_{\rm B}T = N\tau,
\label{Eq:pVIG}
\ee
where throughout the paper we use the shorthand
\be
\tau \equiv k_{\rm B}T.
\ee
\ese
For an ideal gas containing molecules with no internal degrees of freedom, the Helmholtz free energy is
\bse
\be
F(\tau,V,N) =-N\tau\left\{\ln\left[\frac{n_QV}{N}\right] + 1\right\},
\label{Eq:IGF}
\ee
where the ``quantum concentration'' $n_Q$ is given by\cite{Kittel1980}
\be
n_Q = \left(\frac{m\tau}{2\pi\hbar^2}\right)^{3/2},
\label{Eq:nQ}
\ee
\ese
$m$ is the mass of a molecule and $\hbar$ is Planck's constant divided by $2\pi$.  Other authors instead use an expression containing the ``thermal wavelength'' $\lambda_{\rm T}$  defined by $n_Q = \lambda_{\rm T}^{-3}$.  The entropy~$S$ is
\be
\frac{S}{k_{\rm B}} = -\left(\frac{\partial F}{\partial \tau}\right)_{V,N} = N\left[\ln\left(\frac{n_QV}{N}\right) + \frac{5}{2}\right].
\label{Eq:sigmaIG}
\ee 
This equation is known as the Sackur-Tetrode equation.  The internal energy~$U$ is
\be
U = F + TS = \frac{3}{2}N\tau
\label{Eq:UIG}
\ee
and the heat capacity at constant volume~$C_{\rm V}$ is
\be
C_{\rm V} = k_{\rm B}\left(\frac{\partial U}{\partial\tau}\right)_{V,N} = \frac{3}{2}Nk_{\rm B}.
\label{Eq:CVID}
\ee
The enthalpy is
\be
H=U+pV = \frac{5}{2}N\tau = \frac{5}{2}Nk_{\rm B}T.
\label{Eq:HIG}
\ee

The isothermal compressibility is
\be
\kappa_{\rm T} = -\frac{1}{V}\left(\frac{\partial V}{\partial p}\right)_T = \frac{1}{p},
\label{Eq:kappTIG}
\ee
and the volume thermal expansion coefficient is
\be
\alpha = \frac{1}{V}\left(\frac{\partial V}{\partial T}\right)_p = \frac{Nk_{\rm B}}{pV} = \frac{1}{T}.
\label{Eq:alphaDef}
\ee
The heat capacity at constant pressure is
\be
C_{\rm p} = C_{\rm V} + \frac{TV\alpha^2}{\kappa_{\rm T}} = \frac{3}{2}Nk_{\rm B} + \frac{pV}{T} = \frac{5}{2}Nk_{\rm B}.
\label{Eq:CpIG}
\ee
Alternatively,
\be
C_{\rm p} = \left(\frac{\partial H}{\partial T}\right)_p
\ee
gives the same result.

The chemical potential is
\be
\mu = \left(\frac{\partial F}{\partial N}\right)_{\tau,V} = \tau\ln\left(\frac{N}{n_QV}\right) = \tau\ln\left(\frac{p}{n_Q\tau}\right),
\ee
where to obtain the last equality we used the ideal gas law~(\ref{Eq:pVIG}).  The Gibbs free energy written in terms of its natural variables $N,\ p$ and~$T$ or $\tau$ is thus
\bse
\be
G(N,p,\tau) = N\mu(p,\tau),
\label{Eq:Gmu}
\ee
where the differential of $G$ is
\be
dG = -S\,dT +V\,dp + \mu\,dN.
\label{Eq:dG}
\ee
\ese

\section{\label{Sec:vdWInteractions} \lowercase{van der} W\lowercase{aals} Intermolecular Interaction Parameters}

\begin{figure}[t]
\includegraphics[width=3.in]{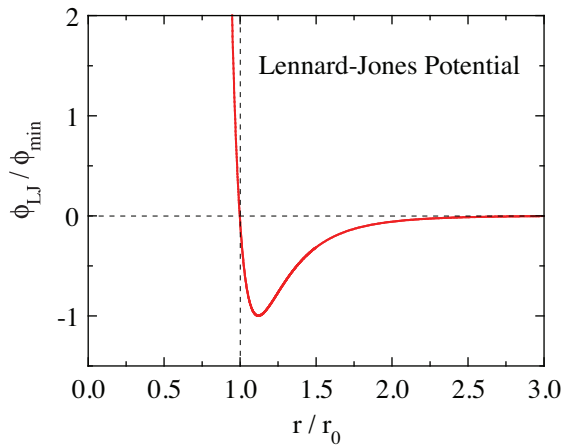}
\caption{(Color online) Lennard-Jones potential energy $\phi_{\rm LJ}$ versus the distance~$r$ between the centers of adjacent molecules from Eq.~(\ref{Eq:VLJ}). The depth of the potential well is $\phi_{\rm min}$ and the zero of the potential energy is at $r = r_0$.}
\label{Fig:Lennard_Jones_Potential}  
\end{figure}

Interactions between neutral molecules or atoms with a center of mass separation~$r$ are often approximated by the so-called Lennard-Jones potential energy $\phi_{\rm LJ}$, given by
\be
\phi_{\rm LJ} = 4\phi_{\rm min}\left[\left(\frac{r_0}{r}\right)^{12}-\left(\frac{r_0}{r}\right)^{6}\right],
\label{Eq:VLJ}
\ee 
where the first term is a short-range repulsive interaction and the second term is a longer-range attractive interaction.  A plot of $\phi_{\rm LJ}/\phi_{\rm min}$ versus~$r/r_0$ is shown in Fig.~\ref{Fig:Lennard_Jones_Potential}.  The value $r=r_0$ corresponds to $\phi_{\rm LJ}=0$, and the minimum value of $\phi_{\rm LJ}$ is $\phi_{\rm LJ}/\phi_{\rm min} = -1$ at
\be
r_{\rm min}/r_0 = 2^{1/6}\approx 1.122.
\label{Eq:rmin}
\ee
By the definition of potential energy, the force between a molecule and a neighbor in the radial direction from the first molecule is $F_r = -d\phi_{\rm LJ}/dr$, which is positive (repulsive) for $r < r_{\rm min}$ and negative (attractive) for $r > r_{\rm min}$.

In the vdW theory of a fluid (gas and/or liquid) discussed in this paper, one ignores possible internal degrees of freedom of the molecules and assumes that the interatomic distance between molecules cannot be smaller  than a molecular diameter, a situation called a ``hard-core repulsion'' where two molecules cannot overlap.  Therefore the  minimum intermolecular distance from center to center is equal to the diameter~$d$ of one molecule.  In terms of the Lennard-Jones potential, we set
\be
d = r_{\rm min},
\ee
rather than $d=r_0$, because the Lennard-Jones interaction between two molecules is repulsive out to a separation of $r_{\rm min}$ as shown in Fig.~\ref{Fig:Lennard_Jones_Potential}.  In the vdW theory, the volume of a molecule (``excluded volume'') is denoted by the variable $b$, so the free volume available for the molecules to move in is $V-Nb$. Thus in the free energy of the ideal gas in Eq.~(\ref{Eq:IGF}) one makes the substitution
\be
V\to V-Nb.
\label{Eq:VvdW}
\ee
In terms of the Lennard-Jones potential, we set
\be
b \equiv d^3 = r_{\rm min}^3 = \sqrt{2}\,r_0^3,
\label{Eq:bSoln}
\ee
where $d = r_{\rm min}$ is a measure of the hard-core diameter of a molecule and we have used Eq.~(\ref{Eq:rmin}).  

For $r > d$ the force between the gas molecules is assumed to be attractive, and the strength of the attraction depends on the distance between the molecules.  In terms of the Lennard-Jones potential this occurs for $r>r_{\rm min}$ according to Eq.~(\ref{Eq:VLJ}) and Fig.~\ref{Fig:Lennard_Jones_Potential}.  One takes into account this attractive part of the interaction in an average way as follows, which is a ``mean-field'' approximation where one ignores local fluctuations in the number density of molecules and short range correlations between their positions.  The number density of molecules is $N/V$.  The number $dN$ of molecules that are at a distance between $r$ and $r+dr$ from the central molecule is $dN=(N/V)dV$, where an increment of volume a distance~$r$ from the center of the central molecule is $dV = 4\pi r^2dr$.  Thus the total average attractive potential energy summed over these molecules, $\phi_{\rm ave}$, is
\be
\phi_{\rm ave} = \left(\frac{N}{V}\right)\frac{1}{2}\int_{r_{\rm min}}^\infty \phi(r)dV =\left(\frac{N}{V}\right) \frac{4\pi}{2}\int_{r_{\rm min}}^\infty \phi(r)r^2dr, 
\label{Eq:phiAve}
\ee
where the prefactor of 1/2 arises because the potential energy of interaction between a molecule and a neighboring molecule is shared equally between them.  In the van der Waals theory, one writes the average potential energy per molecule as
\be
\phi_{\rm ave} = -\left(\frac{N}{V}\right)a
\label{Eq:PhiAve}
\ee
where the parameter $a\geq0$ is an average value of the potential energy per unit concentration, given here using Eq.~(\ref{Eq:phiAve}) by
\be
a = -2\pi\int_{r_{\rm min}}^\infty \phi(r)r^2dr.
\label{Eq:a}
\ee

One can obtain an expression for $a$ in terms of the Lennard-Jones potential.  Substituting the Lennard-Jones potential in Eq.~(\ref{Eq:VLJ}) into~(\ref{Eq:a}), one has
\be
a = -8\pi \phi_{\rm min}\int_{r_{\rm min}}^\infty \left[\left(\frac{r_0}{r}\right)^{12}-\left(\frac{r_0}{r}\right)^{6}\right]r^2dr.
\ee
Changing variables to $x = r/r_0$ and using Eq.~(\ref{Eq:rmin}) gives
\be
a = -8\pi \phi_{\rm min}r_0^3\int_{2^{1/6}}^\infty \left(\frac{1}{x^{10}}-\frac{1}{x^4}\right)dx.
\ee
The integral is $-5/(18\sqrt{2})$, yielding
\be
a = \frac{20\pi r_0^3\phi_{\rm min}}{9\sqrt{2}}.
\label{Eq:aSoln}
\ee
From Eqs.~(\ref{Eq:bSoln}) and~(\ref{Eq:aSoln}), $a$ and $b$ per molecule are related to each other according to
\be
\frac{a}{b} = \frac{5\pi\phi_{\rm min}}{9}.
\label{Eq:abratio}
\ee
This illustrates the important feature that the ratio $a/b$ for a given van der Waals fluid is a fixed value that depends on the intermolecular potential function.

\section{\label{Sec:Fab} Helmholtz Free Energy and Critical Parameters in Terms of the \lowercase{van der} W\lowercase{aals} Interaction Parameters}

The change in the internal energy due to the attractive part of the intermolecular interaction is the potential energy $N\phi_{\rm ave}$ and from Eq.~(\ref{Eq:PhiAve}) one obtains
\be
\Delta U = N\phi_{\rm ave} = -\frac{N^2a}{V}.
\label{Eq:DeltaU}
\ee 
When one smoothly turns on interactions in a thought experiment, effectively one is doing work on the system and this does not transfer thermal energy.  Therefore the potential energy represented by the parameter~$a$ introduces no entropy change and hence the change in the free energy is $\Delta F = \Delta U - \Delta(\tau\sigma)= \Delta U$.  The attractive part of the intermolecular potential energy that results in a change in $F$ compared to the free energy of the ideal gas is then given by
\be
F\to F+\Delta U = F -\frac{N^2a}{V}.
\label{Eq:FDU}
\ee
Making the changes in Eqs.~(\ref{Eq:VvdW}) and~(\ref{Eq:FDU}) to the free energy of the ideal gas in Eq.~(\ref{Eq:IGF}) gives the free energy of the van der Waals gas as
\be
F(\tau,V,N) = -N\tau\left\{\ln\left[\frac{n_Q(V - Nb)}{N}\right] + 1\right\}-\frac{N^2a}{V}.
\label{Eq:FvdW2}
\ee
This is a quantum mechanical expression because $\hbar$ is present in $n_Q$.  However, we will see that the thermodynamic properties of the vdW fluid are classical, where $\hbar$ does not appear in the final calculations.  In the limit $N/V\to0$ or equivalently $a,\ b\to0$, the Helmholtz free energy becomes that of the ideal gas in Eq.~(\ref{Eq:IGF}).  

The critical pressure $p_{\rm c}$, the critical volume $V_{\rm c}$ and critical temperature $\tau_{\rm c}\equiv k_{\rm B}T_{\rm c}$ define the critical point of the van der Waals fluid as discussed later.  These are given in terms of the parameters $a$, $b$ and~$N$ as
\bse
\label{Eqs:RedVars}
\be
p_{\rm c} = \frac{a}{27b^2},\qquad V_{\rm c} = 3Nb,\qquad \tau_{\rm c} = \frac{8a}{27b}.
\label{Eq:CritPars}
\ee
The product of the first two expressions gives an energy scale
\be
p_{\rm c}V_{\rm c} = \frac{Na}{9b} = \frac{3N\tau_{\rm c}}{8} = \frac{3Nk_{\rm B}T_{\rm c}}{8},
\label{Eq:pcVc}
\ee
yielding the universal ratio called the critical ``compression factor'' $Z_{\rm c}$ as
\be
Z_{\rm c} \equiv \frac{p_{\rm c}V_{\rm c}}{N\tau_{\rm c}} = \frac{3}{8}.
\label{Eq:pcVcTc}
\ee
The critical temperatures, pressures and volumes of representative gases are shown in Table~\ref{Tab:vdWPars}.  One sees from the table that the experimental values of $Z_{\rm c} $ are $\sim30$\% smaller that the value of 3/8 predicted by the vdW theory in Eq.~(\ref{Eq:pcVcTc}), indicating that the theory does not accurately describe real gases.  One can solve Eqs.~(\ref{Eq:CritPars}) for $a$, $b$ and $N$ in terms of the critical variables, yielding
\be
a=\frac{27 \tau_{\rm c}^2}{64p_{\rm c}} = 27b^2p_{\rm c},\quad b = \frac{\tau_{\rm c}}{8p_{\rm c}},\quad N = \frac{8p_{\rm c}V_{\rm c}}{3\tau_{\rm c}}.
\label{Eq:abN}
\ee
\ese

\begin{table*}
\caption{\label{Tab:vdWPars} Experimental data for representative gases obtained from the {\it Handbook of Chemistry and Physics} (CRC Press, Cleveland, 2013).  Shown are the molecular weights (MW), critical temperature $T_{\rm c}$, critical pressure $p_{\rm c}$, critical volume $V_{\rm c}$ and the dimenionless critical compression factor $Z_{\rm c} \equiv p_{\rm c}V_{\rm c}/(RT_{\rm c})$ where $R$ is the molar gas constant.  The value of $V_{\rm c}$ predicted by the van der Waals theory is $Z_{\rm c} = 3/8 = 0.375$ according to Eq.~(\ref{Eq:pcVcTc}), which is $\sim30$\% larger than the observed factors listed in the table.  Also shown are the van der Waals parameters $a$ and~$b$ per molecule derived from $T_{\rm c}$ and~$p_{\rm c}$ using Eqs.~(\ref{Eq:abN}), where $a$ is a mean-field measure  of the attractive force between two molecules and $b$ is the excluded volume per molecule due to the molecular hard cores.  A measure of the van der Waals hard-core molecular diameter is defined here as $d\equiv b^{1/3}$.  Assuming a Lennard-Jones potential between molecules, the depth $\phi_{\rm min}$ of the potential well in Fig.~\ref{Fig:Lennard_Jones_Potential} is calculated from $a/b$ using Eq.~(\ref{Eq:abratio}).}
\begin{ruledtabular}
\begin{tabular}{ccccccccccc}
  Gas && MW & $T_{\rm c}$ & $p_{\rm c}$ & $V_{\rm c}$ & $Z_{\rm c} $ & $a$ & $b$ & $d$ & $\phi_{\rm min}$\\	
Name & formula & (g/mol) & (K) & (kPa) & (cm$^3$/mol) &  &(eV\,\AA$^3$) & (\AA$^3$) & (\AA) & (meV)\\
\hline
{\bf Noble gases}\\
Helium & He	&	4.0030	&	5.1953	&	227.46	&	57	&	0.300	&	0.05956	&	39.418		&	3.4033	&	0.8657	\\
Neon & Ne		&	20.183	&	44.490	&	2678.6	&	42	&	0.304	&	0.37090	&	28.665		&	3.0604	&	7.414	\\
Argon & Ar	&	39.948	&	150.69	&	4863		&	75	&	0.291	&	2.344	&	53.48		&	3.768	&	25.11	\\
Krypton & Kr	&	83.800	&	209.48	&	5525		&	91	&	0.289	&	3.987	&	65.43		&	4.030	&	34.91	\\
Xenon & Xe	&	131.30	&	289.73	&	5842		&	118	&	0.286	&	7.212	&	85.59		&	4.407	&	48.28	\\
{\bf Diatomic gases}\\
Hydrogen& H$_2$	&	2.0160	&	33.140	&	1296.4	&	65	&	0.306	&	0.42521	&	44.117	&	3.5335	&	5.5223	\\
Hydrogen fluoride&  HF	&	20.006	&	461.00	&	6480	&	69	&	0.117	&	16.46	&	122.8	&	4.970	&	76.82	\\
Nitrogen & N$_2$	&	28.014	&	126.19	&	3390		&	90	&	0.291	&	2.358	&	64.24	&	4.005	&	21.03	\\
Carbon monoxide & CO	&	28.010	&	132.86	&	3494	&	93	&	0.294	&	2.536	&	65.62	&	4.034	&	22.14	\\
Nitric Oxide & NO	&	30.010	&	180.00	&	6480		&	58	&	0.251	&	2.510	&	47.94	&	3.633	&	29.99	\\
Oxygen & O$_2$	&	32.000		&	154.58	&	5043		&	73	&	0.286	&	2.378	&	52.90	&	3.754	&	25.76	\\
Hydrogen chloride & HCl	&	36.461	&	324.70	&	8310	&	81	&	0.249	&	6.368	&	67.43	&	4.070	&	54.11	\\
Fluorine & F$_2$	&	37.997	&	144.41	&	5172.4	&	66	&	0.284	&	2.024	&	48.184	&	3.6389	&	24.06	\\
Chlorine & Cl$_2$	&	70.910	&	417.00	&	7991		&	123	&	0.284	&	10.92	&	90.06	&	4.482	&	69.49	\\
{\bf Polyatomic gases}\\
Ammonia & NH$_3$	&	17.031	&	405.56	&	11357	&	69.9	&	0.235	&	7.2692	&	61.629	&	3.9500	&	67.581	\\
Water & H$_2$O	&	18.015	&	647.10	&	22060		&	56	&	0.230	&	9.5273	&	50.624	&	3.6993	&	107.83	\\
Carbon dioxide & CO$_2$	&	44.010	&	304.13	&	7375	&	94	&	0.274	&	6.295	&	71.17	&	4.144	&	50.68	\\
Nitrous oxide & N$_2$O	&	44.013	&	309.52	&	7245	&	97	&	0.273	&	6.637	&	73.73	&	4.193	&	51.58	\\
Carbon oxysulfide & COS	&	60.074	&	375.00	&	5880	&	137	&	0.258	&	12.00	&	110.1	&	4.792	&	62.49	\\
{\bf Alkanes}\\
Methane & CH$_4$	&	16.043	&	190.56	&	4600		&	99	&	0.287	&	3.962	&	71.49	&	4.150	&	31.75	\\
Ethane & C$_2$H$_6$	&	30.070	&	305.36	&	4880		&	146	&	0.281	&	9.591	&	108.0	&	4.762	&	50.88	\\
Propane & C$_3$H$_8$	&	44.097	&	369.9	&	4250	&	199	&	0.275	&	16.16	&	150.2	&	5.316	&	61.64	\\
Butane & C$_4$H$_{10}$	&	55.124	&	425.2	&	3790	&	257	&	0.276	&	23.94	&	193.6	&	5.785	&	70.85	\\
Pentane & C$_5$H$_{12}$	&	72.151	&	469.7	&	3370	&	310	&	0.268	&	32.86	&	240.5	&	6.219	&	78.27	\\
Hexane & C$_6$H$_{14}$	&	86.178	&	507.5	&	3030	&	366	&	0.263	&	42.67	&	289.1	&	6.612	&	84.57	\\
Heptane & C$_7$H$_{16}$	&	100.21	&	540.1	&	2740	&	428	&	0.261	&	53.44	&	340.2	&	6.981	&	90.00	\\
\end{tabular}
\end{ruledtabular}
\end{table*}

Shown in Table~\ref{Tab:vdWPars} are the van der Waals parameters $a$ and~$b$ per molecule derived from the measured values of $T_{\rm c}$ and~$p_{\rm c}$ using the first two of Eqs.~(\ref{Eq:abN}).    The listed values of $a$ and~$b$ are expressed in units associated with a molecule such as eV and \AA, which are more physically relevant to the molecules composing the fluid than the common units of these quantities, which are, e.g., bar\,(L/mol)$^2$ and L/mol, respectively.  Thus the parameter $b$ is the excluded volume per molecule expressed in units of \AA$^3$, from which the effective diameter per molecule~$d$ in~\AA\ is obtained here as $d = b^{1/3}$ as shown in the table.  From Eq.~(\ref{Eq:CritPars}), the critical volume per molecule is $V_{\rm c}/N = 3b$, which is only a factor of three larger than the excluded volume of a molecule itself.  Shown in the last column of Table~\ref{Tab:vdWPars} is the effective Lennard-Jones intermolecular potential well depth~$\phi_{\rm min}$ in Fig.~\ref{Fig:Lennard_Jones_Potential} calculated from $a$ and~$b$ using Eq.~(\ref{Eq:abratio}).  The values of $\phi_{\rm min}$ are seen to be smallest for He and H$_2$ and largest for H$_2$O and the alkanes.

\begin{figure}[t]
\includegraphics[width=3.in]{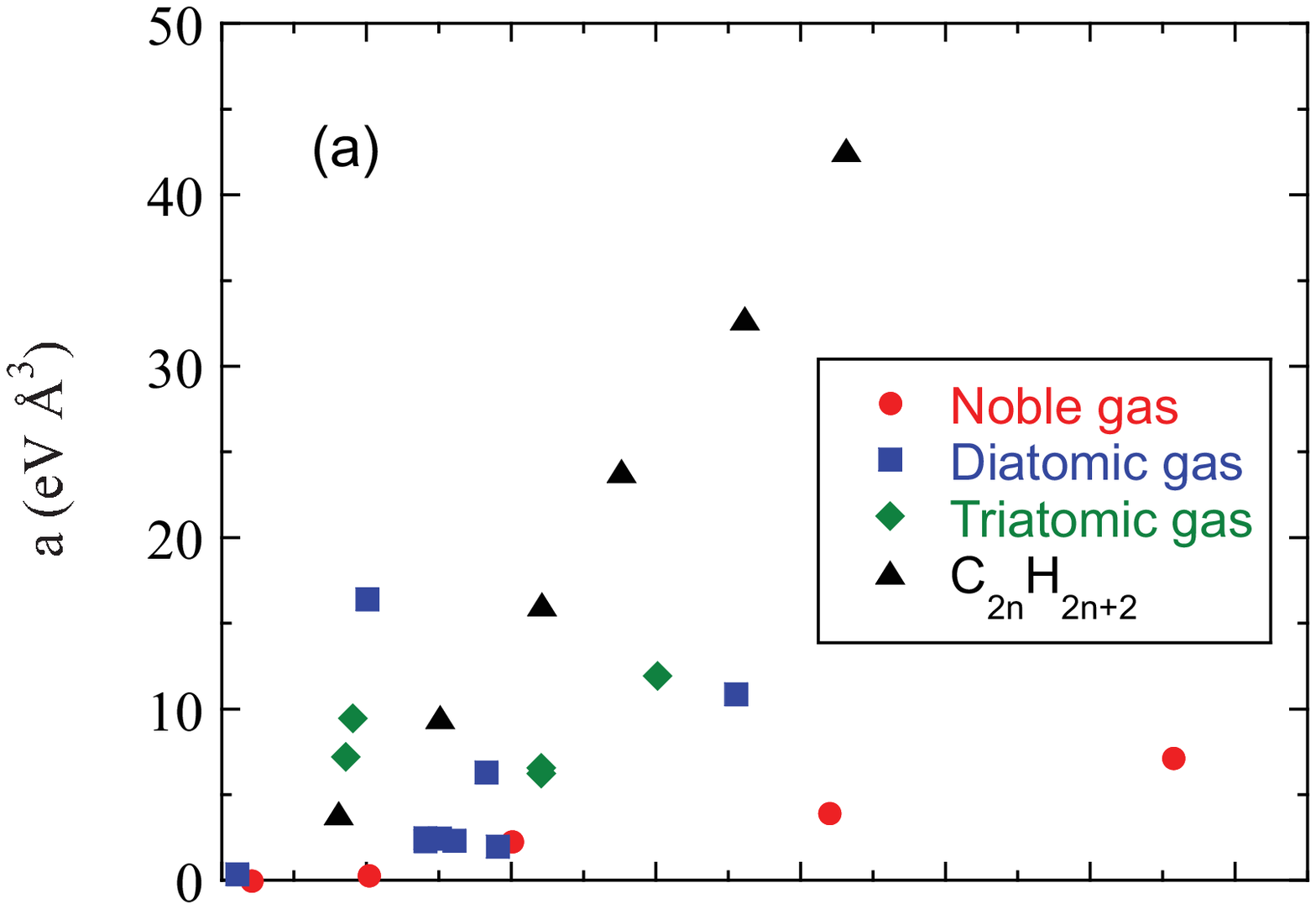}\vspace{-0.21in}
\includegraphics[width=3in]{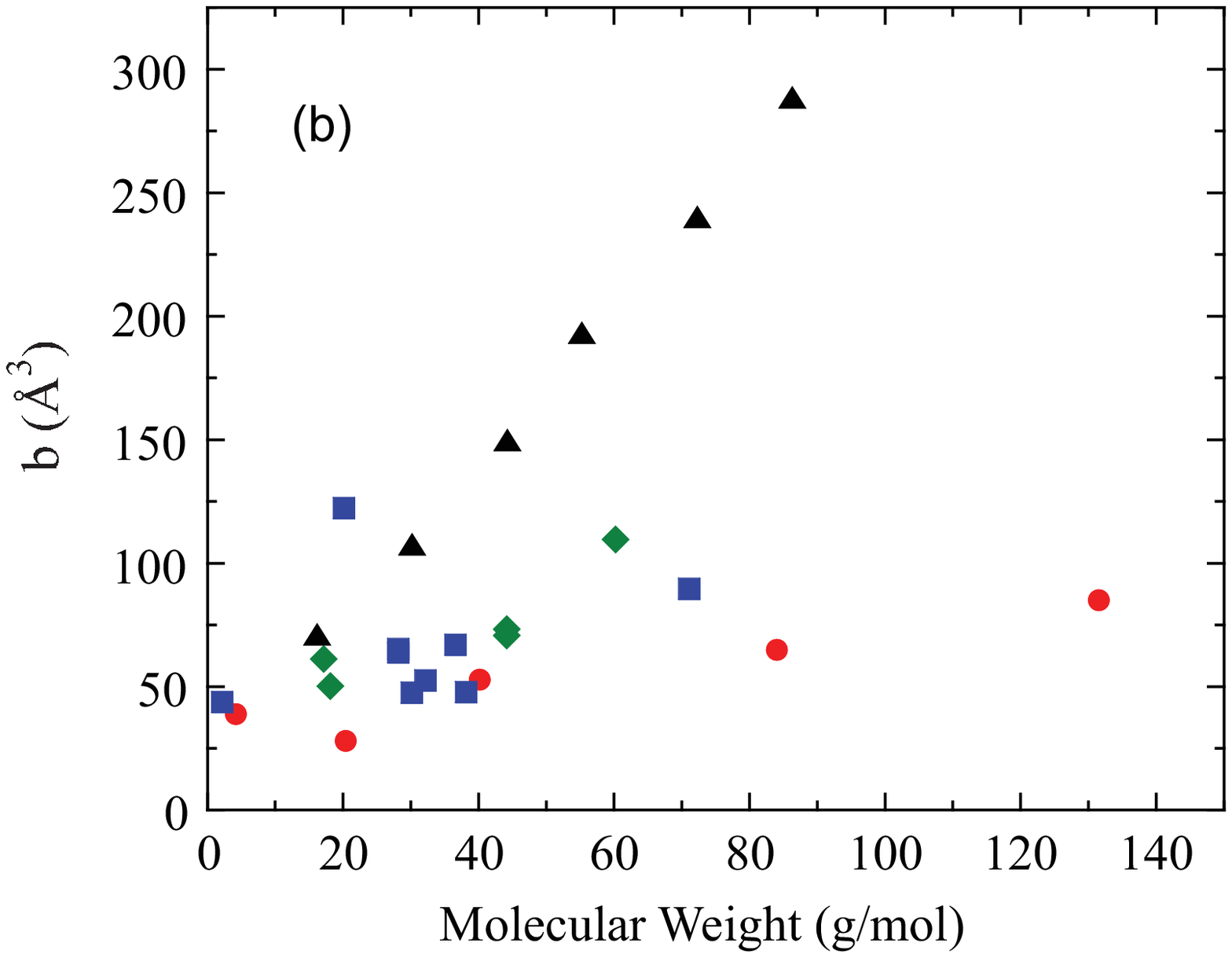}
\caption{(Color online) van der Waals parameters per molecule (a) $a$ and (b) $b$ versus the molecular weight of noble (monatomic) gases, diatomic, triatomic/polyatomic and alkane (C$_n$H$_{2n+2}$) gases in Table~\ref{Tab:vdWPars}. }
\label{Fig:vdW_b_vs_MW}  
\end{figure}

The dependences of~$a$ and~$b$ on the molecular weight (MW) of the various gases in Table~\ref{Tab:vdWPars} are shown in Figs.~\ref{Fig:vdW_b_vs_MW}(a) and \ref{Fig:vdW_b_vs_MW}(b), respectively.  For fluids of one of the types shown, the two parameters~$a$ and~$b$ do not increase monotonically with molecular weight except for the alkanes.

\section{\label{Sec:SUCV} Entropy, Internal Energy and Heat Capacity at Constant Volume}

The entropy of the vdW fluid is calculated using Eq.~(\ref{Eq:FvdW2}) to be
\be
\frac{S}{k_{\rm B}} = -\left(\frac{\partial F}{\partial \tau}\right)_{V,N} = N\left\{\ln\left[\frac{n_Q(V - Nb)}{N}\right] + \frac{5}{2}\right\},
\label{Eq:sigmavdW}
\ee
which is smaller than that of the ideal gas in Eq.~(\ref{Eq:sigmaIG}) because the entropy scales with the free volume, which is smaller in the van der Waals fluid.  In the limits $V\to\infty$ or $b\to0$, Eq.~(\ref{Eq:sigmavdW}) becomes identical to~(\ref{Eq:sigmaIG}).

The internal energy is obtained using Eqs.~(\ref{Eq:FvdW2}) and~(\ref{Eq:sigmavdW}) as
\be
U = F + TS = \frac{3}{2}N\tau -\frac{N^2a}{V},
\label{Eq:UvdV}
\ee
which is lower than that of the ideal gas in Eq.~(\ref{Eq:UIG}) by the attractive potential energy in the second term on the right.  However, because the interaction parameter $a$ is independent of temperature, it does not contribute to the temperature dependence of $U$ given by the first term on the right side of Eq.~(\ref{Eq:UvdV}) which is the same as for the ideal gas in Eq.~(\ref{Eq:UIG}).

Since the temperature dependence of the internal energy of the vdW gas is the same as for the ideal gas,  the heat capacity at constant volume is
\be
C_{\rm V} = \left(\frac{\partial U}{\partial T}\right)_{V,N} = \frac{3}{2}Nk_{\rm B},
\label{Eq:CVvdW}
\ee
which is the same as for the ideal gas in Eq.~(\ref{Eq:CVID}).  This heat capacity  is independent of $T$, so the van der Waals gas is in the classical limit of a quantum Fermi or Bose gas.  Furthermore, the forms of the thermodynamic functions are the same for the pure gas and pure liquid phases of the van der Waals fluid, which only differ in the temperature, pressure and volume regions in which they occur.  Therefore, in particular, the gas and liquid phases discussed below have the same constant value of $C_{\rm V}$.

\section{\label{Sec:PH} Pressure and Enthalpy}

The pressure~$p$ is obtained from the free energy in Eq.~(\ref{Eq:FvdW2}) as
\be
p = -\left(\frac{\partial F}{\partial V}\right)_{V,N} = \frac{N\tau}{V-Nb}- \frac{N^2a}{V^2},
\label{Eq:vdWp}
\ee
As discussed above, the volume $Nb$ is the excluded volume of the incompressible molecules and $V- Nb$ is the free volume in which the molecules can move.  With decreasing volume~$V$, the pressure diverges when $Nb = V$ because then all of the volume is occupied by the total excluded volume of the molecules themselves, and the incompressible hard cores of the molecules are touching.  Therefore the minimum possible volume of the system is $V_{\rm min}=Nb$.  Hence the first term on the right is always positive and the second term negative.  The competition between these two terms in  changing the pressure of the gas, compared to that of an ideal gas at the same temperature and volume, is discussed in Sec.~\ref{Sec:Pvsab} below.

Plots of $p(V)$ at constant temperature using Eq.~(\ref{Eq:vdWp}) have the shapes shown in Fig.~\ref{Fig:vdW_p_vs_V} below.  At the critical point $\tau =\tau_{\rm c}$, $ p = p_{\rm c}$ and $V = V_{\rm c}$, $p(V)$ shows an inflection point where the slope $(\partial p/\partial V)_{\tau}$ and the curvature $(\partial^2 p/\partial V^2)_{\tau}$ are both zero.  From these two conditions one can solve for the the critical temperature $\tau_{\rm c}$ and pressure $p_{\rm c}$ in terms of the van der Waals parameters $a$ and $b$, and then from the equation of state one can solve for the critical volume~$V_{\rm c}$ in terms of $a$, $b$ and~$N$ as given above in Eq.~(\ref{Eq:CritPars}).

Using Eqs.~(\ref{Eq:UvdV}) and~(\ref{Eq:vdWp}), the enthalpy is
\be
H=U+pV = N\left(\frac{3\tau}{2}+ \frac{\tau V}{V - Nb}- \frac{2Na}{V}\right).
\label{Eq:Enthalpy}
\ee
In the limit of large volume~$V$ or small interaction parameters~$a$ and~$b$, one obtains Eq.~(\ref{Eq:HIG}) for the enthalpy of the ideal gas.

\section{\label{Sec:EOSRedCorr}The \lowercase{vd}W Equation of State, Reduced Variables and Laws of Corresponding States}

Equation~(\ref{Eq:vdWp}) can be written
\be
\left(p + \frac{N^2a}{V^2}\right)(V-Nb) = N\tau.
\label{Eq:vdWEoS}
\ee
This is the van der Waals equation of state, which reduces to the ideal gas equation of state (the ideal gas law) $pV=N\tau$ when the molecular interaction parameters $a$ and~$b$ are zero.

Using Eqs.~(\ref{Eq:abN}), one can write Eq.~(\ref{Eq:vdWEoS}) as
\be
\left[\frac{p}{p_{\rm c}} + \frac{3}{(V/V_{\rm c})^2}\right]\left(\frac{V}{V_{\rm c}} - \frac{1}{3}\right) = \frac{8\tau}{3\tau_{\rm c}}.
\label{Eq:vdWRed1}
\ee
Note that $N$ has disappeared as a state variable from this equation.  Following the notation in Ref.~\onlinecite{Kittel1980}, we define the reduced variables
\be
\hat{p} \equiv \frac{p}{p_{\rm c}},\qquad \widehat{V} \equiv \frac{V}{V_{\rm c}},\qquad \hat{\tau} \equiv \frac{\tau}{\tau_{\rm c}} = \frac{T}{T_{\rm c}}.
\label{Eq:RedVar}
\ee 
Then Eq.~(\ref{Eq:vdWRed1}) becomes
\be
\left(\hat{p} + \frac{3}{\widehat{V}^2}\right)\left(3\widehat{V} - 1\right) = 8\hat{\tau},
\label{Eq:vdWRed2}
\ee
which is the vdW equation of state written in reduced variables.  When two fluids are in ``corresponding states'', they have the same set of three reduced parameters $\hat{p}$, $\widehat{V}$ and~$\hat{\tau}$.  The differences between $p_{\rm c}$, $V_{\rm c}$ and $\tau_{\rm c}$ of different fluids are subsumed into the reduced parameters $\hat{p},\ \widehat{V}$  and~$\hat{\tau}$.  Therefore Eq.~(\ref{Eq:vdWRed2}) is an example of a ``law of corresponding states'' which is obeyed by all van der Waals fluids.  Many other laws of corresponding states are derived below for the vdW fluid.  From Eq.~(\ref{Eq:vdWRed2}), the pressure versus volume and temperature is expressed in reduced variables as
\be
\hat{p} = \frac{8\hat{\tau}}{3\widehat{V} - 1}- \frac{3}{\widehat{V}^2}.
\label{Eq:RedpVsRedV}
\ee
Thus with decreasing $\widehat{V}$, $\hat{p}$ diverges at $\widehat{V} = 1/3$, which is the reduced volume at which the entire volume occupied by the fluid is filled with the hard-core molecules with no free volume remaining.

\begin{figure}[t]
\includegraphics[width=3.3in]{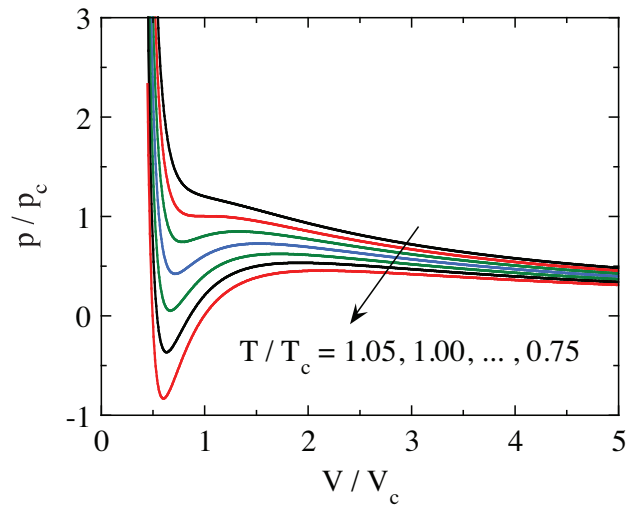}
\caption{(Color online) Reduced pressure $\hat{p}\equiv p/p_{\rm c}$ versus reduced volume $\widehat{V} \equiv V/V_{\rm c}$ at several values of reduced temperature $\hat{\tau} \equiv \tau/\tau_{\rm c} = T/T_{\rm c}$ according to Eq.~(\ref{Eq:RedpVsRedV}).  The region on the far left corresponds to the liquid phase and the region to the far right corresponds to the gas or fluid phase depending on the temperature.  The regions of negative pressure and positive $d\hat{p}/d\widehat{V}$ are unphysical and correspond to regions of coexistence of the gas and liquid phases that are not taken into account in this figure.}
\label{Fig:vdW_p_vs_V}  
\end{figure}

\begin{figure}[t]
\includegraphics[width=3.3in]{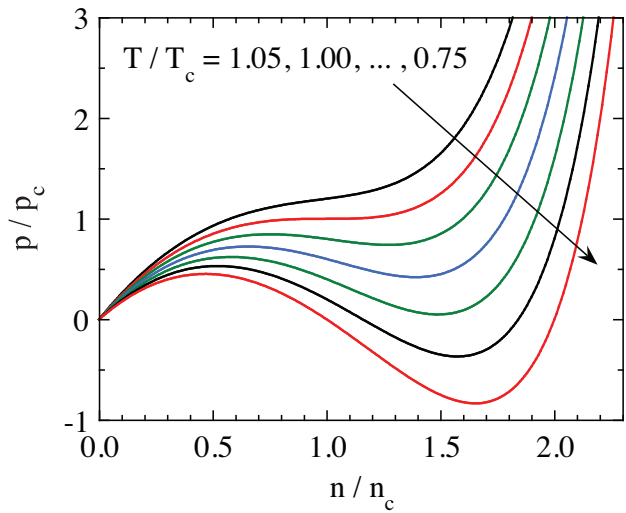}
\caption{(Color online) Reduced pressure $\hat{p}\equiv p/p_{\rm c}$ versus reduced number density $\hat{n} \equiv n/n_{\rm c}$ at several values of reduced temperature $\hat{\tau}$ according to Eq.~(\ref{Eq:RedpVsRedn}).  The region on the far left corresponds to the gas or fluid phase and the region to the far right corresponds to the liquid phase depending on the temperature.  As in Fig.~\ref{Fig:vdW_p_vs_V}, the regions of negative pressure and negative $d\hat{p}/d\hat{n}$ are unphysical for a homogeneous fluid and correspond to regions of coexisting gas and liquid.}
\label{Fig:vdW_p_vs_n}  
\end{figure}

\begin{figure}[t]
\includegraphics[width=3.3in]{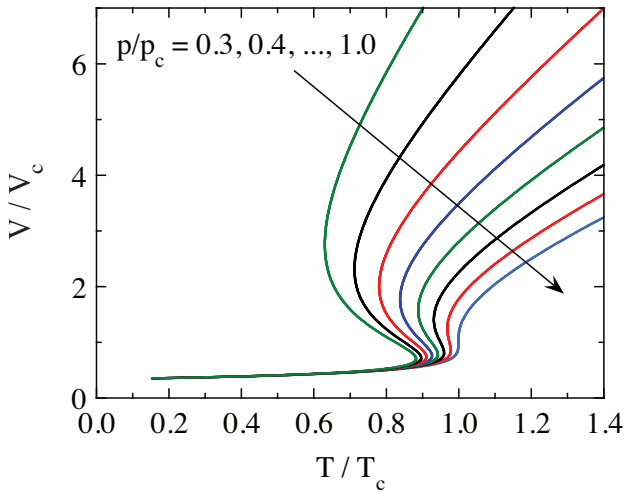}
\caption{(Color online) Reduced volume $\widehat{V}=V/V_{\rm c}$ versus reduced temperature $\hat{\tau}$ at several values of reduced pressure $\hat{p}$  according to Eq.~(\ref{Eq:RedpVsRedV}).  The regions of negative $d\widehat{V}/d\hat{\tau}$ are unphysical and correspond to regions of coexisting gas and liquid phases.}
\label{Fig:vdW_V_T_isobars}  
\end{figure}

Using Eq.~(\ref{Eq:RedpVsRedV}), $\hat{p}$ versus~$\widehat{V}$ isotherms at several temperatures~$\hat{\tau}$ are plotted in Fig.~\ref{Fig:vdW_p_vs_V}.  One notices that for $\hat{\tau}>1$ ($\tau>\tau_{\rm c}$), the pressure monotonically decreases with increasing volume.  This temperature region corresponds to a ``fluid'' region where gas and liquid cannot be distinguished.  At $\tau < \tau_{\rm c}$ the isotherms show unphysical (unstable) behaviors in which the pressure increases with increasing volume over a certain range of $p$ and $V$\@.  This unstable region forms part of the volume region where liquid and gas coexist in equilibrium as further discussed below.

The order parameter for the liquid-gas phase transition is the difference in the number density $n = N/V$ between the liquid and gas phases.\cite{Kadanoff1967, Stanley1971} Using Eq.~(\ref{Eq:abN}), one has
\bse
\label{Eqs:nDefs}
\be
n = \frac{N}{V} = \frac{8p_{\rm c}V_{\rm c}}{3T_{\rm c}V} = \frac{8p_{\rm c}}{3T_{\rm c}\widehat{V}},
\label{Eq:nDef}
\ee
where Eq.~(\ref{Eq:RedVar}) was used to obtain the last equality.  The value of $n_{\rm c}$ at the critical point is obtained by setting $\widehat{V} = 1$, yielding
\be
n_{\rm c} = \frac{8p_{\rm c}}{3T_{\rm c}}.
\label{Eq:ncDef}
\ee
\ese
The reduced form of the number density analogous to those in Eq.~(\ref{Eq:RedVar}) is obtained from Eqs.~(\ref{Eqs:nDefs}) as
\be
\hat{n}\equiv \frac{n}{n_{\rm c}} = \frac{1}{\widehat{V}}.
\label{Eq:nhatDef}
\ee
Using this expression, one can write Eq.~(\ref{Eq:RedpVsRedV}) in terms of the reduced fluid number density as
\be
\hat{p} = \frac{8\hat{\tau}\hat{n}}{3 - \hat{n}}- 3\hat{n}^2,
\label{Eq:RedpVsRedn}
\ee
with the restriction $\hat{n}<3$ due to the excluded volume of the fluid.  Isotherms of $\hat{p}$ versus $\hat{n}$ are shown in Fig.~\ref{Fig:vdW_p_vs_n}.  The unphysical regions where $\hat{p}<0$ and $d\hat{p}/d\hat{n}<0$ correspond to similar regions in Fig.~\ref{Fig:vdW_p_vs_V}.

Volume versus temperature isobars are shown in Fig.~\ref{Fig:vdW_V_T_isobars}.  Some of these show unphysical regions as in Figs.~\ref{Fig:vdW_p_vs_V} and~\ref{Fig:vdW_p_vs_n} that are associated with coexisting gas and liquid phases as discussed in Sec.~\ref{Sec:ChemPot}.

\subsection{\label{Sec:Pvsab} Influence of the vdW Interactions on the Pressure of the Gas/Fluid Phase}

There has been much discussion in the literature and books about whether the interactions between the molecules in the vdW gas increase the pressure or decrease the pressure of the gas compared to that of a (noninteracting) ideal gas at the same temperature and volume.  For example Stanley\cite{Stanley1971} and Berberan-Santos et al.\cite{Berberan-Santos2008} state that the pressure decreases below that of the ideal gas due to the attactive interaction~$a$, whereas Kittel and Kroemer\cite{Kittel1980} claim that the pressure increases.  Others give no clear opinion.\cite{Schroeder2000}  Tuttle has reviewed the history of this controversy, including a quote from van der Waals himself who evidently claimed that the pressure decreases.\cite{Tuttle1975} Implicit in these statements is that the temperature and volume of the gas are not relevant to the argument as long as one is in the gas-phase or supercritical fluid part of the phase diagram.  Here we show quantitatively that the same vdW interactions $a$ and~$b$ can both increase and decrease the pressure in the same vdW gas compared to the ideal gas, depending on the temperature and volume of the gas.

The compression factor~$Z$ of a gas is defined in Eq.~(\ref{Eq:pcVcTc}) above as
\be
Z \equiv \frac{pV}{N\tau}.
\ee
For the ideal gas one has $Z_{\rm IG}  = 1$.  Using Eq.~(\ref{Eq:vdWp}), the compression factor of the vdW gas is
\be
Z = \frac{1}{1-Nb/V} - \frac{Na}{V\tau}.
\ee
The deviation of $Z$ from $Z_{\rm IG}$ is then
\be
Z - 1 = \frac{Nb/V}{1-Nb/V} - \frac{Na}{V\tau}.
\label{Eq:z-1}
\ee
In the present discussion the temperature and volume are constant as the vdW interactions are turned on and the right side of Eq.~(\ref{Eq:z-1}) becomes nonzero.  One sees from Eq.~(\ref{Eq:z-1}) that increasing~$b$ increases the pressure and increasing~$a$ decreases the pressure, where $a/b$ is a fixed value for a given gas according to Eq.~(\ref{Eq:abratio}).  Therefore a competition occurs between these two effects on the pressure as the interactions are turned on.  In reduced variables, Eqs.~(\ref{Eq:abN}) give
\be
\frac{Nb}{V} = \frac{1}{3\widehat{V}},\qquad \frac{Na}{V\tau} = \frac{9}{8\widehat{V}\hat{\tau}}. 
\ee
Inserting these expressions into Eq.~(\ref{Eq:z-1}) gives
\be
Z - 1 = \frac{\frac{1}{3\widehat{V}}}{1-\frac{1}{3\widehat{V}}} - \frac{9}{8\widehat{V}\hat{\tau}}.
\label{Eq:z-1B}
\ee
One has the limits $0 < 1/\widehat{V} < 3$.  We recall that the first term on the right side of Eq.~(\ref{Eq:z-1B}) arises from the parameter $b$ and the second one from $a$, rewritten in terms of reduced variables. The right side is zero if $a=b=0$.  

\begin{figure}[t]
\includegraphics[width=3.3in]{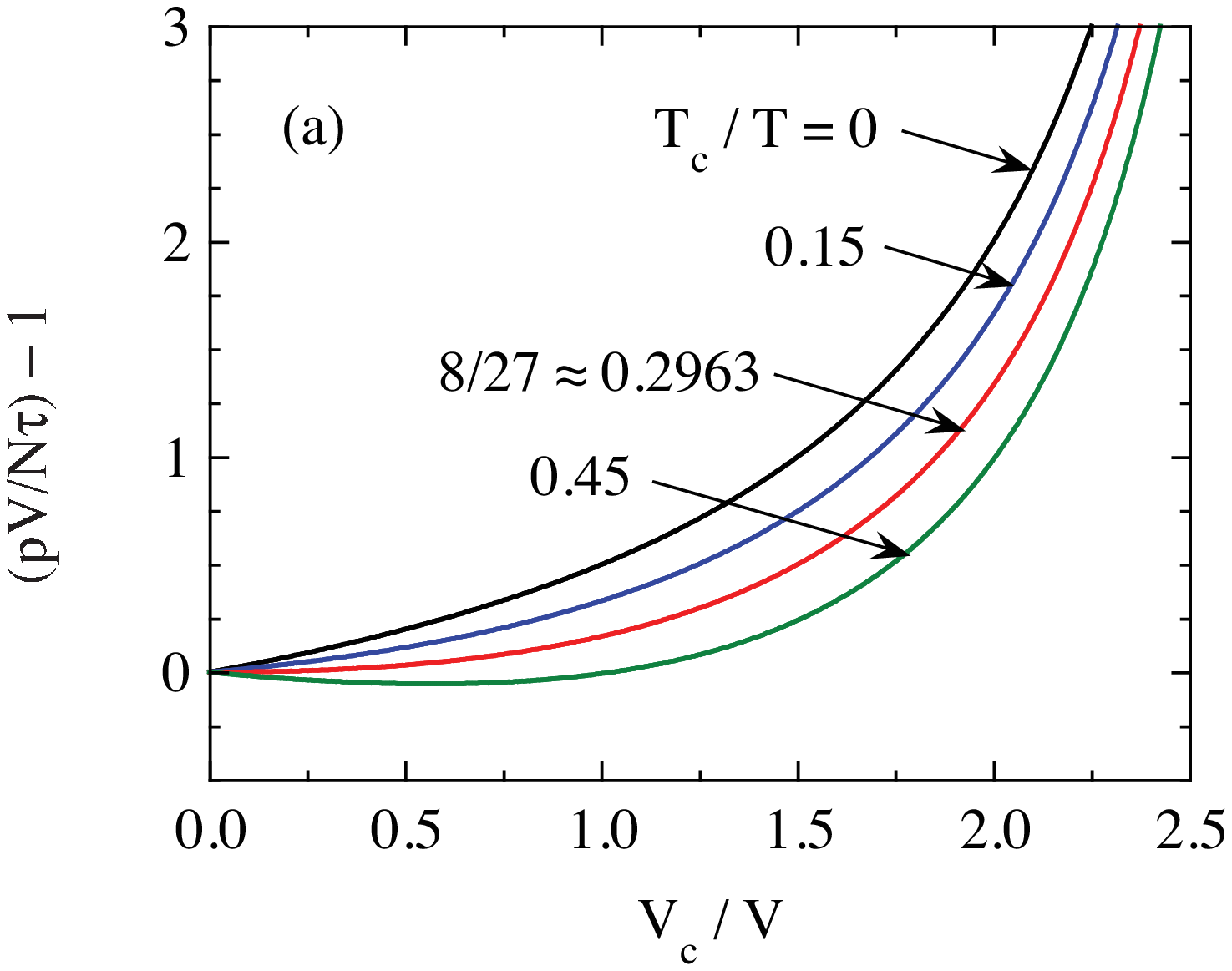}
\includegraphics[width=3.3in]{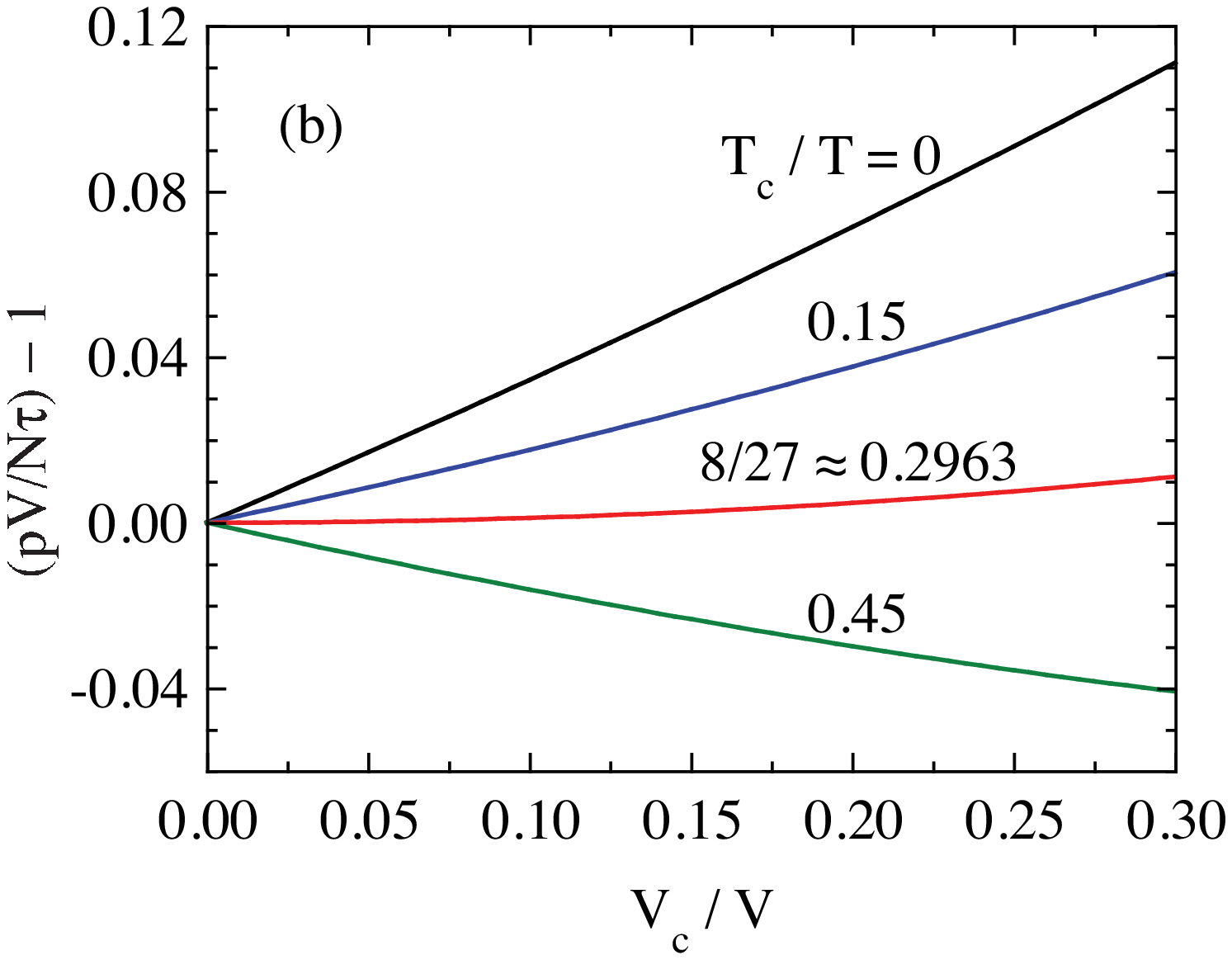}
\caption{(Color online) (a) Isotherms of the quantity $(pV/N\tau) - 1$ versus inverse reduced volume $1/\widehat{V} = V_{\rm c}/V$ at the inverse temperatures $1/\hat{\tau} = T_{\rm c}/T$ as indicated.  (b) Expanded plots at small values of $1/\widehat{V}$ for the same values of $1/\hat{\tau}$ as in~(a).}
\label{Fig:vdW_p_vs_ab}  
\end{figure}

Shown in Fig.~\ref{Fig:vdW_p_vs_ab}(a) are isotherms of $Z-1$ versus $1/\widehat{V}$ plotted using Eq.~(\ref{Eq:z-1B}) at several values of $1/\hat{\tau}$ as indicated.  Expanded isotherms at low values of $1/\widehat{V}$ and~$1/\hat{\tau}$ are shown in Fig.~\ref{Fig:vdW_p_vs_ab}(b).  One sees from Fig.~\ref{Fig:vdW_p_vs_ab}(b) that if $\hat{\tau}$ is below a certain value $\hat{\tau}_{\rm max}$, the pressure of the vdW gas is smaller than that of the ideal gas for a range of inverse volumes.  There is a crossover at $\hat{\tau}_{\rm max}$ where the initial slope $\partial(Z-1)/\partial(1/\widehat{V}) $ goes from positive to negative with decreasing values of $\hat{\tau}$.  By solving $\partial(Z-1)/\partial(1/\widehat{V}) = 0$ using Eq.~(\ref{Eq:z-1B}), the crossover occurs at
\be
\hat{\tau}_{\rm max} = \frac{27}{8},
\label{Eq:tau0}
\ee
as indicated in Fig.~\ref{Fig:vdW_p_vs_ab}.  Thus if the temperature of a vdW gas is less than $\hat{\tau}_{\rm max}$, there is a range of inverse volumes over which the molecular interactions cause the pressure to be less than that of the ideal gas, whereas if the temperature is greater than~$\hat{\tau}_{\rm max}$, the interactions increase the pressure irrespective of the value of the inverse volume.  To find this maximum value of the inverse volume  versus~$\hat{\tau}$, one can set $Z-1=0$ in Eq.~(\ref{Eq:z-1B}) and solve for $(1/\widehat{V})_{\rm max}$, yielding
\be
\left(\frac{1}{\widehat{V}}\right)_{\rm max} = 3 - \frac{8\hat{\tau}}{9}\quad(\hat{\tau} < \hat{\tau}_{\rm max}).
\label{Eq:Vmin}
\ee
If $\hat{\tau}<1$, the possibility of liquifaction of the gas exists as discussed below, so the discussion here refers only to the gas phase in this temperature range.

We conclude that the same vdW interaction parameters can give rise to either an increase or a decrease in the pressure of a vdW gas or supercritical fluid relative to that of an ideal gas at the same volume and temperature, depending on the values of the volume and temperature.

\subsection{\label{Sec:Boyle} Boyle Temperature}

From Eq.~(\ref{Eq:z-1B}), the temperature~$\hat{\tau}_{\rm B}$  at which $Z - 1 = 0$, at which the net effect of the molecular interactions on the compression factor compared to that of the ideal gas is zero, is
\be
\hat{\tau}_{\rm B} = \frac{9}{8}\left(3 - \frac{1}{\widehat{V}}\right),
\label{Eq:TBoyle}
\ee
where $\hat{\tau}_{\rm B}$ is known as the Boyle temperature.  At large volumes the Boyle temperature approaches the limit $\hat{\tau}_{\rm B\,max} = \hat{\tau}_{\rm max} = 27/8$ in Eq.~(\ref{Eq:tau0}) and it decreases monotonically from there with decreasing volume.  For the minimum value of $\widehat{V}$ of 1/3 (at which the free volume goes to zero), Eq.~(\ref{Eq:TBoyle}) gives the minimum value of the Boyle temperature as $\hat{\tau}_{\rm B\,min}=0$.

It is sometimes stated that the Boyle temperature is the temperature at which a gas with molecular interactions  behaves like an ideal gas.  This definition is misleading, because it only applies to the compression factor and not to thermodynamic properties like the heat capacity at constant pressure~$C_{\rm p}$, the isothermal compressibility~$\kappa_{\rm T}$ or the coefficient of volume expansion~$\alpha$.  We show in the following Sec.~\ref{Eq:PropsRedVars} that the molecular interactions have nonzero influences on these thermodynamic properties at all finite temperatures and volumes of the vdW fluid.

\subsection{\label{Eq:PropsRedVars} Internal Energy, Helmholtz Free Energy, Entropy, Isothermal Compressibility, Thermal Expansion Coefficient, Heat Capacity at Constant Pressure and Latent Heat of Vaporization Expressed in Reduced Variables}

One can write the internal energy in Eq.~(\ref{Eq:UvdV}) in terms of the reduced variables in Eqs.~(\ref{Eqs:RedVars}) and also in terms of $\hat{n}=1/\widehat{V}$ defined in Eq.~(\ref{Eq:nhatDef}) as  
\be
\frac{U}{p_{\rm c}V_{\rm c}} = 4\hat{\tau}-\frac{3}{\widehat{V}} = 4\hat{\tau}-3\widehat{n}.
\label{Eq:URed}
\ee
At the critical point, one obtains
\be
\frac{U_{\rm c}}{p_{\rm c}V_{\rm c}} = 1\qquad(\hat{\tau}=\widehat{V}=\hat{p} = 1).
\label{Eq:Ucrit}
\ee
It is also useful to write $F$ in terms of reduced variables.  We first write the quantum concentration in Eq.~(\ref{Eq:nQ}) as
\be
n_{Q} = n_{Q{\rm c}}\, \hat{\tau}^{3/2},
\label{Eq:nQ2}
\ee
where 
\be
n_{Q{\rm c}} \equiv \left(\frac{m\tau_{\rm c}}{2\pi\hbar^2}\right)^{3/2}.
\ee

In terms of the reduced variables, the Helmholtz free energy in Eq.~(\ref{Eq:FvdW2}) becomes
\bse
\label{Eq:FRed}
\be
\frac{F}{p_{\rm c}V_{\rm c}} = -\frac{8\hat{\tau}}{3} \left\{ \ln\left[ x_{\rm c}\,\hat{\tau}^{3/2}\left(3\widehat{V}-1\right)\right]+1\right\} - \frac{3}{\widehat{V}}
\ee
where the dimensionless variable $x_{\rm c}$ is
\be
x_{\rm c}\equiv \frac{n_{Q{\rm c}} \tau_{\rm c}}{8p_{\rm c}},
\ee
\ese
the entropy in Eq.~(\ref{Eq:sigmavdW}) becomes
\bea
\frac{S}{Nk_{\rm B}} &=& \ln\left[x_{\rm c}\,\hat{\tau}^{3/2}\left(3\widehat{V}-1\right)\right] + \frac{5}{2}\label{Eq:sigmaRed}\\*
&=& \ln\left[x_{\rm c}\,\hat{\tau}^{3/2}\left(3-\hat{n}\right)/\hat{n}\right] + \frac{5}{2}\nonumber,
\eea
and the enthalpy in Eq.~(\ref{Eq:Enthalpy}) becomes
\be
\frac{H}{p_{\rm c}V_{\rm c}} = \frac{4\hat{\tau}\left(5\widehat{V}-1\right)}{3\widehat{V}-1} - \frac{6}{\widehat{V}} = \frac{4\hat{\tau}\left(5-\hat{n}\right)}{3-\hat{n}} - 6\hat{n}.
\label{Eq:EnthalpyRed}
\ee
The entropy diverges to $-\infty$ at $\hat{\tau}\to0$, which violates the third law of thermodynamics and thus shows that the vdW fluid is in the classical regime just as the ideal gas is.  At the critical point $\hat{\tau}=\hat{p}=\widehat{V}=\hat{n}=1$, the enthalpy~$H_{\rm c}$ is given by Eq.~(\ref{Eq:EnthalpyRed}) as
\be
\frac{H_{\rm c}}{p_{\rm c}V_{\rm c}} = 2 \quad{\rm (at\ the\ critical\ point)}.
\label{Eq:Hcrit}
\ee

Equations~(\ref{Eq:URed}) and~(\ref{Eq:EnthalpyRed}) are laws of corresponding states.  However Eqs.~(\ref{Eq:FRed}) and~(\ref{Eq:sigmaRed}) are not because they explicitly depend on the mass~$m$ of the molecules in the particular fluid considered.  On the other hand, the change in entropy per particle $\Delta S/(Nk_{\rm B})$ from one reduced state of a vdW fluid to another is a law of corresponding states.  Taking the reference state to be the critical point at which $\hat{\tau}^{3/2}\left(3\widehat{V}-1\right)=2$, Eq.~(\ref{Eq:sigmaRed}) yields
\bea
\frac{\Delta S}{Nk_{\rm B}} &\equiv& \frac{S(\hat{\tau},\widehat{V}) - S(1,1)}{Nk_{\rm B}}\label{Eq:DeltasigmaRed}\\*
&=& \ln\left[\hat{\tau}^{3/2}\left(3\widehat{V}-1\right)/2\right]\nonumber\\*
&=& \ln\left[\hat{\tau}^{3/2}\left(3-\hat{n}\right)/(2\hat{n})\right].\nonumber
\eea

The isothermal compressibility $\kappa_{\rm T}$ is given by Eq.~(\ref{Eq:kappTIG}).  In the reduced units in Eq.~(\ref{Eq:RedVar}) one obtains
\bse
\be
\frac{1}{\kappa_{\rm T} p_{\rm c}} = -\widehat{V}\left(\frac{\partial \hat{p}}{\partial \widehat{V}}\right)_{\hat{\tau}}.
\label{Eq:kappaTRed}
\ee
One can write the partial derivative on the right side as\cite{Stanley1971}
\be
\left(\frac{\partial \hat{p}}{\partial \widehat{V}}\right)_{\hat{\tau}} = -\left(\frac{\partial \hat{p}}{\partial \hat{\tau}}\right)_{\widehat{V}}\left(\frac{\partial \hat{\tau}}{\partial \widehat{V}}\right)_{\hat{p}},
\ee
so Eq.~(\ref{Eq:kappaTRed}) can also be written
\be
\frac{1}{\kappa_{\rm T} p_{\rm c}} = \widehat{V}\left(\frac{\partial \hat{p}}{\partial \hat{\tau}}\right)_{\widehat{V}}\left(\frac{\partial \hat{\tau}}{\partial \widehat{V}}\right)_{\hat{p}}.
\label{Eq:kappaTRed23}
\ee
\ese

Utilizing the expression for the reduced pressure for the van der Waals fluid in Eq.~(\ref{Eq:RedpVsRedV}), Eq.~(\ref{Eq:kappaTRed}) gives
\bse
\be
\kappa_{\rm T} p_{\rm c} = \frac{(3\widehat{V}-1)^2\widehat{V}^2/6}{4\hat{\tau}\widehat{V}^3 - (3\widehat{V}-1)^2}.
\label{Eq:kappaTRedvdW}
\ee
In terms of $\hat{n}=1/\widehat{V}$, Eq.~(\ref{Eq:kappaTRedvdW}) becomes
\be
\kappa_{\rm T} p_{\rm c} = \frac{(3-\hat{n})^2/(6\hat{n})}{4\hat{\tau} - \hat{n}(3-\hat{n})^2}.
\label{Eq:kappaTRedvdW2}
\ee
Using Eq.~(\ref{Eq:pcVcTc}), a Taylor series expansion of Eq.~(\ref{Eq:kappaTRedvdW}) in powers of $1/\widehat{V}$ gives
\be
\kappa_{\rm T} = \frac{3\widehat{V}}{8\hat{\tau}p_{\rm c}}\left[1+\frac{27-8\hat{\tau}}{12\hat{\tau}\widehat{V}}+ {\cal O}\left(\frac{1}{\widehat{V}^2}\right)\right],
\label{Eq:Kappa1/VExpand}
\ee
where the prefactor is
\be
\frac{3\widehat{V}}{8\hat{\tau}p_{\rm c}} = \frac{V}{Nk_{\rm B}T},
\label{Eq:kappaBigV}
\ee
\ese
which is the result for the ideal gas in Eq.~(\ref{Eq:kappTIG}) that $\kappa_T=1/p = V/(Nk_{\rm B}T)$.  Thus in the limit $\widehat{V}\to\infty$ one obtains the expression for the ideal gas.  

The volume thermal expansion coefficient $\alpha$ is defined in Eq.~(\ref{Eq:alphaDef}).  In reduced units one has
\be
\frac{\alpha\tau_{\rm c}}{k_{\rm B}} = \frac{1}{\widehat{V}}\left(\frac{\partial \widehat{V}}{\partial \hat{\tau}}\right)_{\hat{p}}.
\label{Eq:alphaRed}
\ee
Comparing Eqs.~(\ref{Eq:kappaTRed23}) and~(\ref{Eq:alphaRed}) shows that 
\be
\frac{\alpha\tau_{\rm c}/k_{\rm B}}{\kappa_{\rm T} p_{\rm c}} = \left(\frac{\partial \hat{p}}{\partial \hat{\tau}}\right)_{\widehat{V}}.
\label{Eq:alphaOVERkappaT}
\ee
Utilizing the expression for the reduced pressure of the van der Waals fluid in Eq.~(\ref{Eq:RedpVsRedV}), Eq.~(\ref{Eq:alphaRed}) gives
\bse
\be
\frac{\alpha\tau_{\rm c}}{k_{\rm B}} = \frac{4(3\widehat{V}-1)\widehat{V}^2/3}{4\hat{\tau}\widehat{V}^3 - (3\widehat{V}-1)^2}.
\label{Eq:alphavdW}
\ee
In terms of the reduced number density $\hat{n}=1/\widehat{V}$ one obtains
\be
\frac{\alpha\tau_{\rm c}}{k_{\rm B}} = \frac{4(3-\hat{n})/3}{4 \hat{\tau}- \hat{n}(3-\hat{n})^2}.
\label{Eq:alphavdW2}
\ee
A Taylor series expansion of Eq.~(\ref{Eq:alphavdW}) in powers of $1/\widehat{V}$ gives
\be
\frac{\alpha\tau_{\rm c}}{k_{\rm B}} = \frac{1}{\hat{\tau}}\left[1 + \frac{27-4\hat{\tau}}{12\hat{\tau}\widehat{V}} + {\cal O}\left(\frac{1}{\widehat{V}^2}\right)\right].
\label{Eq:alphaExpand1/V}
\ee
In the limit of large volumes or small concentrations one obtains
\be
\frac{\alpha\tau_{\rm c}}{k_{\rm B}} = \frac{1}{\hat{\tau}}\qquad (\widehat{V}\to\infty),
\label{Eq:alphavdWtoInfty}
\ee
\ese
which agrees with the ideal gas value for the thermal expansion coefficient in Eq.~(\ref{Eq:alphaDef}).  

Comparing Eqs.~(\ref{Eq:alphavdW}) and~(\ref{Eq:kappaTRedvdW}) shows that the dimensionless reduced values of $\kappa_T$ and~$\alpha$ are simply related according to
\be
\frac{\alpha\tau_{\rm c}/k_{\rm B}}{\kappa_{\rm T} p_{\rm c}} = \frac{8}{3\widehat{V}-1} = \frac{8\hat{n}}{3-\hat{n}},
\label{Eq:alphakapparat}
\ee
which using the expression~(\ref{Eq:RedpVsRedV}) for the pressure is seen to be in agreement with the general Eq.~(\ref{Eq:alphaOVERkappaT}).

The heat capacity at constant pressure $C_{\rm p}$ and at constant volume $C_{\rm V}$ are related according to the thermodynamic relation in Eq.~(\ref{Eq:CpIG}).  In reduced units, this equation becomes
\bse
\be
C_{\rm p} - C_{\rm V} = k_{\rm B}\left(\frac{p_{\rm c}V_{\rm c}}{\tau_{\rm c}}\right)\frac{\hat{\tau}\widehat{V}(\alpha\tau_{\rm c}/k_{\rm B})^2}{(\kappa_T p_{\rm c})}.
\ee
Using Eq.~(\ref{Eq:pcVcTc}) one then obtains
\be
\frac{C_{\rm p} - C_{\rm V}}{Nk_{\rm B}} = \frac{3\hat{\tau}\widehat{V}(\alpha\tau_{\rm c}/k_{\rm B})^2}{8(\kappa_T p_{\rm c})}.
\label{Eq:CpCVvdW}
\ee
The expression for $C_{\rm V}$ in Eq.~(\ref{Eq:CVvdW}) then gives
\be
\frac{C_{\rm p} }{Nk_{\rm B}} = \frac{3}{2} + \frac{3\hat{\tau}\widehat{V}(\alpha\tau_{\rm c}/k_{\rm B})^2}{8(\kappa_T p_{\rm c})}.
\label{Eq:CpvdW}
\ee
\ese

Utilizing Eqs.~(\ref{Eq:alphavdW}) and~(\ref{Eq:alphakapparat}), the heat capacity at constant pressure in Eq.~(\ref{Eq:CpvdW}) for the vdW fluid simplifies to
\bse
\be
\frac{C_{\rm p} }{Nk_{\rm B}} = \frac{3}{2} + \frac{1}{1 - \frac{(3\widehat{V} -1)^2}{4\hat{\tau}\widehat{V}^3}}.
\label{Eq:CpvdW2}
\ee
The $C_{\rm p}$ can be written in terms of the reduced number density $\hat{n}=1/\widehat{V}$ as
\be
\frac{C_{\rm p} }{Nk_{\rm B}} = \frac{3}{2} + \frac{4\hat{\tau}}{4\hat{\tau} - \hat{n}(3-\hat{n})^2}.
\label{Eq:CpvdW3}
\ee
A Taylor series expansion of Eq.~(\ref{Eq:CpvdW2}) in powers of $1/\widehat{V}$ gives
\be
\frac{C_{\rm p} }{Nk_{\rm B}} = \frac{5}{2} + \frac{9}{4\hat{\tau}\widehat{V}} + {\cal O}\left(\frac{1}{\widehat{V}^2}\right).
\label{Eq:CpvdW1/VExpand}
\ee
\ese
In the limit $\widehat{V}\to\infty$, this equation gives the ideal gas expression for $C_{\rm p}$ in Eq.~(\ref{Eq:CpIG}). 

As one approaches the critical point with $\hat{p}\to1$, $\widehat{V}\to1$, $\hat{n}\to1$ and $\hat{\tau}\to1$, one obtains $\kappa_{\rm T},\ \alpha,\ C_{\rm p}\to\infty$.  These critical behaviors will be quantitatively discussed in Sec.~\ref{Sec:CritExps} below.  

The latent heat (enthalpy) of vaporization~$L$ is defined as
\be
L = T\Delta S_{\rm X},
\ee
where $\Delta S_{\rm X}$ is the change in entropy of the system when liquid is completely converted to gas at constant temperature.  Using Eqs.~(\ref{Eq:pcVcTc}) and~(\ref{Eq:RedVar}), $L$ can be written in dimensionless reduced form as
\be
\frac{L}{p_{\rm c}V_{\rm c}} = \left(\frac{8\hat{\tau}}{3}\right)\frac{\Delta S_{\rm X}}{Nk_{\rm B}}.
\label{Eq:ReducedL}
\ee

\section{\label{Sec:ChemPot} Chemical Potential}

\begin{figure}[t]
\includegraphics[width=3.3in]{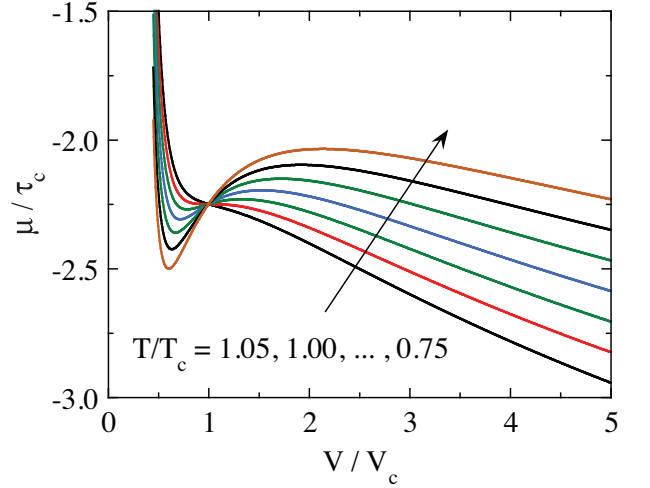}
\caption{(Color online) Reduced chemical potential $\mu/\tau_{\rm c}$ versus reduced volume $\widehat{V} \equiv V/V_{\rm c}$ at several values of reduced temperature $\hat{\tau} \equiv \tau/\tau_{\rm c}$ according to Eq.~(\ref{Eq:muNorm2}).  The region on the far left corresponds to the liquid phase and the region to the far right corresponds to the gas phase, with a region of coexistence between them.}
\label{Fig:vdW_mu_versus_V}  
\end{figure}

\begin{figure}[t]
\includegraphics[width=3.3in]{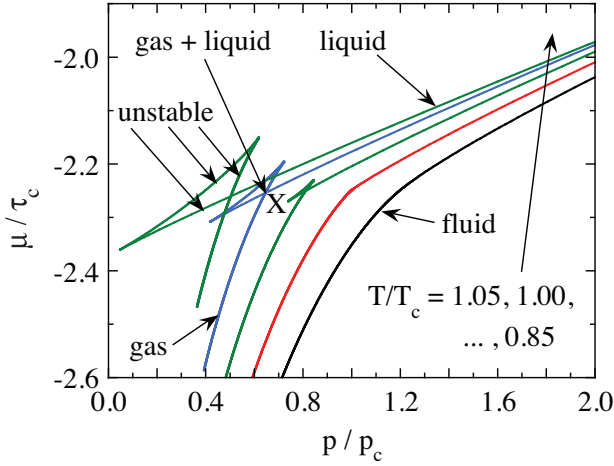}
\caption{(Color online) Reduced chemical potential $\mu/\tau_{\rm c}$ versus reduced pressure $\hat{p} \equiv p/p_{\rm c}$, with the volume of the system as an implicit parameter, at several values of reduced temperature $\hat{\tau} \equiv \tau/\tau_{\rm c}=T/T_{\rm c}$.  The low pressure and high volume pure gas region is at the lower left and the high-pressure and low volume pure liquid region is at the upper right.  As the volume decreases on moving upwards along an isotherm, the pressure increases until the system encounters the intersection with an unstable or metastable triangle-shaped part of the isotherm, labeled as the point X\@.  At this point the chemical potentials of the gas and liquid are the same.  As the volume decreases further, if the system is in equilibrium the pressure remains constant at this point until all the gas is converted to liquid.  Then the pressure starts to increase again when the gas is completely converted to liquid as the volume of liquid decreases.  The fluid phase, where liquid and gas cannot be distinguished, occurs at temperatures above the critical temperature $T/T_{\rm c}>1$.}
\label{Fig:vdW_mu_versus_p}  
\end{figure}

\begin{figure}[t]
\includegraphics[width=3.in]{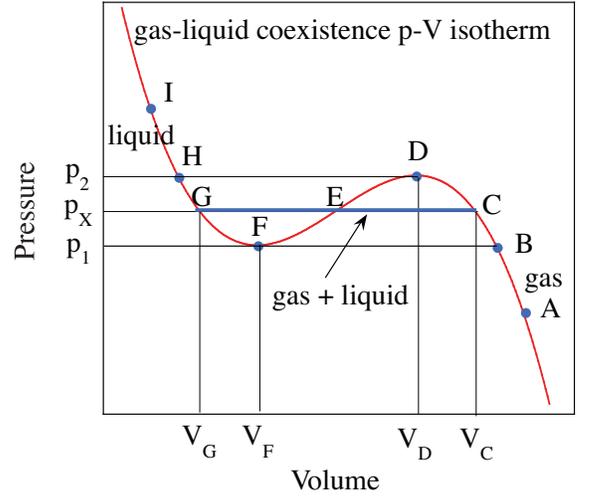}\vspace{0.1in}
\includegraphics[width=2.5in]{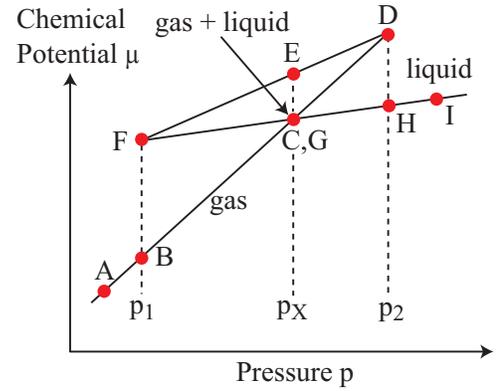}
\caption{(Color online) (Top panel) A schematic pressure versus volume ($p$-$V$) isotherm showing coexistence between the gas and liquid phases such as in Fig.~\ref{Fig:vdW_p_vs_V}.  Various points on the isotherm are labeled.  With decreasing volume, in equilibrium the system follows the path A-B-C-E-G-H-I\@.  One starts with pure gas at point~A\@. Liquid starts to form at point~C and all the gas is converted to liquid at point~G.  The system is pure liquid along the path G-H-I\@. (Bottom panel) A schematic chemical potential $\mu$ versus pressure isotherm such as shown in Fig.~\ref{Fig:vdW_mu_versus_p}, with points labeled as in the top panel. Starting at low pressure (large volume), in equilibrium the system follows the path with the lowest chemical potential, i.e., A-B-(C,G)-H-I\@.  The paths with decreasing volume C-D and F-G are metastable regions and the region D-E-F is unconditionally unstable to phase separation.  The temperatures of the isotherms in the two panels are the same.}
\label{Fig:Cubic_Eqn}  
\end{figure}

The vdW fluid contains attractive interactions, and one therefore expects that it may liquify at sufficiently low temperature and/or sufficiently high pressure.  The liquid ($l$) phase is more stable than the gas~($g$) phase when the liquid  and gas chemical potentials satisfy $\mu_l < \mu_g$, or equivalently, when the Gibbs free energy satisfies $G_l < G_g$ and the gas phase is more stable when $\mu_l > \mu_g$ or $G_l > G_g$.  The two phases can coexist if $\mu_l = \mu_g$ or $G_l = G_g$.  For calculations of $\mu$ of the vdW fluid, it is most convenient to calculate it from the Helmholtz free energy in Eq.~(\ref{Eq:FvdW2}), yielding
\bea
\mu(\tau,V,N) &=& \left(\frac{\partial F}{\partial N}\right)_{\tau,V}\label{Eq:muFromF} \\*
&&\hspace{-0.5in}= -\tau\ln\left[\frac{n_Q(V - Nb)}{N}\right] + \frac{Nb\tau}{V-Nb} -\frac{2Na}{V}\nonumber\\*
&&\hspace{-0.5in}=-\tau\ln\left[\frac{V - Nb}{N}\right] + \frac{Nb\tau}{V-Nb} -\frac{2Na}{V}-\tau \ln n_Q.\nonumber
\eea
Using Eqs.~(\ref{Eq:abN}), one can express $\mu(\tau,V,N)$ in reduced variables as
\bse
\bea
\frac{\mu}{\tau_{\rm c}} &=& -\hat{\tau}\ln\left(3\widehat{V} - 1\right) + \frac{\hat{\tau}}{3\widehat{V} - 1}\label{Eq:muNorm0}\\*
&&\hspace{0.6in} -\ \frac{9}{4\widehat{V}} - \hat{\tau}\ln X,\nonumber\\
X &\equiv& \frac{\tau_{\rm c}n_Q}{8p_{\rm c}}.
\eea
In terms of the reduced number density $\hat{n}=1/\widehat{V}$, Eq.~(\ref{Eq:muNorm0}) becomes
\bea
\frac{\mu}{\tau_{\rm c}} &=& -\hat{\tau}\ln\left(\frac{3 - \hat{n}}{\hat{n}}\right) + \frac{\hat{\tau}\hat{n}}{3 - \hat{n}}\label{Eq:muNorm10}\\*
&&\hspace{0.6in} -\ \frac{9\hat{n}}{4} - \hat{\tau}\ln X,\nonumber
\eea
\ese
The last term in $X$ depends on the particular gas being considered.

We add and subtract $\ln(2e^{-1/2})$ from the right side of Eq.~(\ref{Eq:muNorm0}), yielding
\bea
\frac{\mu}{\tau_{\rm c}} &=& -\hat{\tau}\ln\left(\frac{3\widehat{V} - 1}{2e^{-1/2}}\right) + \frac{\hat{\tau}}{3\widehat{V} - 1} - \frac{9}{4\widehat{V}}\label{Eq:muNorm1}\\*
&&\hspace{0.5in} -\ \hat{\tau}\ln\left(2e^{-1/2}X\right).\nonumber
\eea
For processes at constant~$\hat{\tau}$, the last (gas-dependent) term in Eq.~(\ref{Eq:muNorm1}) just has the effect of shifting the origin of $\mu/\tau_{\rm c}$ as $\hat{\tau}$ is changed.  When plotting $\mu/\tau_{\rm c}$ versus $p$ or $\mu/\tau_{\rm c}$ versus $V$ isotherms, we set that constant to zero, yielding
\be
\frac{\mu}{\tau_{\rm c}} =  -\hat{\tau}\ln\left(\frac{3\widehat{V} - 1}{2e^{-1/2}}\right) + \frac{\hat{\tau}}{3\widehat{V} - 1} - \frac{9}{4\widehat{V}}.
\label{Eq:muNorm2}
\ee
Including the factor of $2e^{-1/2}$ in Eq.~(\ref{Eq:muNorm2}) causes the $\mu/\tau_{\rm c}$ versus $\widehat{V}$ isotherms to all cross at $\widehat{V} = 1$.  Isotherms of $\mu/\tau_{\rm c}$ versus $\widehat{V}$ obtained using Eq.~(\ref{Eq:muNorm2}) are plotted in Fig.~\ref{Fig:vdW_mu_versus_V} for the critical temperature and several adjacent temperatures.  

\section{\label{Sec:EquilpVTBpT} Equilibrium Pressure-Volume, Temperature-Volume and Pressure-Temperature Phase Diagrams}

By combining the data in Figs.~\ref{Fig:vdW_p_vs_V} and~\ref{Fig:vdW_mu_versus_V}, one can plot $\mu$ versus $p$ isotherms with $V$ as an implicit parameter, as shown in Fig.~\ref{Fig:vdW_mu_versus_p}.  Since here $\mu=\mu(\tau,p)$ as in Eq.~(\ref{Eq:Gmu}), in equilibrium the state occurs with the lowest Gibbs free energy and therefore also the lowest chemical potential.

Following Reif,\cite{Reif1965} certain points on a $p$-$V$ isotherm at $\hat{\tau} < 1$ are shown in the top panel of Fig.~\ref{Fig:Cubic_Eqn} and compared with the corresponding points on a plot of $\mu$ versus $p$ in the bottom panel.  Starting from the bottom left of the bottom panel of Fig.~\ref{Fig:Cubic_Eqn}, at low pressure the stable phase is seen to be the gas phase.  As the pressure increases, a region occurs at which the chemical potential of the gas and liquid become the same, at the point~X at pressure $p_{\rm X}$, which signals entry into a triangle-shaped unstable region of the plot which the system does not enter in thermal equilibrium.  The pressure~$p_{\rm X}$ is a constant pressure part of the $p$-$V$ isotherm at which the gas and liquid coexist as indicated by the horizontal line in the top panel.  The system remains at constant pressure at the point~X in the bottom panel as the system volume decreases until all the gas is converted to liquid.  At higher pressures, the pure liquid has the lower chemical potential as indicated in the bottom panel.

Essential variables of the calculations of the  thermodynamic properties of the van der Waals fluid are the reduced equilibrium pressure $\hat{p}_{\rm X}$ for coexistence of gas and liquid phases at a given $\hat{\tau}$ and the associated reduced volumes $\widehat{V}_{\rm C}$, $\widehat{V}_{\rm D}$, $\widehat{V}_{\rm F}$, and $\widehat{V}_{\rm G}$ in Fig.~\ref{Fig:Cubic_Eqn}.  Using these values and the equations for the thermodynamic variables and properties, the equilibrium and nonequilibrium properties versus temperature, volume or pressure can be calculated and the various phase diagrams constructed.  The condition for the coexistence of the liquid and gas phases is that their chemical potentials $\mu(\hat{\tau},\hat{p})$, temperatures and pressures must be the same at their respective volumes $\widehat{V}_{\rm G}$ and $\widehat{V}_{\rm C}$ in Fig.~\ref{Fig:Cubic_Eqn}, where $\mu(\hat{\tau},\widehat{V})/\tau_{\rm c}$ is given in Eq.~(\ref{Eq:muNorm2}).  This requirement allows $\hat{p}_{\rm X}$ to be determined.

To determine the values of $\hat{p}_{\rm X}$, $\widehat{V}_{\rm C}$, $\widehat{V}_{\rm D}$, $\widehat{V}_{\rm F}$, and $\widehat{V}_{\rm G}$ in Fig.~\ref{Fig:Cubic_Eqn} where gas and liquid phases coexist in equilibrium, one can use a parametric solution in which $\hat{p}$ and $\mu/\tau_{\rm c}$ are calculated at fixed $\hat{\tau}$ using $\widehat{V}$ as an implicit parameter and thereby express $\mu/\tau_{\rm c}$ versus~$\hat{p}$ at fixed~$\hat{\tau}$.  From the numerical $\mu(\hat{\tau},\hat{p})$ data, one can then determine the values of the above four reduced volumes and then the value of $p_{\rm X}$ from $\widehat{V}_{\rm C}$ or $\widehat{V}_{\rm G}$ and the vdW equation of state.  The following steps are carried out for each specified value of $\hat{\tau}$ to implement this sequence of calculations.
\begin{enumerate}
\item{The two volumes $\widehat{V}_{\rm D}$ and $\widehat{V}_{\rm F}$ at the maximum and minimum of the S-shaped region of the $p$-$V$ plot in Fig.~\ref{Fig:Cubic_Eqn} are determined by solving Eq.~(\ref{Eq:RedpVsRedV}) for the two volumes at which $d\hat{p}/d\widehat{V}=0$. These two volumes enclose the unstable region of phase separation of the gas and liquid phases since the isothermal compressibility is negative in this region.}
\item{The pressure $\hat{p}_2$ at the volume $\widehat{V}_{\rm D}$ is determined from the equation of state~(\ref{Eq:RedpVsRedV}).}
\item{The volume $\widehat{V}_{\rm H}$ is determined by solving for the two volumes at pressure $\hat{p}_2$ (the other one, $\widehat{V}_{\rm D}$, is already calculated in Step~1).  These two volumes are needed to set the starting values of the numerical calculations of the volumes $\widehat{V}_{\rm C}$ and $\widehat{V}_{\rm G}$ in the next step.}
\item{The volumes $\widehat{V}_{\rm C}$ and $\widehat{V}_{\rm G}$ are determined by solving two simultaneous equations which equate the pressure and chemical potential of the gas and liquid phases at these two volumes at a fixed set temperature, respectively:
\bse
\label{Eqs:pEqmuEq}
\bea
\hat{p}(\hat{\tau},\widehat{V}_{\rm G}) &=& \hat{p}(\hat{\tau},\widehat{V}_{\rm C}), \\*
\frac{\mu(\hat{\tau},\widehat{V}_{\rm G})}{\tau_{\rm c}} &=& \frac{\mu(\hat{\tau},\widehat{V}_{\rm C})}{\tau_{\rm c}}.
\eea
\ese 
The {\tt FindRoot} utility of {\tt Mathematica} is fast and accurate in finding the solutions for $\widehat{V}_{\rm G}$ and~$\widehat{V}_{\rm C}$ if appropriate starting values for these parameters are given.  The starting values we used were $0.97\widehat{V}_{\rm H}$ and $1.1\widehat{V}_{\rm D}$, respectively, where the volumes $\widehat{V}_{\rm H}$ and~$\widehat{V}_{\rm D}$ are obtained from Steps~3 and~1, respectively. }
\item{The pressure $\hat{p}_{\rm X}$ at which the gas and liquid are in equilibrium at a given temperature is calculated from either $\widehat{V}_{\rm C}$ or $\widehat{V}_{\rm G}$ using the equation of state~(\ref{Eq:RedpVsRedV}).}
\end{enumerate}
Representative values of the above reduced parameters calculated versus reduced temperature are given in Table~\ref{Tab:CoexistData} of Appendix~\ref{App:Tables}.

\begin{figure}[t]
\includegraphics[width=3.3in]{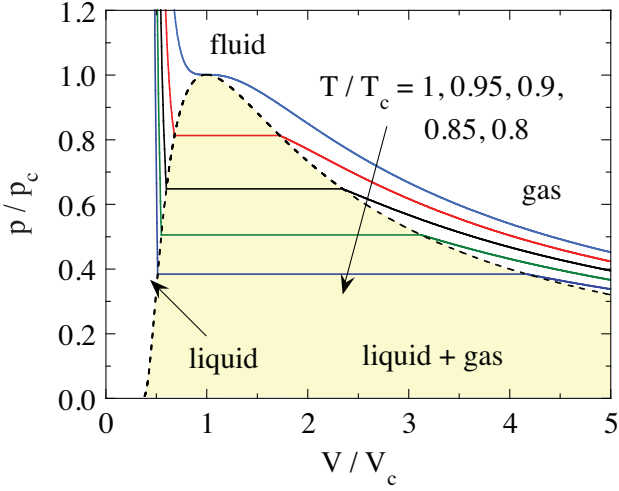}
\caption{(Color online) Equilibrium pressure versus volume isotherms showing the pure gas regime on the right, the liquid-gas coexistence region in the middle and the pure liquid regime on the left. The dashed curve is the boundary between the regions of pure liquid, liquid + gas, and pure gas in the equilibrium $p$-$V$ phase diagram.  The yellow-shaded region bounded from above by the dashed curve is the region of coexistence of the liquid and gas phases in the $p$-$V$ phase diagram in Fig.~\ref{Fig:p_T_vs_V_phase_diags}(a).  For temperatures above the critical temperature $T/T_{\rm c} = 1$, there is no phase separation and no physical distinction between gas and liquid phases; this undifferentiated phase is denoted as the fluid phase.}
\label{Fig:vdW_p_vs_V_equilib}  
\end{figure}

\begin{figure}[t]
\includegraphics[width=3.in]{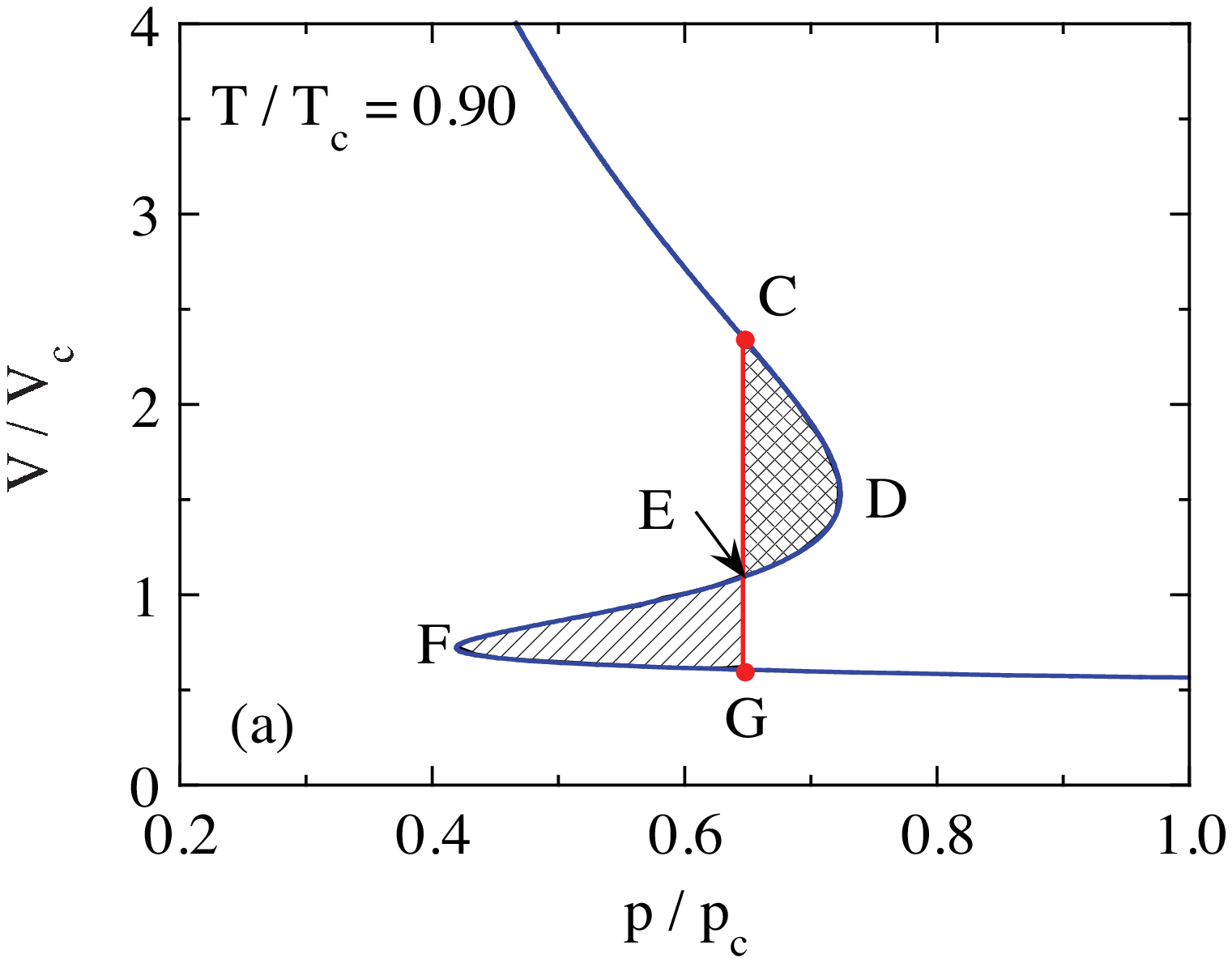}
\includegraphics[width=3.in]{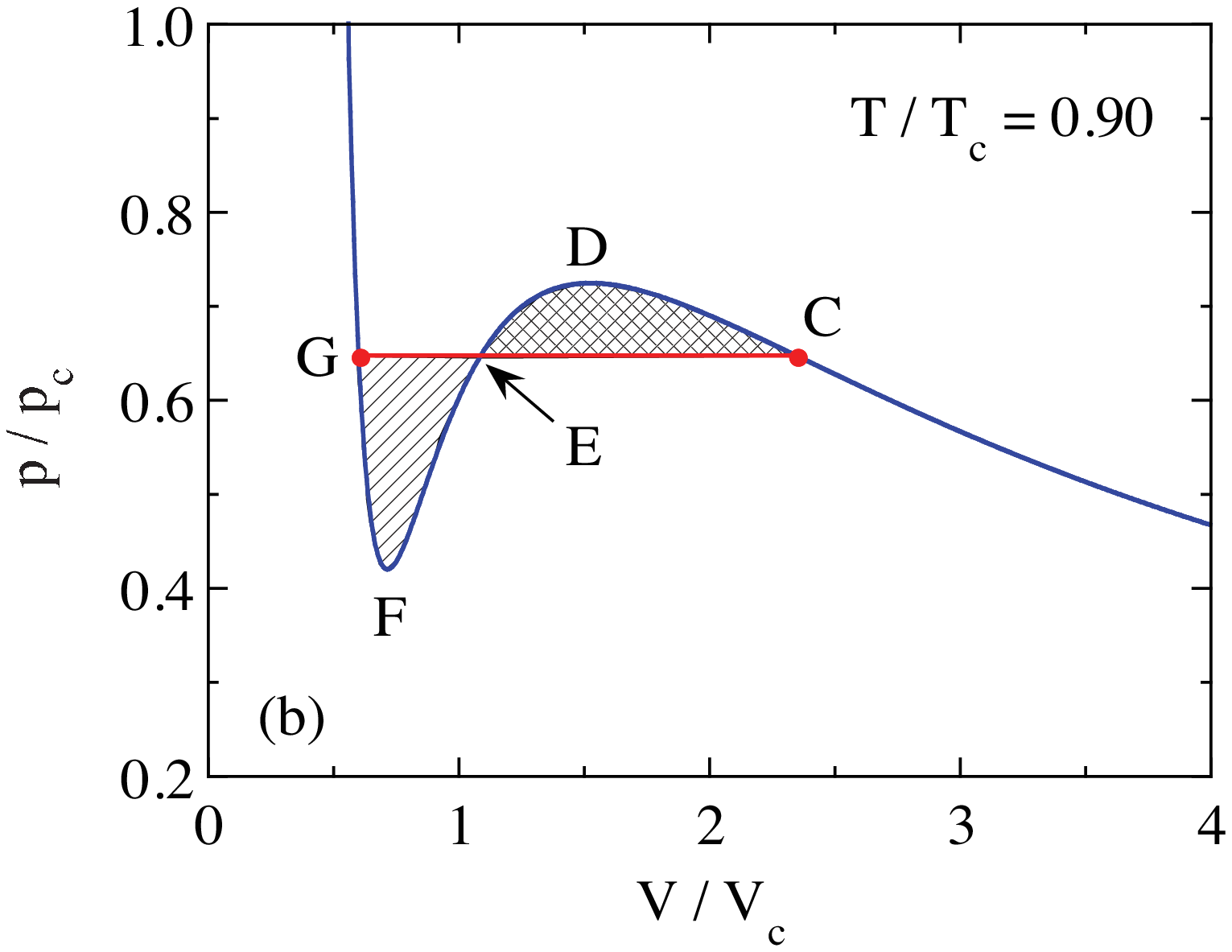}
\caption{(Color online) (a) Reduced volume~$\widehat{V} = V/V_{\rm c}$ versus reduced pressure~$\hat{p} = p/p_{\rm c}$ at reduced temperature $\hat{\tau} = T/T_{\rm c} = 0.90$, for which $\widehat{V}_{\rm G} = 0.6034$ and~$\widehat{V}_{\rm C} = 2.3488$.  The letter designations of points on the curve are the same as in Fig.~\ref{Fig:Cubic_Eqn}.  The vertical red line at pressure $\hat{p}_{\rm X} = 0.6470$ is the equilibrium pressure of liquid-gas coexistence at $\hat{\tau} = 0.90$.  According to Eq.~(\ref{Eq:DeltaG}), the net area under the curve from point~C, where the fluid is pure gas, to~G, where the fluid is pure liquid, is proportional to the change $\Delta\mu$ in the chemical potential between these two points, which is equal to zero for the region of liquid-gas coexistence from~C to~G\@.  This $\Delta\mu$ is proportional to the net area of $\widehat{V}$ versus~$\hat{p}$ along the path from point~C to point~G, which in turn is the sum of the positive hatched area to the left of the path C-D-E, and the negative hatched area to the right of the path E-F-G\@.  Therefore the magnitudes of the first and second areas must be the same.  This requirement is drawn on the corresponding $p$-$V$ diagram in (b), where the magnitudes of the two hatched areas above and below the horizontal line at $\hat{p}_{\rm X}=0.6470$ must be the same.  This is called Maxwell's construction.}
\label{Fig:Maxwell_Construction}  
\end{figure}

\begin{figure}[t]
\includegraphics[width=3.3in]{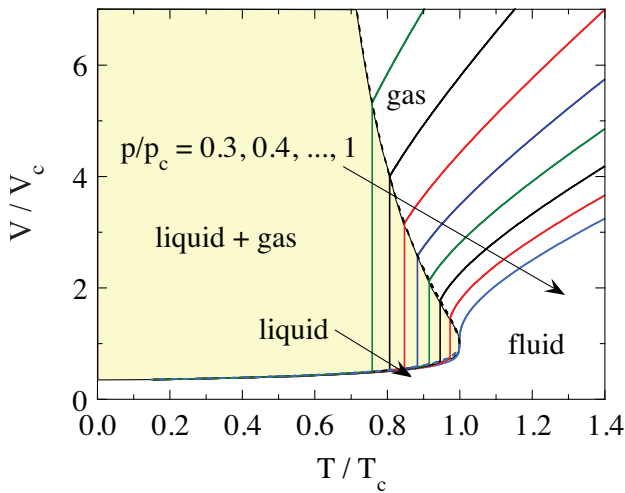}
\caption{(Color online) Equilibrium volume versus temperature isobars at the indicated pressures.  The dashed curve separates the pure liquid, liquid + gas and pure gas regions.  The yellow-shaded region is the region of liquid + gas coexistence in the $V$-$T$ phase diagram.}
\label{Fig:vdW_V_T_isotherms}  
\end{figure}

\begin{figure}[t]
\includegraphics[width=3.3in]{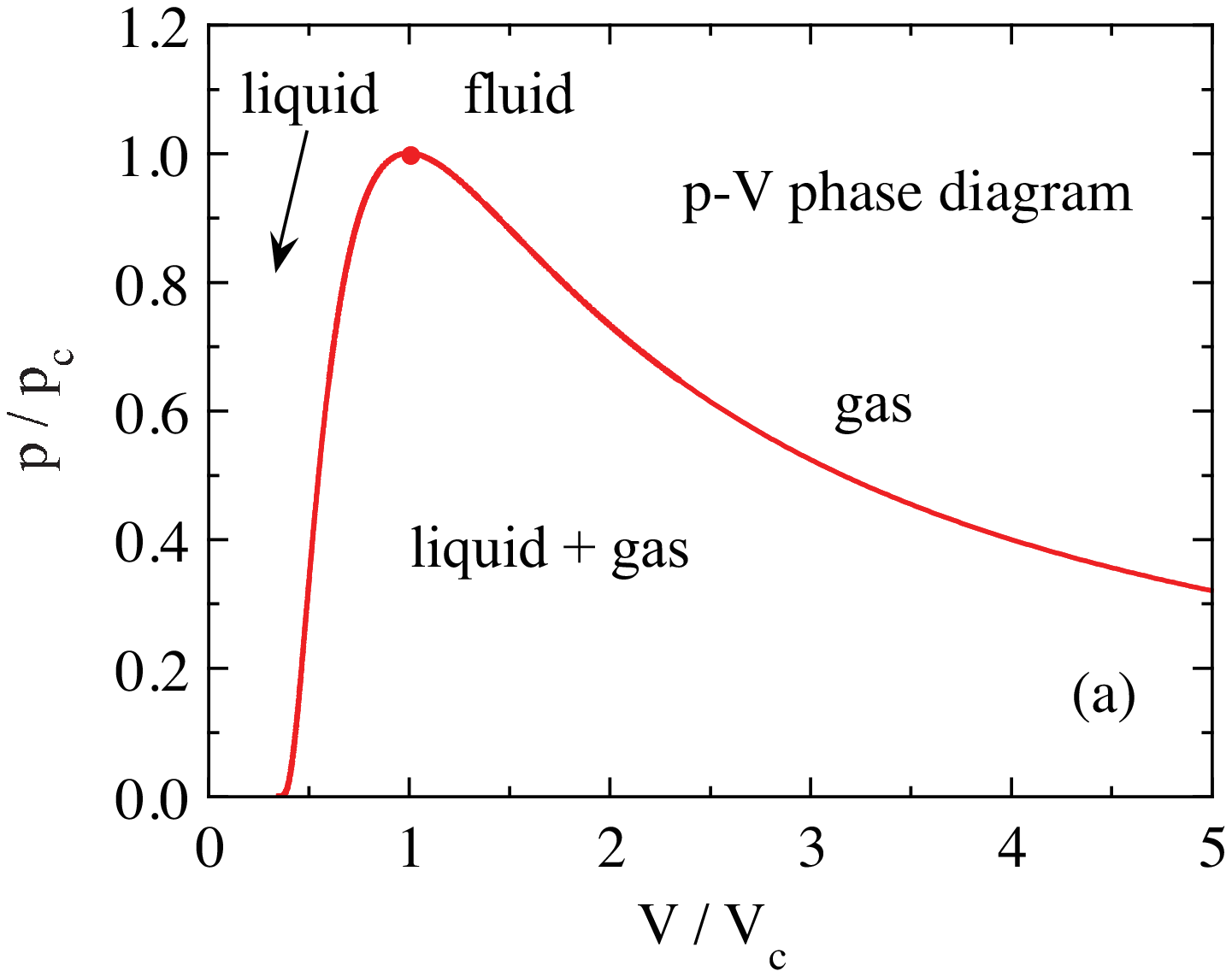}\vspace{-0.1in}
\includegraphics[width=3.3in]{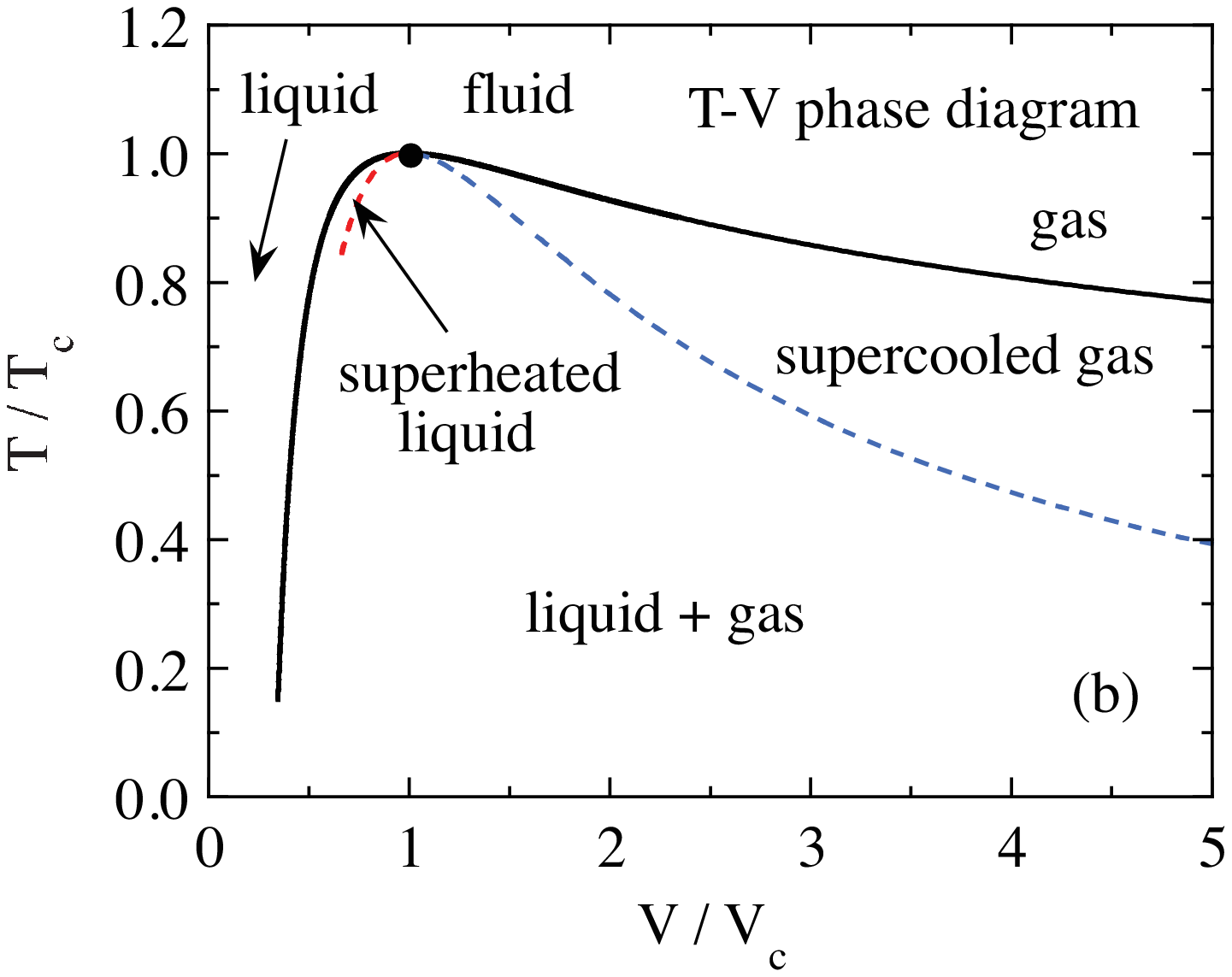}\vspace{-0.1in}
\caption{(Color online) Equilibrium phase diagrams of the van der Waals fluid in the (a) pressure-volume ($p$-$V$) plane and in the (b) temperature-volume ($T$-$V$) plane.  The critical point is denoted by a filled circle in each panel.  In (b), the regions of metastable superheated liquid and supercooled gas, derived as in Fig.~\ref{Fig:Volume_Hysteresis_p_0.3}, are also shown with boundaries from the numerical data in Table~\ref{Tab:CoexistData} that are defined by the solid black equilibrium curve and the respective colored dashed metastable curves as shown.}
\label{Fig:p_T_vs_V_phase_diags}  
\end{figure}

\begin{figure}[t]
\includegraphics[width=3.3in]{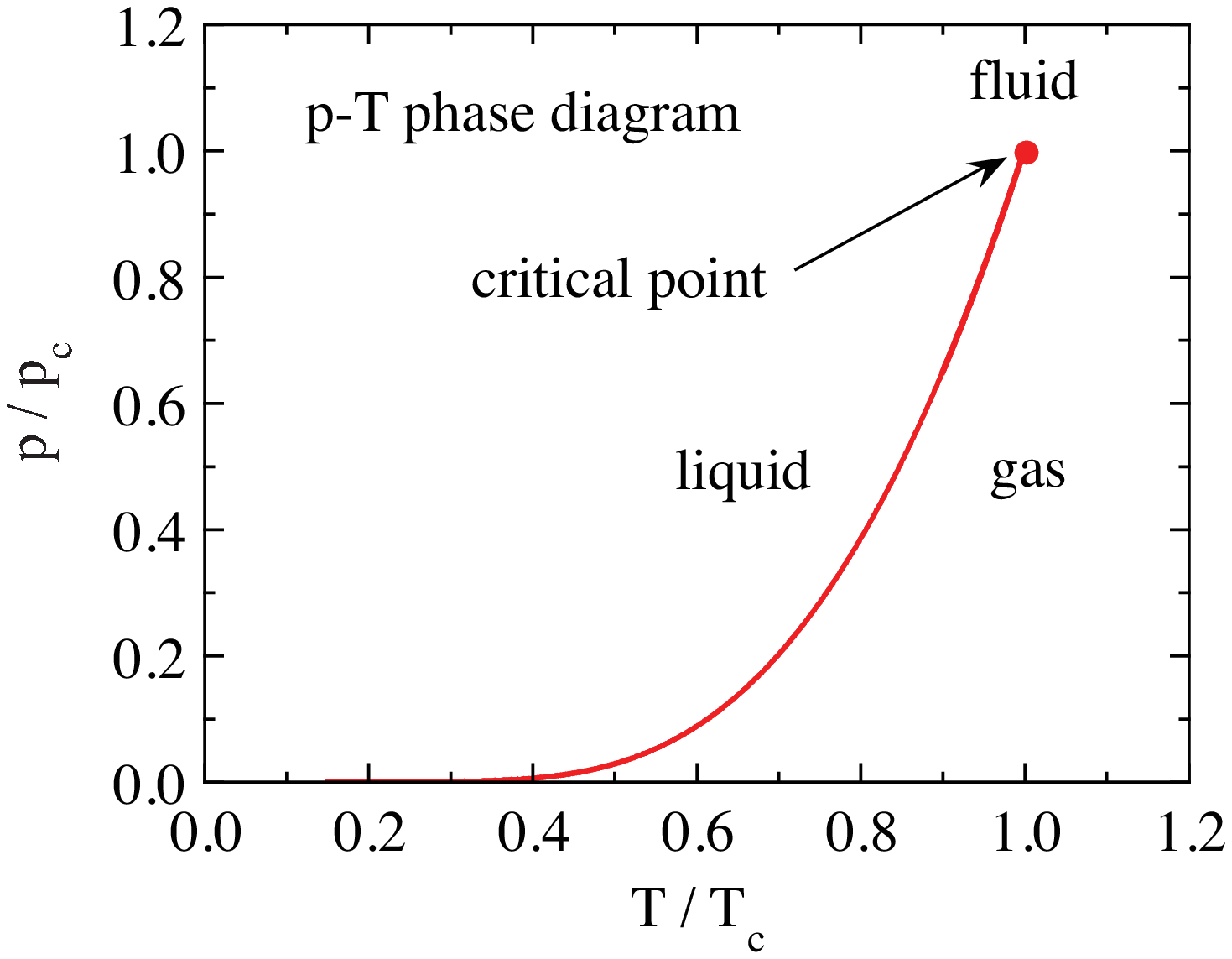}
\caption{(Color online) Phase diagram in the reduced pressure-temperature ($\hat{p}$-$\hat{\tau}$) plane.  The coexistence curve between liquid and gas phases is indicated, which terminates at the critical point with $p=p_{\rm c}$, $T=T_{\rm c}$, $V=V_{\rm c}$.  On crossing the curve from left to right or top to bottom, the system transforms from pure liquid to pure gas.  The liquid-gas coexistence curve is the function $\hat{p}_{\rm X}(\hat{\tau})$, where $\hat{p}_{\rm X}$ is defined in Fig.~\ref{Fig:Cubic_Eqn}. A fit of the coexistence curve is given in Eq.~(\ref{Eq:FitpX}) with coefficients~$c_n$ in Table~\ref{Tab:FitpX1/t}.  At supercritical  temperatures above $T_{\rm c}$, the liquid and gas phases are physically indistinguishable and the substance is termed a fluid in this region.}
\label{Fig:vdW_p_vs_T_phase_diag}  
\end{figure}

By solving for the pressure $p_{\rm X}$ versus temperature at which the gas and liquid phases coexist as described above, one can derive equilibrium pressure-volume isotherms.   Representative isotherms are shown in Fig.~\ref{Fig:vdW_p_vs_V_equilib} for $T/T_{\rm c} = 1$~(critical temperature), 0.95, 0.90, 0.85 and 0.80.  
The pressure $p_{\rm X}$ at which the horizontal two-phase line occurs in the upper panel of Fig.~\ref{Fig:Cubic_Eqn} can be shown to satisfy the so-called Maxwell construction as follows.  According to Eq.~(\ref{Eq:dG}) with constant $\tau$ and~$N$, the differential of the Gibbs free energy is $dG = V\,dp$, so the difference in $G$ between two pressures along a path in the $V$-$p$ plane is
\be
\Delta G = N\Delta\mu = G(p_2,V_2) - G(p_,V_1) =\int_{p_1}^{p_2}V(p)\,dp.
\label{Eq:DeltaG}
\ee
As shown in Fig.~\ref{Fig:Maxwell_Construction}(a), this integral is the integral along the path from point~C to point~G\@.  The part of the area beneath the curve from~C to~D that lies below the curve from~D to~E is cancled out because the latter area is negative.  Similarly, part of the negative area from~E to~F that lies below the path from~F to~G is canceled out by the positive area below the path from~F to G\@.  Therefore the net area from~C to~G is the sum of the positive hatched area to the right of the vertical line and the negative hatched area to the left of the vertical line that are shown in Fig.~\ref{Fig:Maxwell_Construction}(a).  Since the vertical line represents equilibrium between the gas and liquid phases, for which the chemical potentials and Gibbs free energies are the same, one has $\Delta G = 0$ and hence the algebraic sum of the two hatched areas is zero.  That means the magnitudes of the two hatched areas have to be the same.  Transferring this information to the corresponding $p$-$V$ diagram in Fig.~\ref{Fig:Maxwell_Construction}(b), one requires that the magnitudes of the same two hatched areas shown in that figure have to be equal.  This is Maxwell's construction.  In terms of the numerical integral of a $\hat{p}$ versus~$\widehat{V}$ isotherm over the two-phase region at temperature~$\hat{\tau}$, Maxwell's construction states that
\be
\int_{\widehat{V}_{\rm G}}^{\widehat{V}_{\rm C}}[\hat{p}(\hat{\tau},\widehat{V}) - \hat{p}_{\rm X}(\hat{\tau})]\,d\widehat{V} = 0,
\ee
or equivalently
\be
\int_{\widehat{V}_{\rm G}}^{\widehat{V}_{\rm C}}\hat{p}(\hat{\tau},\widehat{V}) \,d\widehat{V} = \hat{p}_{\rm X}(\hat{\tau})(\widehat{V}_{\rm G} - \widehat{V}_{\rm C}).
\ee

Equilibrium volume versus temperature isobars are shown in Fig.~\ref{Fig:vdW_V_T_isotherms} for reduced pressures $\hat{p} = 1.1,\ 1.0,\ \ldots,\ 0.5$.  The isobars illustrate the first-order increase in volume at the liquid to gas transition temperature for each pressure with $\hat{\tau}<1$.  For $\hat{\tau}\geq 1$ only the undifferentiated fluid phase occurs.  The coexistence curves of pressure versus volume and temperature versus volume as shown in Figs.~\ref{Fig:p_T_vs_V_phase_diags}(a) and~\ref{Fig:p_T_vs_V_phase_diags}(b), respectively.  Also included in Fig.~\ref{Fig:p_T_vs_V_phase_diags}(b) are regions in which metastable superheated liquid and supercooled gas occur, as discussed below in Sec.~\ref{Sec:Hysteresis}.  It may seem counterintuitive that both dashed lines lie {\it below} the equilibrium curve.  However, in Fig.~\ref{Fig:Volume_Hysteresis_p_0.3}(b) below, it is shown that the superheated liquid has a larger volume without much change in temperature, resulting in the superheated metastable region in Fig.~\ref{Fig:p_T_vs_V_phase_diags}(b) being below the equilibrium curve.

\begin{table}
\caption{\label{Tab:FitpX1/t} Coefficients $c_n$ for the parametrization of the reduced pressure~$\hat{p}_{\rm X}$ versus reduced temperature $\hat{\tau}$ by the ninth-order polynomial in Eq.~(\ref{Eq:FitpX}).  }
\begin{ruledtabular}
\begin{tabular}{crc}
\hspace{0.8in}$c_n$ & value \hspace{0.25in} \\	
\hline
\hspace{0.8in}$c_0$ & 5.66403835e00 \\
\hspace{0.8in}$c_1$ & $-$8.73724257e00 \\
\hspace{0.8in}$c_2$ & 5.14022974e00 \\
\hspace{0.8in}$c_3$ & $-$2.92538942e00 \\
\hspace{0.8in}$c_4$ & 1.09108819e00 \\
\hspace{0.8in}$c_5$ & $-$2.74194800e-1 \\
\hspace{0.8in}$c_6$ & 4.59922654e-2 \\
\hspace{0.8in}$c_7$ & $-$4.92809927e-3 \\
\hspace{0.8in}$c_8$ & 3.04520105e-4 \\
\hspace{0.8in}$c_9$ & $-$8.24218733e-6 \\
\end{tabular}
\end{ruledtabular}
\end{table}

The pressure-temperature phase diagram derived from the above numerical data is shown in Fig.~\ref{Fig:vdW_p_vs_T_phase_diag}.  Here there are no metastable, unstable or hysteretic regions.  The gas-liquid coexistence curve has positive slope everywhere along it and terminates in the critical point at $p=p_{\rm c}$, $T=T_{\rm c}$, $V=V_{\rm c}$ above which the gas and liquid phases cannot be distinguished.  We obtained an analytic parametrization of the coexistence curve as follows.  The $\ln(\hat{p}_{\rm X})$ versus~$1/\hat{\tau}$ data were fitted by the ninth-order polynomial
\be
\ln(\hat{p}_{\rm X}) = \sum_{n=0}^9 c_n(1/\hat{\tau})^n,
\label{Eq:FitpX}
\ee
where the fitted $c_n$ coefficients are listed in Table~\ref{Tab:FitpX1/t}.  To compare the fit with the $\hat{p}_{\rm X}$ versus~$\hat{\tau}$ calculated data one exponentiates both sides of Eq.~(\ref{Eq:FitpX}).  The fitted $\hat{p}_{\rm X}$ values agree with the calculated values to $\lesssim 0.01$\% of~$\hat{p}_{\rm X}$ over the temperature range $0.15 \leq \hat{\tau} \leq 1$ of the fit.

\section{\label{Sec:Lekner} Lekner's Parametric Solution of the Coexistence Curve and Associated Properties}

Lekner provided an elegant and very useful alternative parametric solution for the coexistence curve in Fig.~\ref{Fig:vdW_p_vs_T_phase_diag} and some properties associated with it that is also based on solving Eqs.~(\ref{Eqs:pEqmuEq}).\cite{Lekner1982}  This solution allows exact calculations of $\hat{p}_{\rm X}$ versus $\hat{\tau}$ and associated properties to be easily carried out for both $\hat{\tau}\to 0$ and $\hat{\tau}\to 1$ as well as numerical calculations in the intermediate temperature regime.  He calculated some critical exponents for $\hat{\tau}\to 1$.\cite{Lekner1982}  Berberan-Santos~et al.\ extended the calculations to additional properties of the vdW fluid for both $\hat{\tau}\to 0$ and $\hat{\tau}\to 1$.\cite{Berberan-Santos2008}  Here we describe and significantly extend this parametric solution and express the predictions from it in terms of our dimensionless reduced variables in Eqs.~(\ref{Eq:CritPars}), (\ref{Eq:RedVar}) and~(\ref{Eq:nhatDef}).

Lekner expressed the solutions to all properties of the coexistence curve in terms of the parameter $y \equiv \Delta s/2$, where $\Delta s$ is the entropy difference per particle between the gas and liquid phases in units of $k_{\rm B}$.  He defined two functions of~$y$ as
\bse
\bea
f(y) &=& \frac{2e^y\left[(y-1)e^{2y} + y + 1\right]}{e^{4y}-4y\, e^{2y}-1}\\*
&=& \frac{y\cosh y - \sinh y}{\sinh y \cosh y - y},\\*
g(y) &=& 1 + 2f(y)\cosh y +f^2(y).
\eea
\ese
He then expressed the following properties of the coexistence curve in terms of $y$, $f(y)$ and~$g(y)$, which we augment and write in terms of the critical parameters in Eq.~(\ref{Eqs:RedVars}) and reduced variables in Eq.~(\ref{Eq:RedVar}) with subscript~X where appropriate which specifies that the quantity is associated with the coexistence curve.  Subscripts $g$ and~$l$ refer to the coexisting gas and liquid phases, respectively.  A symbol $\hat{z}$ means the value of $z$ divided by its value at the critical point.  The symbols are: $\Delta S_{\rm X}$: difference in entropy between the pure gas and liquid phases at the two edges of the coexistence region in Fig.~\ref{Fig:vdW_p_vs_V_equilib}; $\hat{\tau}_{\rm X}$: temperature on the coexistence curve; $\hat{p}_{\rm X}$: pressure on the coexistence curve; $\widehat{V}_g$ and $\widehat{V}_l$: volumes of the respective coexisting phases; and $\hat{n}_g = 1/\widehat{V}_g$ and $\hat{n}_l = 1/\widehat{V}_l$: number densities of the respective coexisting phases.  The expressions are
\bse
\label{Eqs:ParaSolns}
\bea
\frac{\Delta S_{\rm X}}{Nk_{\rm B}} &\equiv& \frac{S_g - S_l}{Nk_{\rm B}} = 2y,\label{Eq:Entropyvsy}\\
\hat{\tau}_{\rm X} &=& \frac{27 f(y)\left[f(y)+\cosh y\right]}{4g^2(y)},\label{Eq:tauFromy}\\
\hat{p}_{\rm X} &=& \frac{27 f^2(y)\left[1-f^2(y)\right]}{g^2(y)},\\
\frac{d\hat{p}_{\rm X}}{d\hat{\tau}} \equiv\frac{d\hat{p}_{\rm X}}{d\hat{\tau}_{\rm X}} &=& \frac{16 y [y\,{\rm coth}(y) -1] }{\sinh(2 y)-2 y},\label{Eq:dpXdX}\\
\widehat{V}_g &=& \frac{1}{3}\left[1 + \frac{e^y}{f(y)}\right],\\
\widehat{V}_l &=& \frac{1}{3}\left[1 + \frac{e^{-y}}{f(y)}\right],\\
\widehat{V}_{\rm X} = \widehat{V}_g+\widehat{V}_l &=& \frac{2}{3}\left[1 + \frac{\cosh y}{f(y)}\right],\label{Eq:DeltaVX}\\
\Delta\widehat{V}_{\rm X} = \widehat{V}_g-\widehat{V}_l &=& \frac{2\sinh y}{3f(y)},\label{Eq:DeltaVX}\\
\hat{n}_g &=& \frac{3f(y)}{e^y + f(y)},\label{Eq:ngasGen}\\
\hat{n}_l &=& \frac{3f(y)}{e^{-y} + f(y)},\label{Eq:nliquidGen}\\
\Delta\hat{n}_{\rm X} = \hat{n}_l-\hat{n}_g &=& \frac{6f(y)\sinh y}{g(y)},\\
\hat{n}_{\rm ave} = \frac{\hat{n}_l+\hat{n}_g}{2} &=& \frac{3f(y)[f(y)+\cosh(y)]}{g(y)}.\label{Eq:nave}
\eea
\ese
Since $y\equiv \Delta s/2$, $\Delta s(\hat{\tau}_{\rm X}=1)=0$ and $\Delta s(\hat{\tau}_{\rm X}=0)=\infty$ as shown below in Sec.~\ref{Sec:LatentHeat}, the implicit variable~$y$ runs from~0 to~$\infty$. Hence one can easily calculate the above properties including $\hat{\tau}_{\rm X}$ as functions of $y$ numerically, and then using $y$ as an implicit parameter evaluate the other ones as a function of $\hat{\tau}_{\rm X}$ or in terms of each other.  Our result for $\hat{p}_{\rm X}$ versus $\hat{\tau}$ (i.e., $\hat{\tau}_{\rm X}$) obtained from the parametric solution is of course the same as already plotted using a different numerical solution in Fig.~\ref{Fig:vdW_p_vs_T_phase_diag}.  However, Lekner's parametrization allows properties to be accurately calculated to lower temperatures than the conventional parametrization in Sec.~\ref{Sec:EquilpVTBpT} using the volume as the implicit parameter.

Expressions for quantitites derived from the above fundamental ones as a function of~$y$, along with references to the above equations originally defining them, are
\bse
\label{Eqs:DerivedQuantities}
\bea
\frac{U_g}{p_{\rm c}V_{\rm c}} &=& 4\hat{\tau}_{\rm X} - 3\hat{n}_g,\hspace{0.93in}(\ref{Eq:URed})\\
\frac{U_l}{p_{\rm c}V_{\rm c}} &=& 4\hat{\tau}_{\rm X} - 3\hat{n}_l,\hspace{0.95in}(\ref{Eq:URed})\\
\frac{U_g-U_l}{p_{\rm c}V_{\rm c}} &=& 3(\hat{n}_l - \hat{n}_g) = 3\Delta\hat{n}_{\rm X},\\
\frac{H_g}{p_{\rm c}V_{\rm c}} &=& \frac{4\hat{\tau}_{\rm X}(5-\hat{n}_g)}{3-\hat{n}_g} - 6\hat{n}_g, \hspace{0.4in}(\ref{Eq:EnthalpyRed})\\
\frac{H_l}{p_{\rm c}V_{\rm c}} &=& \frac{4\hat{\tau}_{\rm X}(5-\hat{n}_l)}{3-\hat{n}_l} - 6\hat{n}_l, \hspace{0.45in}(\ref{Eq:EnthalpyRed})\\
\kappa_{{\rm T}g} p_{\rm c} &=& \frac{(3-\hat{n}_g)^2/(6\hat{n}_g)}{4\hat{\tau}_{\rm X} - \hat{n}_g(3-\hat{n}_g)^2}, \hspace{0.38in}(\ref{Eq:kappaTRedvdW2})\\
\kappa_{{\rm T}l} p_{\rm c} &=& \frac{(3-\hat{n}_l)^2/(6\hat{n}_l)}{4\hat{\tau}_{\rm X} - \hat{n}_l(3-\hat{n}_l)^2}, \hspace{0.41in}(\ref{Eq:kappaTRedvdW2})\\
\frac{\alpha_g\tau_{\rm c}}{k_{\rm B}} &=& \frac{4(3-\hat{n}_g)/3}{4 \hat{\tau}_{\rm X}- \hat{n}_g(3-\hat{n}_g)^2}, \hspace{0.37in}(\ref{Eq:alphavdW2})\\
\frac{\alpha_l\tau_{\rm c}}{k_{\rm B}} &=& \frac{4(3-\hat{n}_l)/3}{4 \hat{\tau}_{\rm X}- \hat{n}_g(3-\hat{n}_l)^2},  \hspace{0.4in}(\ref{Eq:alphavdW2})\\
\frac{C_{{\rm p}g} }{Nk_{\rm B}} &=& \frac{3}{2} + \frac{4\hat{\tau}_{\rm X}}{4\hat{\tau}_{\rm X} - \hat{n}_g(3-\hat{n}_g)^2}, \hspace{0.1in}(\ref{Eq:CpvdW3})\\
\frac{C_{{\rm p}l} }{Nk_{\rm B}} &=& \frac{3}{2} + \frac{4\hat{\tau}_{\rm X}}{4\hat{\tau}_{\rm X} - \hat{n}_l(3-\hat{n}_l)^2}, \hspace{0.17in}(\ref{Eq:CpvdW3})\\
\frac{L}{p_{\rm c}V_{\rm c}} &=& \frac{H_g-H_l}{p_{\rm c}V_{\rm c}}  = \frac{16\, y\,\hat{\tau}_{\rm X}}{3}, \hspace{0.37in}(\ref{Eq:ReducedL})\label{Eq:LatentHeatvsy}
\eea
\ese
where $U$ is the internal energy, $H$ is the enthalpy,  $L$ is the latent heat (enthalpy) of vaporization on crossing the coexistence curve in Fig.~\ref{Fig:vdW_p_vs_T_phase_diag}, $\kappa_{\rm T}$ is the isothermal compressibility, $\alpha$~is the volume thermal expansion coefficient and $C_{\rm p}$ is the heat capacity at constant pressure.  

Because a first-order transition occurs on crossing the coexistence curve at $\hat{p},\ \hat{\tau}<1$ in Fig.~\ref{Fig:vdW_p_vs_T_phase_diag}, there are discontinuities in $\kappa_{\rm T}$, $\alpha$ and $C_{\rm p}$ on crossing the curve.  One can calculate the values of these discontinuities versus $\hat{\tau}_{\rm X}$ using Eqs.~(\ref{Eqs:DerivedQuantities}) and the parametric solutions for $\hat{n}_g,\ \hat{n}_l$ and $\hat{\tau}_{\rm X}$ with $y$ an implicit parameter.  Our analytic expressions for the discontinuities $\Delta\kappa_{\rm T}p_{\rm c} \equiv (\kappa_{{\rm T}g} - \kappa_{{\rm T}l})p_{\rm c}$, $\Delta\alpha \tau_{\rm c}/k_{\rm B} \equiv (\alpha_g -  \alpha_l)\tau_{\rm c}/k_{\rm B}$ and $\Delta C_{\rm p}/(Nk_{\rm B})\equiv (C_{{\rm p}l} - C_{{\rm p}g})/(Nk_{\rm B})$ in terms of~$y$ derived from Eqs.~(\ref{Eqs:ParaSolns}) and~(\ref{Eqs:DerivedQuantities}) are given in Appendix~\ref{Sec:Discont}.

\subsection{Thermodynamic Behaviors as $\hat{\tau}_{\rm X}\to1^-$}

To solve for  the above properties versus temperature for small deviations of $\hat{\tau}_{\rm X}$ from~1 ($y\to0$) or~0 ($y\to\infty$) requires the  solution to $y(\hat{\tau}_{\rm X})$ obtained from Eq.~(\ref{Eq:tauFromy}) in the respective limit to some order of approximation as discussed in this and the following section, respectively.

In this section the relevant quantities are the values of the parameters minus their values at the critical point.  We define 
\be
t_{\rm 0}\equiv 1-\hat{\tau}, \qquad t_{\rm 0X}\equiv 1-\hat{\tau}_{\rm X},
\label{Eq:t0XDef}
\ee
which are positive for $\hat{\tau},\ \hat{\tau}_{\rm X}<1$.   Taylor expanding Eq.~(\ref{Eq:tauFromy}) to $6^{\rm th}$ order in $y$ gives
\bse
\be
t_{\rm 0X} = \frac{y^2}{9} - \frac{y^4}{75} + \frac{946 y^6}{637\,875}.
\ee
Solving for $y(t_{\rm 0X})$ to lowest orders gives
\be
y = 3 \,{t_{\rm 0X}}^{1/2} + \frac{81 \,{t_{\rm 0X}}^{3/2}}{50} + \frac{50\,403 \,{t_{\rm 0X}}^{5/2}}{35\,000}.
\label{Eq:y(dtX)Soln}
\ee
\ese

Taylor expanding Eqs.~(\ref{Eqs:ParaSolns}) about $y=0$, substituting Eq.~(\ref{Eq:y(dtX)Soln}) into these Taylor expansions and simplifying gives the $y\to0$ and $t_{\rm 0X}\to0$ behaviors of the quantities in Eqs.~(\ref{Eqs:ParaSolns}) to lowest orders as
\begin{widetext}
\bse
\label{Eqs:PropsHighT}
\bea
\frac{\Delta S_{\rm X}}{Nk_{\rm B}} \equiv 2y  &=& 6 \,{t_{\rm 0X}}^{1/2} + \frac{81 \,{t_{\rm 0X}}^{3/2}}{25} + \frac{50\,403 \,{t_{\rm 0X}}^{5/2}}{17\,500},\\
\frac{L}{p_{\rm c}V_{\rm c}} = \frac{16 \hat{\tau}_{\rm X}(y)\,y}{3} = \frac{16 y}{3} - \frac{16 y^3}{27} + \frac{16 y^5}{225} &=& 16 \,{t_{\rm 0X}}^{1/2} - \frac{184 \,{t_{\rm 0X}}^{3/2}}{25} - \frac{4198 \,{t_{\rm 0X}}^{5/2}}{4375}	,\\
p_{\rm 0X} \equiv \hat{p}_{\rm X} - 1 = -\frac{4 y^2}{9} + \frac{76 y^4}{675} - \frac{13672 y^6}{637\,875} &=& -4 \,{t_{\rm 0X}} + \frac{24 \,{t_{\rm 0X}}^2}{5} - \frac{816 \,{t_{\rm 0X}}^3}{875}	,\\
\frac{d\hat{p}_{\rm X}}{d\hat{\tau}_{\rm X}} = 4 - \frac{16 y^2}{15} + \frac{256 y^4}{1575} - \frac{64 y^6}{3375} &=& 4 - \frac{48 \,{t_{\rm 0X}}}{5} + \frac{2448 \,{t_{\rm 0X}}^2}{875} + \frac{56\,832 \,{t_{\rm 0X}}^3}{21\,875} ,\\
v_{0g} \equiv \widehat{V}_g -1 = \frac{2 y}{3} + \frac{2 y^2}{5} + \frac{8 y^3}{45}  &=& 2 \,{t_{\rm 0X}}^{1/2} + \frac{18 \,{t_{\rm 0X}}}{5} + \frac{147 \,{t_{\rm 0X}}^{3/2}}{25}	,\\
v_{0l} \equiv \widehat{V}_l-1  = -\frac{2 y}{3} + \frac{2 y^2}{5} - \frac{8 y^3}{45} &=& -2 \,{t_{\rm 0X}}^{1/2} + \frac{18 \,{t_{\rm 0X}}}{5} - \frac{147 \,{t_{\rm 0X}}^{3/2}}{25}	,\\
v_{\rm 0X} \equiv v_{0g}-v_{0l} = \frac{4 y}{3} + \frac{16 y^3}{45} + \frac{32 y^5}{1575} &=& 4 \,{t_{\rm 0X}}^{1/2} + \frac{294 \,{t_{\rm 0X}}^{3/2}}{25} + \frac{196\,081 \,{t_{\rm 0X}}^{5/2}}{8750}	,\\
\widehat{V}_{\rm X} = 2 + v_{0g} + v_{0l} = 2 + \frac{4 y^2}{5} + \frac{68 y^4}{525} + \frac{32 y^5}{1575} &=& 2 + \frac{36 \,{t_{\rm 0X}}}{5} + \frac{15\,984 \,{t_{\rm 0X}}^2}{875} + \frac{864 \,{t_{\rm 0X}}^{5/2}}{175} ,\\
n_{0g} \equiv \hat{n}_g -1 = -\frac{2 y}{3} + \frac{2 y^2}{45} + \frac{8 y^3}{135} &=& -2 \,{t_{\rm 0X}}^{1/2} + \frac{2 \,{t_{\rm 0X}}}{5} + \frac{13 \,{t_{\rm 0X}}^{3/2}}{25}	,\\
n_{0l} \equiv \hat{n}_l -1  = \frac{2 y}{3} + \frac{2 y^2}{45} - \frac{8 y^3}{135} &=& 2 \,{t_{\rm 0X}}^{1/2} + \frac{2 \,{t_{\rm 0X}}}{5} - \frac{13 \,{t_{\rm 0X}}^{3/2}}{25}	,\\
\Delta \hat{n}_{\rm X} \equiv\hat{n}_l-\hat{n}_g = \frac{4 y}{3} - \frac{16 y^3}{135} + \frac{544 y^5}{42\,525} &=& 4 \,{t_{\rm 0X}}^{1/2} - \frac{26 \,{t_{\rm 0X}}^{3/2}}{25} - \frac{1359 \,{t_{\rm 0X}}^{5/2}}{8750}, \label{Eq:DeltanX}\\
\hat{n}_{\rm ave\,X} \equiv \frac{\hat{n}_l+\hat{n}_g}{2} = 1 + \frac{2 y^2}{45} - \frac{2 y^4}{567} + \frac{4 y^6}{18\,225} &=& 1 + \frac{2 \,{t_{\rm 0X}}}{5} + \frac{128 \,{t_{\rm 0X}}^2}{875} + \frac{136 \,{t_{\rm 0X}}^3}{3125}.\label{Eq:naveHighT}
\eea
\ese
\end{widetext}
In these expressions, it is important to remember the definition $t_{\rm 0X}\equiv 1-\hat{\tau}_{\rm X}$ in Eq.~(\ref{Eq:t0XDef}). Thus, $t_{\rm 0X}$ increases as $\hat{\tau}_{\rm X}$ decreases below the critical temperature.  The leading expression in the last equality of each equation is the asymptotic critical behavior of the quantity as $\hat{\tau}\to 1^-$, as further discussed in Sec.~\ref{Sec:CritExps} below.  
\subsection{Thermodynamic Behaviors as $\hat{\tau}_{\rm X}\to0$}

Expanding the hyperbolic functions in the expression for $\hat{\tau}_{\rm X}(y)$ in Eq.~(\ref{Eq:tauFromy}) into their constituent exponentials gives
\be
\hat{\tau}_{\rm X} = \frac{27 \left[1 +  (y-1)e^{2 y} + y\right] \left(e^{4 y} - 4y\, e^{2 y} - 1\right)^2}{4 \left(e^{2 y} - 1\right) \left[( 2 y - 1)e^{4 y}  + (2 - 4 y^2)e^{2 y}  - 2 y -1\right]^2}.
\label{Eq:tXofyLowT}
\ee
The method of determining the behavior of $\hat{\tau}_{\rm X}(y)$ at low temperatures where $y\to\infty$ is the same for all  thermodynamic variables and functions. The behaviors of the numerator and denominator on the right side of Eq.~(\ref{Eq:tXofyLowT}) are dominated by the respective exponential with the highest power of~$y$.  Retaining only those exponentials and their prefactors, Eq.~(\ref{Eq:tXofyLowT}) becomes
\be
\hat{\tau}_{\rm X} = \frac{27 \left[(y-1) e^{2 y} \right] e^{8y}}{4 e^{2 y} \left[( 2 y - 1)e^{4 y} \right]^2} = \frac{27 (y-1)}{4 ( 2 y - 1)^2}.
\label{Eq:tauXVersusyLowT}
\ee
In this case, the exponentials cancel out but for other quantities they do not.  Taylor expanding the expression on the far right of Eq.~(\ref{Eq:tauXVersusyLowT}) in powers of $1/y$ to order $1/y^4$ gives
\be
\hat{\tau}_{\rm X}(y) = \frac{27}{16 y} - \frac{27}{64 y^3} - \frac{27}{64 y^4}\qquad (y\to\infty). 
\ee
Interestingly, the $y^{-2}$ term is zero.  Solving for $y(\hat{\tau}_{\rm X})$ to order ${\hat{\tau}_{\rm X}}^2$ gives
\be
y = \frac{27}{16 \,\hat{\tau}_{\rm X}} - \frac{4 \,\hat{\tau}_{\rm X}}{27} - \frac{64 \,{\hat{\tau}_{\rm X}}^2}{729} \qquad (\hat{\tau}_{\rm X}\to0),
\label{Eq:yoftXLowT}
\ee
where here the ${\hat{\tau}_{\rm X}}^0$ term is zero.  The entropy and latent heat for $\hat{\tau}_{\rm X}\to0$ are obtained by substituting Eq.~(\ref{Eq:yoftXLowT}) into Eqs.~(\ref{Eq:Entropyvsy}) and~(\ref{Eq:LatentHeatvsy}), respectively.  The low-temperature limiting behaviors of the other functions versus $y$ are obtained as above for $\hat{\tau}_{\rm X}(y)$.  If there is an exponential still present after the above reduction, it is of course retained.  In that case, only the leading order term of $y(\hat{\tau}_{\rm X})$ is inserted into the argument of the exponential, Eq.~(\ref{Eq:yoftXLowT}) is inserted for $y$ in the exponential prefactor and then a power series in $1/\hat{\tau}_{\rm X}$ is obtained for the prefactor.  The results for the low-order terms for $1/y\to0$ and $\hat{\tau}_{\rm X}\to0$ are 

\begin{widetext}
\bse
\label{Eqs:PropsLowT}
\bea
\frac{\Delta S_{\rm X}}{Nk_{\rm B}} = 2\,y  &=& \frac{27}{8 \,{\hat{\tau}_{\rm X}}} - \frac{8 \,{\hat{\tau}_{\rm X}}}{27} - \frac{128 \,{\hat{\tau}_{\rm X}}^2}{729} \label{Eq:DeltaSLowT}	,\\
\frac{L}{p_{\rm c}V_{\rm c}} = \frac{16 \,{\hat{\tau}_{\rm X}}\, y}{3} &=& 9 - \frac{64 \,{\hat{\tau}_{\rm X}}^2}{81} - \frac{1024 \,{\hat{\tau}_{\rm X}}^3}{2187} - \frac{8192 \,{\hat{\tau}_{\rm X}}^4}{59\,049},\label{Eq:LatentHeatLowT}\\
\hat{p}_{\rm X} = \frac{108 (y - 1)^2e^{-2 y}}{ (2 y - 1)^2}  &=& \left(\frac{1\,594\,323}{256 \,{\hat{\tau}_{\rm X}}^3} + \frac{98\,415}{64 \,{\hat{\tau}_{\rm X}}^2} - \frac{5103}{4 \,{\hat{\tau}_{\rm X}}} \right) e^{\frac{-27}{8 \,{\hat{\tau}_{\rm X}}}}	,\\
\widehat{V}_g = \frac{e^{2 y}}{6 (y - 1)} &=& \left(\frac{8 \,{\hat{\tau}_{\rm X}}}{81} + \frac{128 \,{\hat{\tau}_{\rm X}}^2}{2187} + \frac{2560 \,{\hat{\tau}_{\rm X}}^3}{59\,049} \right) e^{\frac{27}{8 \,{\hat{\tau}_{\rm X}}}},\\
\widehat{V}_l =  \frac{2 y - 1}{6 (y - 1)} &=& \frac{1}{3} + \frac{8 \,{\hat{\tau}_{\rm X}}}{81} + \frac{128 \,{\hat{\tau}_{\rm X}}^2}{2187} + \frac{2560 \,{\hat{\tau}_{\rm X}}^3}{59\,049},\\
\hat{n}_g = \frac{1}{\widehat{V}_g} = 6 (y - 1) e^{-2 y} &=& 6  \left(\frac{27}{16 \,{\hat{\tau}_{\rm X}}} - 1 - \frac{4 \,{\hat{\tau}_{\rm X}}}{27} - \frac{64 \,{\hat{\tau}_{\rm X}}^2}{729} \right) e^{\frac{-27}{8 \,{\hat{\tau}_{\rm X}}}},\label{Eq:ngas}\\
\hat{n}_l = \frac{1}{\widehat{V}_l} = \frac{6 (y - 1)}{2 y - 1} &=& 3 - \frac{8 \,{\hat{\tau}_{\rm X}}}{9} + \frac{64 \,{\hat{\tau}_{\rm X}}^2}{243} - \frac{1024 \,{\hat{\tau}_{\rm X}}^3}{6561},\label{Eq:nliquid}\\
\hat{n}_{\rm ave\,X} = \frac{\hat{n}_l}{2} = \frac{3 (y - 1)}{2 y - 1} &=& \frac{3}{2} - \frac{4 \,{\hat{\tau}_{\rm X}}}{9} + \frac{32 \,{\hat{\tau}_{\rm X}}^2}{243} - \frac{512 \,{\hat{\tau}_{\rm X}}^3}{6561}.\label{Eq:hatnXLowT}
\eea
\ese
\end{widetext}
In Eq.~(\ref{Eq:hatnXLowT}), we used the fact that $\hat{n}_g(\hat{\tau}_{\rm X})$ approaches zero exponentially for $\hat{\tau}_{\rm X}\to0$ instead of as a power law as does $\hat{n}_l(\hat{\tau}_{\rm X})$.  

\subsection{\label{Sec:Lekner2} Coexisting Liquid and Gas Densities, Transition Order Parameter, Temperature-Density Phase Diagram}

\begin{figure}[t]
\includegraphics[width=3.3in]{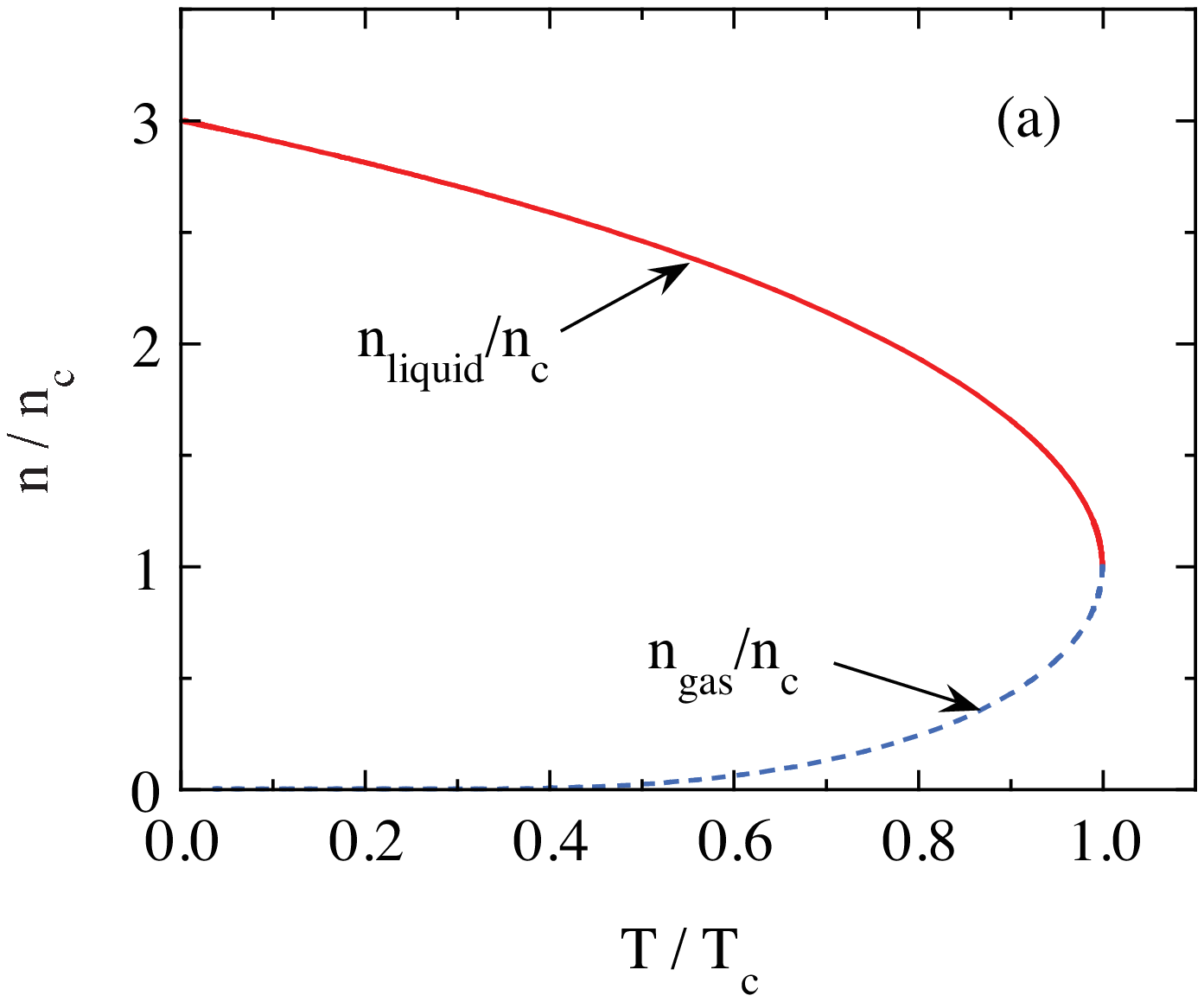}
\includegraphics[width=3.35in]{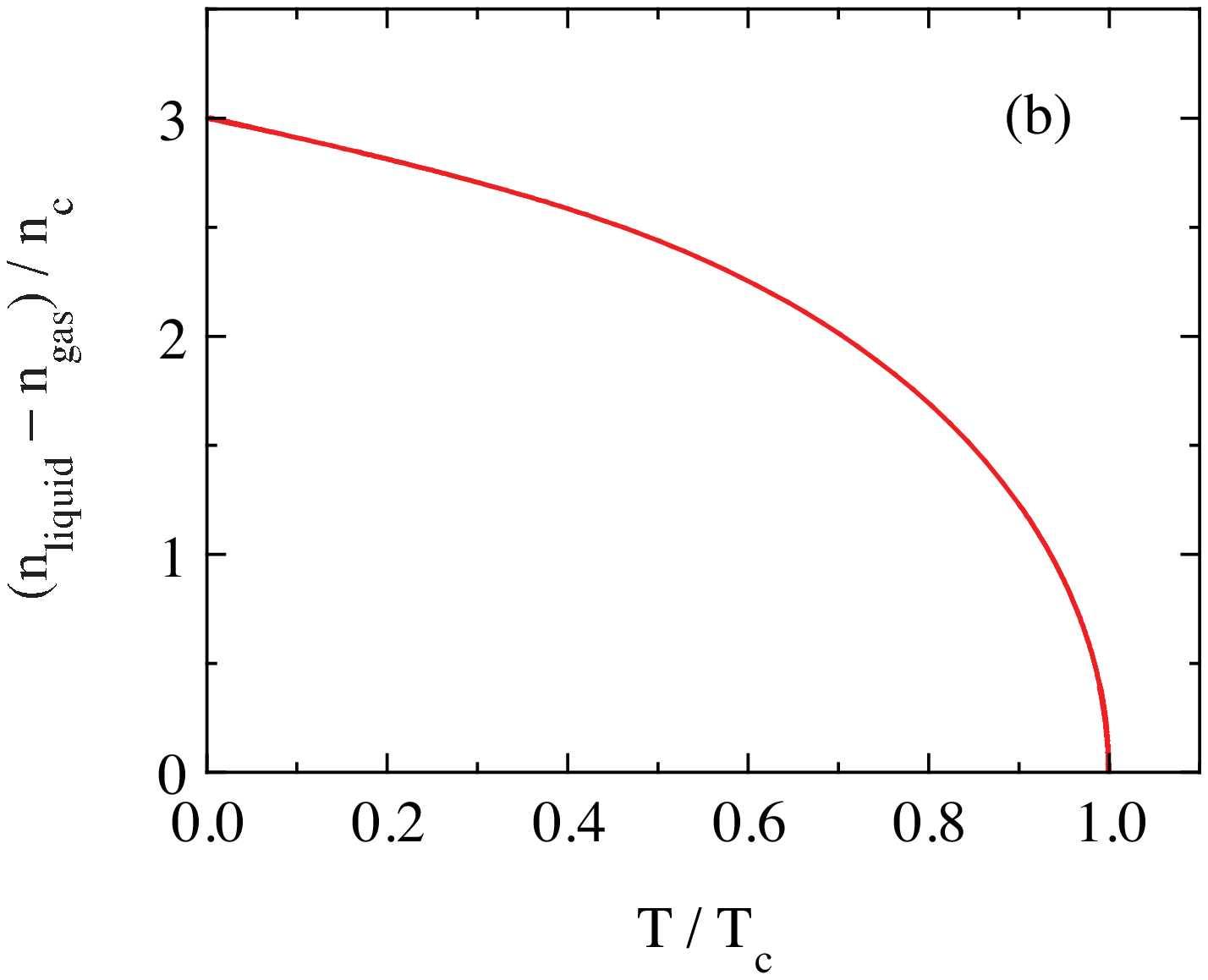}
\caption{(Color online) (a) Densities $n_g/n_{\rm c}$ and $n_l/n_{\rm c}$ of coexisting gas and liquid phases versus temperature $T/T_{\rm c}$ from Eqs.~(\ref{Eq:ngas}) and~(\ref{Eq:nliquid}), respectively. Here $n_{\rm c}$ is the density of the fluid at the critical point.  (b) Order parameter $\Delta \hat{n}_{\rm X}\equiv \hat{n}_l - \hat{n}_g = (n_l - n_g)/n_{\rm c}$ of the liquid-gas phase transition versus $T/T_{\rm c}$.}
\label{Fig:DensityVsTemp}  
\end{figure}

\begin{figure}[t]
\includegraphics[width=3.3in]{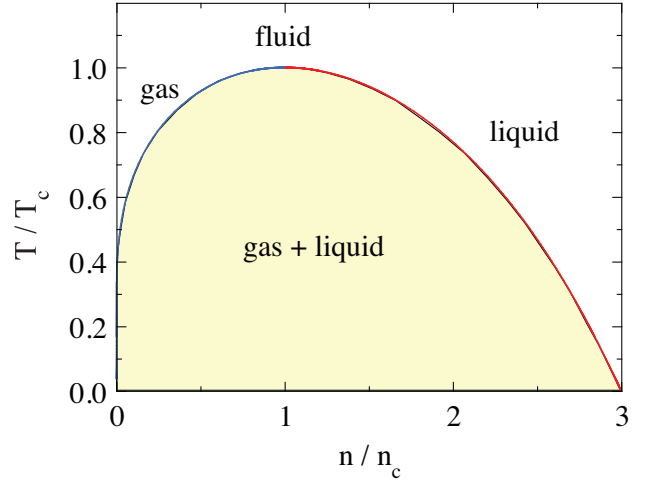}
\caption{(Color online) Reduced temperature $\hat{\tau} = T/T_{\rm c}$ versus reduced density~$\hat{n} = n/n_{\rm c}$ phase diagram for the van der Waals fluid, constructed by reversing the axes of Fig.~\ref{Fig:DensityVsTemp}(a).  This phase diagram is complementary to those in Figs.~\ref{Fig:vdW_p_vs_V_equilib} and~\ref{Fig:vdW_V_T_isotherms}.  The maximum density the system can have is $\hat{n} = 3$, at which there is no free volume left in which the molecules can move.}
\label{Fig:TempVsDensity}  
\end{figure}

\begin{figure}[t]
\includegraphics[width=3.3in]{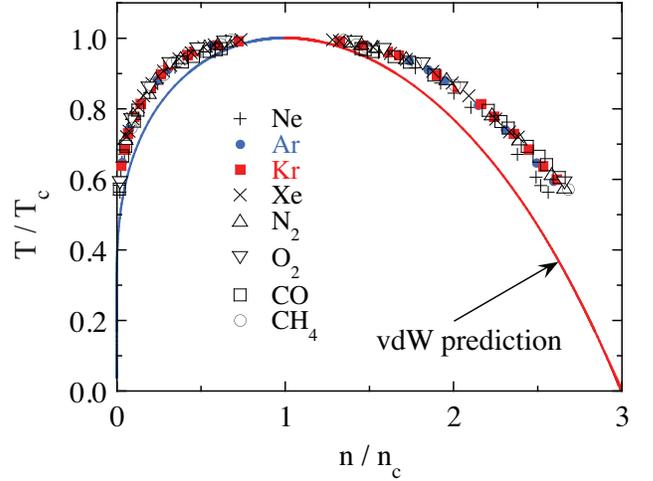}
\caption{(Color online) Reduced temperature $\hat{\tau} = T/T_{\rm c}$ versus reduced densities~$\hat{n} = n/n_{\rm c}$ of the coexisting gas and liquid phases of eight different fluids.\cite{Guggenheim1945}  Also shown as the solid curve is the prediction for the vdW fluid from Fig.~\ref{Fig:TempVsDensity}.  The experimental data follow a law of corresponding states,\cite{Guggenheim1945} but the one predicted for the vdW fluid does not accurately describe the data.}
\label{Fig:TempVsDensityExpt}  
\end{figure}

\begin{figure}[t]
\includegraphics[width=3.3in]{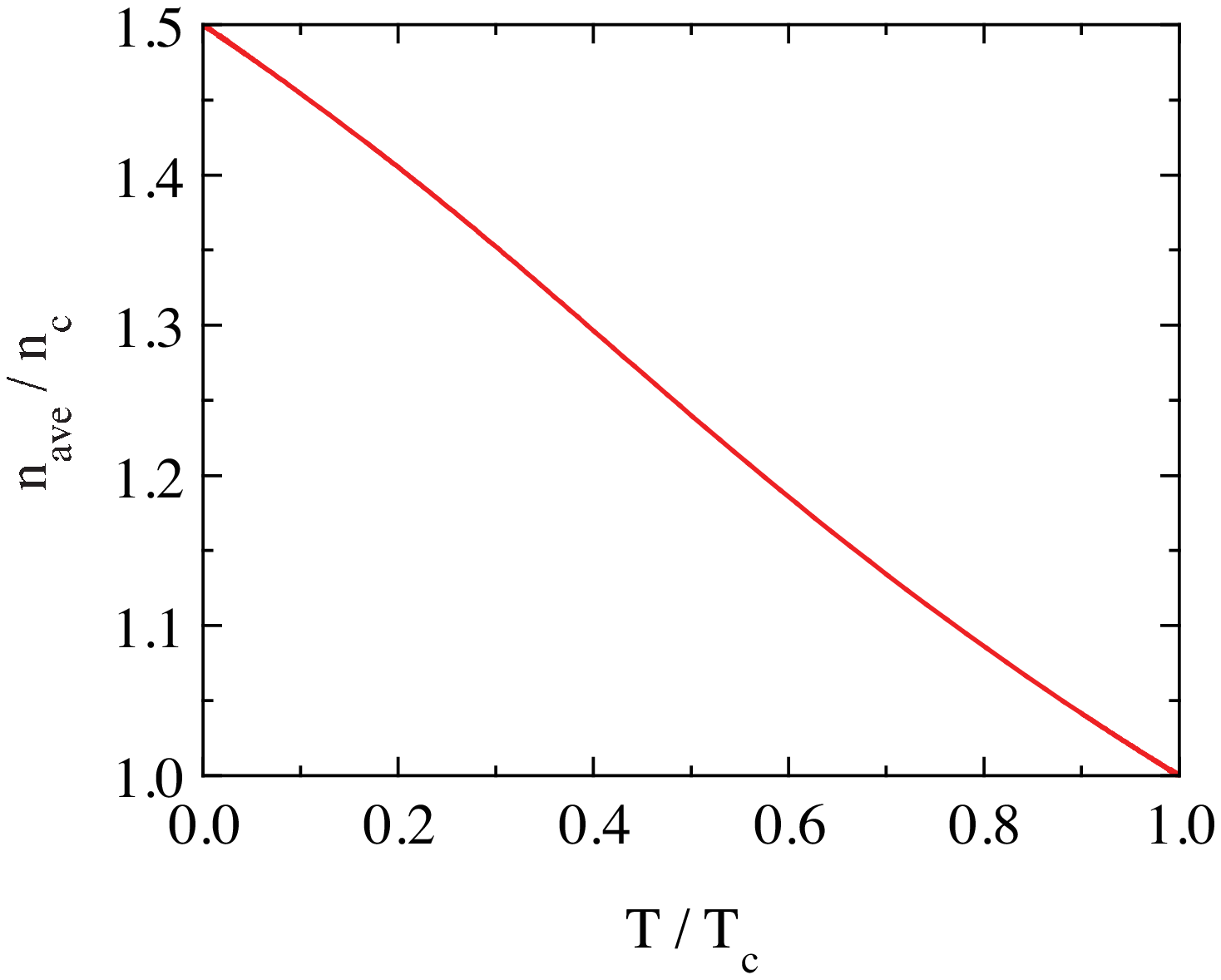}
\caption{(Color online) Average number density of coexisting liquid and gas phases $\hat{n}_{\rm ave} = (\hat{n}_l + \hat{n}_g)/2$ versus reduced temperature $\hat{\tau} = T/T_{\rm c}$ obtained from Eqs.~(\ref{Eq:tauFromy}) and~(\ref{Eq:nave}).  The curve has an S-shape and is hence not rectilinear.}
\label{Fig:DensityAveVsTemp}  
\end{figure}

The densities $\hat{n}_g$ and $\hat{n}_g$ of coexisting gas and liquid phases obtained from Eqs.~(\ref{Eq:ngasGen}) and~(\ref{Eq:nliquidGen}), respectively, together with Eq.~(\ref{Eq:tauFromy}) are plotted versus reduced temperature $T/T_{\rm c}$ in Fig.~\ref{Fig:DensityVsTemp}(a).  At the critical temperature they become the same.  The difference $\Delta\hat{n}_{\rm X}\equiv \hat{n}_l - \hat{n}_g$ is the order parameter of the gas-liquid transition and is plotted versus $T/T_{\rm c}$ in Fig.~\ref{Fig:DensityVsTemp}(b).

Data such as in Fig.~\ref{Fig:DensityVsTemp}(a) are often plotted with reversed axes, yielding the temperature-density phase diagram\cite{Guggenheim1945} in Fig.~\ref{Fig:TempVsDensity}.  The phase diagram and associated temperature dependences of the coexisting densities of the liquid and gas phases experimentally determined for eight different gases are shown in Fig.~\ref{Fig:TempVsDensityExpt}, along with the prediction for the vdW fluid from Fig.~\ref{Fig:TempVsDensity}. The experimental data were digitized from Fig.~2 of Ref.~\onlinecite{Guggenheim1945}.  Interestingly, the experimental data follow a law of corresponding states,\cite{Guggenheim1945} although that law does not quantitively agree with the one predicted for the vdW fluid.

A comparison of the high- and low-temperature limits of the average density $\hat{n}_{\rm ave}$ in Eqs.~(\ref{Eq:naveHighT}) and~(\ref{Eq:hatnXLowT}), respectively, of the coexisting gas and liquid phases shows that $\hat{n}_{\rm ave}$ is not a rectilinear function of temperature, which was noted by Lekner.\cite{Lekner1982}  Shown in Fig.~\ref{Fig:DensityAveVsTemp} is a plot of $\hat{n}_{\rm ave}$ versus $\hat{\tau}$ obtained from Eqs.~(\ref{Eq:tauFromy}) and~(\ref{Eq:nave}), which instead shows an S-shaped behavior.

\subsection{\label{Sec:LatentHeat} Latent Heat and Entropy of Vaporization}

\begin{figure}[t]
\includegraphics[width=2.8in]{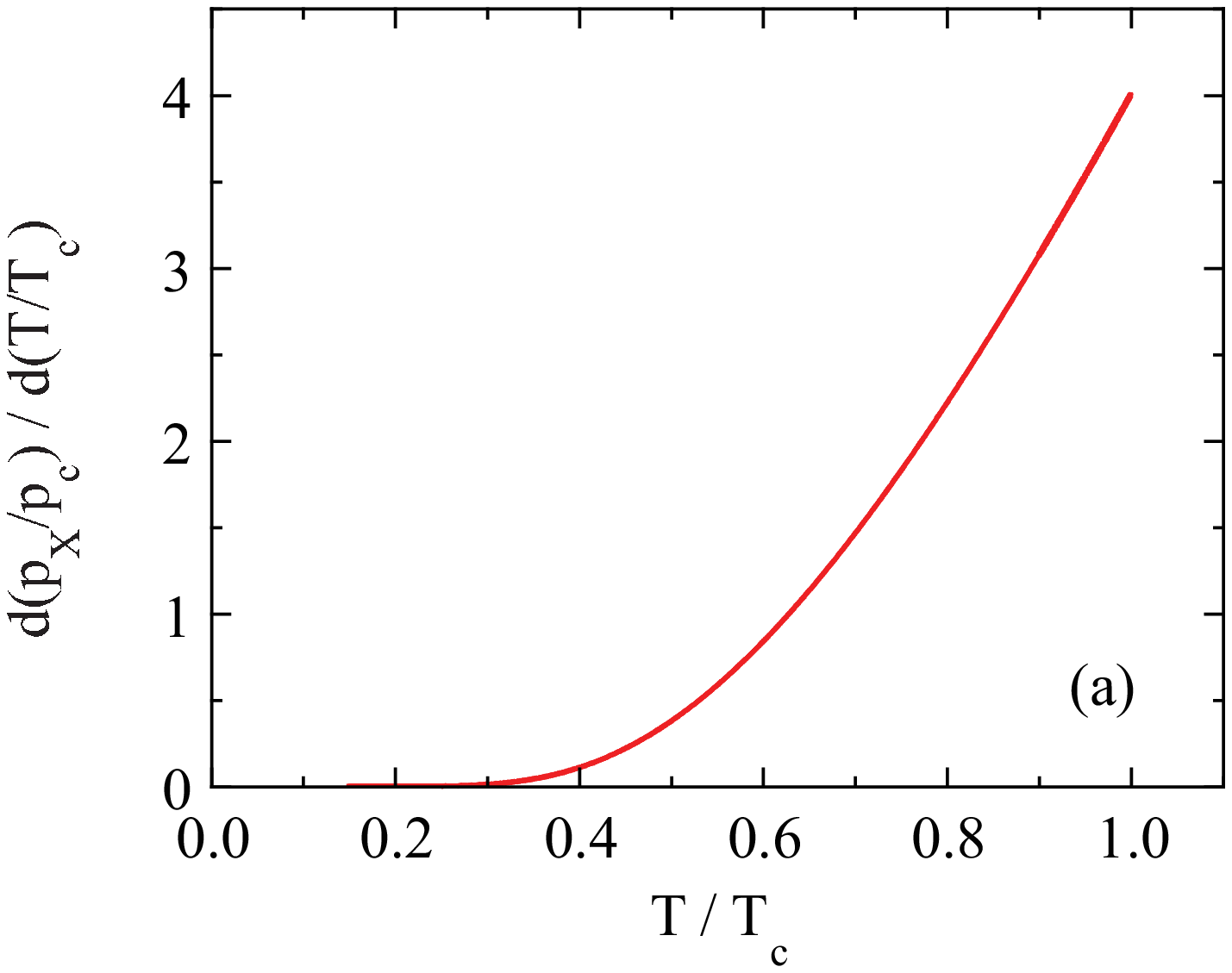}\vspace{-0.1in}
\includegraphics[width=2.8in]{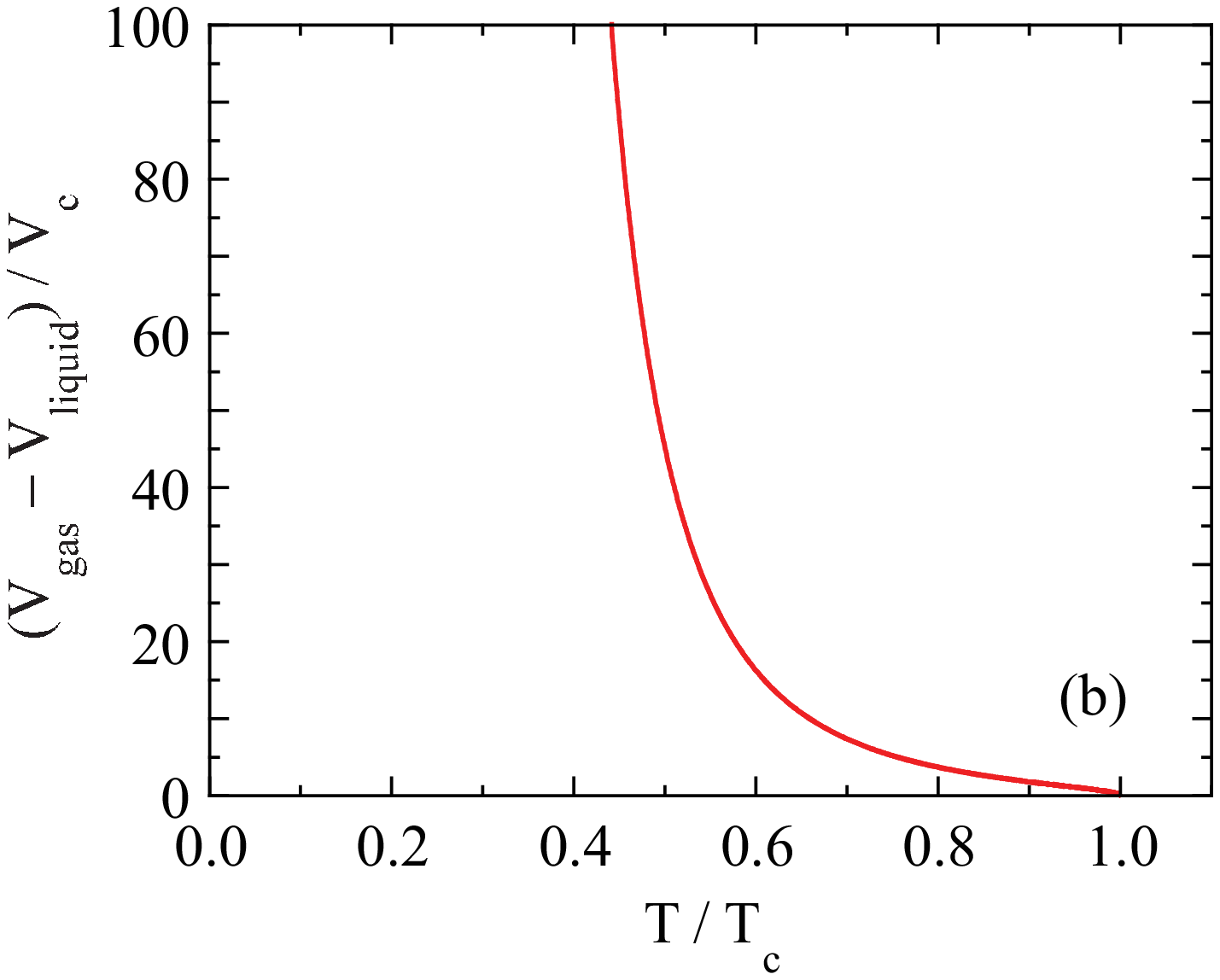}\vspace{-0.1in}
\includegraphics[width=2.8in]{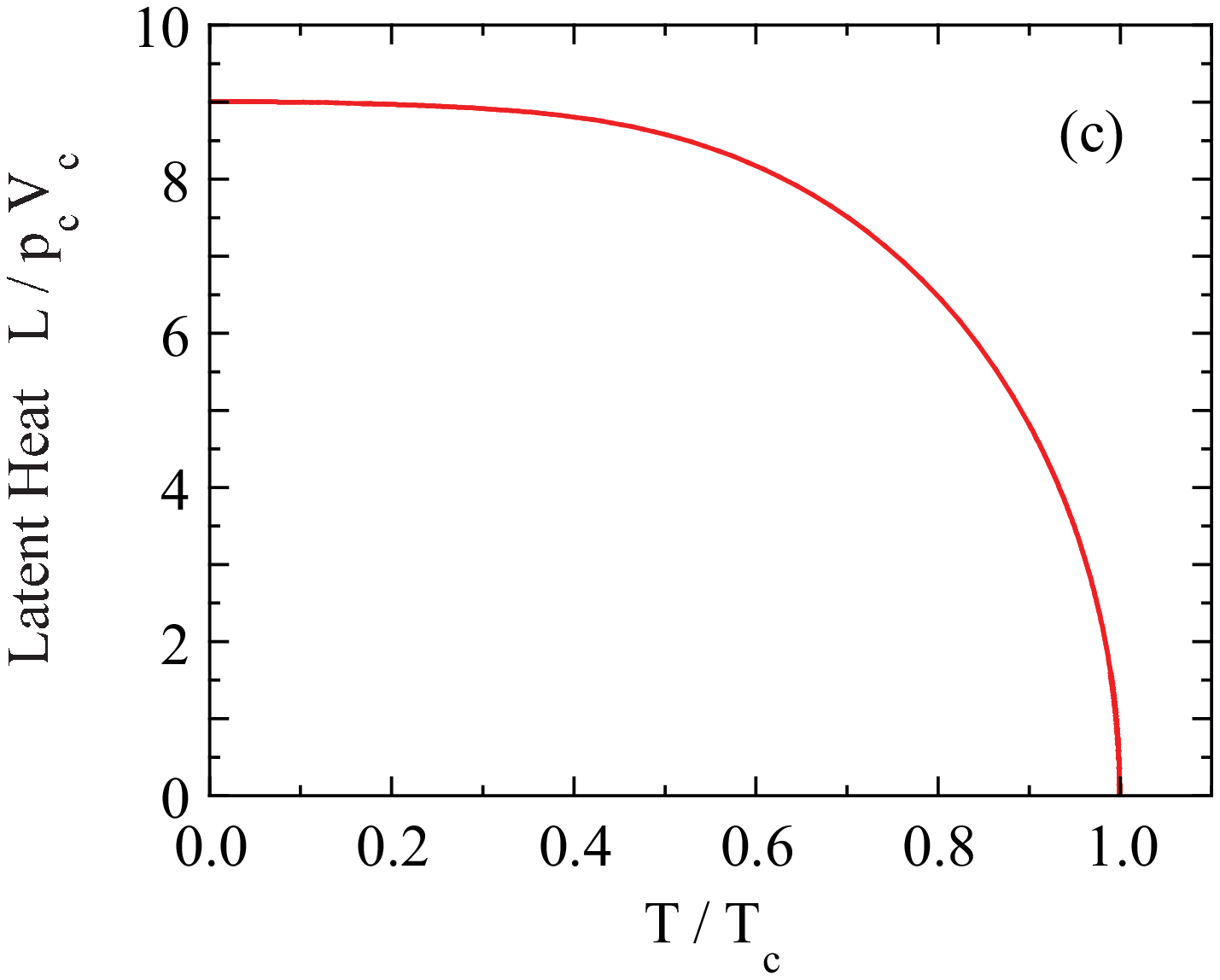}\vspace{-0.1in}
\caption{(Color online) (a) The derivative $d\hat{p}_{\rm X}/d\hat{\tau}$ versus $\hat{\tau}$ of the liquid-gas coexistence curve in Fig.~\ref{Fig:vdW_p_vs_T_phase_diag}.  (b) Difference in normalized volume $\Delta V_{\rm X}/V_{\rm c} \equiv (V_{\rm gas}-V_{\rm liquid})/V_{\rm c}$ on crossing the $p$-$T$ gas-liquid coexistence line in Fig.~\ref{Fig:vdW_p_vs_T_phase_diag} versus $\hat{\tau}$.  As also seen in Table~\ref{Tab:CoexistData}, $\Delta V_{\rm X}/V_{\rm c}$ diverges for $\hat{\tau}\to0$.  (c) Latent heat $L$ versus $\hat{\tau}$ obtained from either Eq.~(\ref{Eq:L}) using data as in panels~(a) and~(b) or from the parametric solution in Eq.~(\ref{Eq:LatentHeatvsy}). Both calculations give identical results to within their respective numerical accuracies.}
\label{Fig:Latent_Heat}  
\end{figure}

\begin{figure}[t]
\includegraphics[width=3.3in]{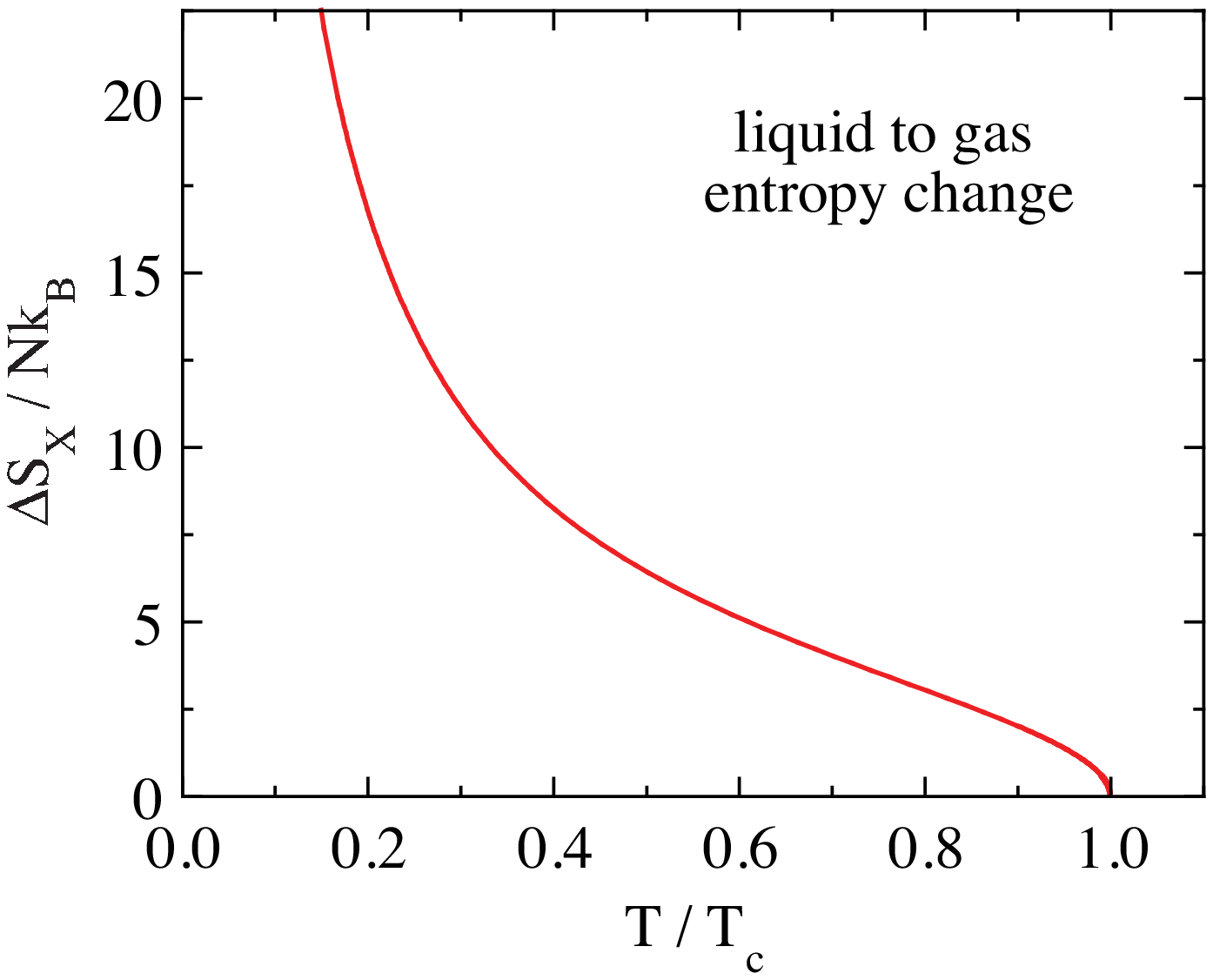}
\caption{(Color online) Entropy change at the first-order liquid-gas transition $\Delta S_{\rm X} \equiv S_{\rm gas}-S_{\rm liquid}$ versus reduced temperature $\hat{\tau} = T/T_{\rm c}$ in Fig.~\ref{Fig:vdW_p_vs_T_phase_diag}, obtained from either the data in Fig.~\ref{Fig:Latent_Heat}(c) using Eq.~(\ref{Eq:DS2}) or from the parametric solution in Eq.~(\ref{Eq:Entropyvsy}).  According to Eq.~(\ref{Eq:DS2}), $\Delta S_{\rm X}$ diverges to~$\infty$ as $T\to0$ because the latent heat of vaporization becomes constant at low temperatures as shown in Fig.~\ref{Fig:Latent_Heat}(c) and Eq.~(\ref{Eq:LatentHeatLowT}).}
\label{Fig:vdW_transition_entropy}  
\end{figure}

The normalized latent heat (or enthalpy) of vaporization $L/(p_{\rm c}V_{\rm c})$ on crossing the liquid-gas coexistence curve in Fig.~\ref{Fig:vdW_p_vs_T_phase_diag} is obtained parametrically versus $\hat{\tau}_{\rm X}$ from Eqs.~(\ref{Eq:tauFromy}) and~(\ref{Eq:LatentHeatvsy}) and is plotted in Fig.~\ref{Fig:Latent_Heat}(c).  The low-temperature behavior agrees with the prediction in Eq.~(\ref{Eq:LatentHeatLowT}).  From Fig.~\ref{Fig:Latent_Heat}(c), one sees that $L\to0$ as $T\to {T_{\rm c}}^-$, which is required because at temperatures at and above the critical temperature, the liquid and gas phases are no longer physically distinguishable.  The normalized entropy of vaporization $\Delta S_{\rm X}/(Nk_{\rm B})$ is obtained from Eqs.~(\ref{Eq:Entropyvsy}) and~(\ref{Eq:tauFromy}) and is plotted versus $\hat{\tau}_{\rm X}$ in Fig.~\ref{Fig:vdW_transition_entropy}.  The entropy difference is seen to diverge for $T\to0$, in agreement with Eq.~(\ref{Eq:DeltaSLowT}.)

From the $\hat{p}_{\rm X}(\hat{\tau})$ data and information about the change in volume $\Delta V_{\rm X} = V_{\rm gas} - V_{\rm liquid}$ across the coexistence line obtained above from numerical calculations, one can also determine~$L$ using the Clausius-Clapeyron equation
\bse
\be
\frac{dp_{\rm X}}{dT} = \frac{L}{T\,\Delta V_{\rm X}},
\ee
or
\be
L = T\,\Delta V_{\rm X} \,\frac{dp_{\rm X}}{dT}.
\label{Eq:CC}
\ee
One can write Eq.~(\ref{Eq:CC}) in terms of the reduced variables in Eq.~(\ref{Eq:RedVar}) as
\be
\frac{L}{p_{\rm c}V_{\rm c}} = \hat{\tau}\Delta\widehat{V}_{\rm X}\,\frac{d\hat{p}_{\rm X}}{d\hat{\tau}}.
\label{Eq:L}
\ee
\ese
The variation of $d\hat{p}_{\rm X}/d\hat{\tau}$ versus $\hat{\tau}$ obtained from Eqs.~(\ref{Eq:tauFromy}) and~(\ref{Eq:dpXdX}) and $\Delta\widehat{V}_{\rm X}$ from Eqs.~(\ref{Eq:tauFromy}) and~(\ref{Eq:DeltaVX}) versus $\hat{\tau}$ are shown in Figs.~\ref{Fig:Latent_Heat}(a) and~\ref{Fig:Latent_Heat}(b), respectively.  These behaviors when inserted into Eq.~(\ref{Eq:L}) give the same $L/(p_{\rm c}V_{\rm c})$ versus $\hat{\tau_{\rm X}}$ behavior as already obtained from Lekner's parametric solution in Fig.~\ref{Fig:Latent_Heat}(c).   

The entropy change $\Delta S_{\rm X}\equiv S_{\rm gas} - S_{\rm liquid}$ on moving left to right across the $p$-$T$ liquid-gas coexistence curve in Fig.~\ref{Fig:vdW_p_vs_T_phase_diag} is given in reduced units by Eq.~(\ref{Eq:ReducedL}) as
\be
\frac{\Delta S_{\rm X}(\hat{\tau})}{Nk_{\rm B}} = \frac{3}{8\,\hat{\tau}}\left[\frac{L(\hat{\tau})}{p_{\rm c}V_{\rm c}}\right].
\label{Eq:DS2}
\ee
The quantity in square brackets on the right side is already plotted in Fig.~\ref{Fig:Latent_Heat}(c).  Using these data and Eq.~(\ref{Eq:DS2}) yields $\Delta S_{\rm X}/(Nk_{\rm B})$ versus $\hat{\tau}$ which is the same as already plotted using Lekner's solution in Fig.~\ref{Fig:vdW_transition_entropy}.  The entropy change goes to zero at the critical point because gas and liquid phases cannot be distinguished at and above the critical temperature.  From Fig.~\ref{Fig:Latent_Heat}(a), the derivative $d\hat{p}_{\rm X}/d\hat{\tau}$ shows no critical divergence.  Therefore, according to Eq.~(\ref{Eq:L}), $\Delta S_{\rm X}$ shows the same critical behavior for $T\to T_{\rm c}$ as does $\Delta \widehat{V}_{\rm X}$ (or $\Delta\hat{n}_{\rm X}$, see Sec.~\ref{Sec:CritExps} below).

Since the latent heat becomes constant at low tempertures according to Fig.~\ref{Fig:Latent_Heat}(c), $\Delta S_{\rm X}$ diverges to~$\infty$ as $T\to0$ according to Eq.~(\ref{Eq:DS2}), as seen in Fig.~\ref{Fig:vdW_transition_entropy}.  This divergence violates the third law of thermodynamics which states that the entropy of a system must tend to a constant value (usually zero) as $T\to0$.  This behavior again demonstrates that like the ideal gas, the vdW fluid is classical.  This means that the predictions of the thermodynamic properties for either gas are only valid in the large-volume classical regime where the number density $N/V$ of the gas is much less than the quantum concentration $n_Q$.  Furthermore, the triple points of materials, where solid, gas and liquid coexist, typically occur at $T/T_{\rm c}\sim 1/2$, so this also limits the temperature range over which the vdW theory is applicable to real fluids.  However, study of the vdW fluid at lower temperatures is still of theoretical interest.

Representative values for the above properties associated with the coexistence curve that we calculated using the parametric equations~(\ref{Eqs:ParaSolns}) and~(\ref{Eqs:DerivedQuantities}) are listed in Table~\ref{Tab:LeknerTableA} in Appendix~\ref{App:Tables}.

\section{\label{Sec:CritExps} Critical Exponents}

\begin{table*}
\caption{\label{Tab:CritExps} Critical exponents for liquid-gas phase transitions.  The parameter $d\hat{\tau} \equiv \hat{\tau}-1 = \frac{T}{T_{\rm c}}-1$ measures the fractional deviation of the temperature from the critical temperature $T_{\rm c}$.  The unprimed critical exponents are for $\tau_0>1$ and the primed ones for $\tau_0<1$.  The prefactors of the powers of $\tau_0$ are the critical amplitudes.  The critical exponent $\alpha$ is for the heat capacity at constant volume, $\beta$ is for the liquid-gas number density difference (order parameter) $\hat{n}_{\rm l} - \hat{n}_{\rm g} = \left(\frac{1}{\widehat{V}_{\rm l}} - \frac{1}{\widehat{V}_{\rm g}}\right)$ on traversing the liquid-gas coexistence line such as in Fig.~\ref{Fig:vdW_p_vs_T_phase_diag}, $\gamma$ and $\gamma_{\rm T}$ are for the isothermal compressibility and $\delta$ is for the critical $p$-$V$ isotherm, which is the $p$-$V$ isotherm such as in Fig.~\ref{Fig:vdW_p_vs_V_equilib} that passes through the critical point $\hat{\tau}=\hat{p}=\widehat{V} = 1$.  The experimental critical exponents of fluids are described well by the exponents for the 3D Ising model as shown.\cite{Sengers2009}  The classical mean-field critical exponents and amplitudes for the van der Waals fluid are in the last two columns.  The definitions of the critical exponents except for $\gamma_{\rm T}$ and $\gamma_{\rm T}^\prime$ are from Ref.~\onlinecite{Stanley1971}, adapted to the definitions of the dimensionless reduced state variables in Eq.~(\ref{Eqs:deltaPars}). }
\begin{ruledtabular}
\begin{tabular}{cclccc}
exponent & definition  &  \hspace{0.05in}thermodynamic & 3D Ising model& van der Waals & van der Waals	\\
&& \hspace{0.35in}path & exponent & exponent  & amplitude 	\\
\hline
$\alpha$				& 	$C_{\rm V}/(Nk_{\rm B}) = a\, \tau_0^{-\alpha}$					&	$\tau_0 > 0$; $\hat{p},\ \widehat{V}=1$	& 0.110(3)	&	0			&  undefined			\\
$\alpha^\prime$		& 	$C_{\rm V}/(Nk_{\rm B}) = a^\prime (-\tau_0)^{-\alpha^\prime}$		&	$\tau_0 < 0$; $\hat{p},\ \widehat{V}=1$	&		&	0			&  undefined			\\
$\beta$				&	$\hat{n}_{\rm l} - \hat{n}_{\rm g} = b (-\tau_0)^\beta$			&	$\tau_0 < 0$; $\hat{p}$-$\hat{\tau}$ coexistence curve&  0.326(2)	&	$\frac{1}{2}$	&	$b = 4$			\\
$\gamma$				&	$\kappa_{\rm T}p_{\rm c} = g\, \tau_0^{-\gamma}$					&	$\tau_0 > 0$; $\widehat{V}=1$&	1.239(2)	&	1			&	$g = \frac{1}{6}$			\\
$\gamma^\prime$		&	$\kappa_{\rm T}p_{\rm c} = g^\prime (-\tau_0)^{-\gamma^\prime}$			&	$\tau_0 < 0$; $\hat{p}$-$\widehat{V}$ coexistence curves&		&	1			&	$g^\prime = \frac{1}{12}$		\\
$\gamma_{\rm p}$		&	$\kappa_{\rm T}p_{\rm c} = g_{\rm p}\, \tau_0^{-\gamma_{\rm T}}$		&	$\tau_0 > 0$; $\hat{p}=1, \widehat{V}\neq1$	&	  &	$\frac{2}{3}$	&	$g_{\rm p} = \frac{1}{3^{1/3}6} $		\\
${\gamma_{\rm p}}^\prime$	&	$\kappa_{\rm T}p_{\rm c} = {g_{\rm p}}^\prime (-\tau_0)^{-{\gamma_{\rm T}}^\prime}$&$\tau_0 < 0$; $\hat{p}=1, \widehat{V}\neq1$&	&$\frac{2}{3}$		&	${g_{\rm p}}^\prime = \frac{1}{3^{1/3}6} $\\
$\delta$				&	$p_0 = d|n_0|^{\delta}{\rm sgn}(n_0)$			&	$\tau_0 = 0$; $p_0,\ n_0\neq0$	&   4.80 (derived)  &		3		&	$d = \frac{3}{2}$			\\
\end{tabular}
\end{ruledtabular}
\end{table*}

We introduce the following notations that are useful when considering the approach to the critical point:
\bea
\tau_0 &\equiv& \hat{\tau}-1,\qquad t_0 \equiv -\tau_0 = 1-\hat{\tau},\qquad v_0 \equiv \widehat{V}-1,\nonumber\\*
n_0 &\equiv& \hat{n}-1\qquad p_0 \equiv  \hat{p}-1, \qquad \mu_0 \equiv \frac{\mu - \mu_{\rm c}}{\tau_{\rm c}},\label{Eqs:deltaPars}
\eea
where $\mu_{\rm c}$ is the chemical potential at the critical point.  The notation $t_0$ was previously introduced in Eq.~(\ref{Eq:t0XDef}) in the context of the coexistence curve.

The asymptotic critical exponents relate the changes in a property of a system to an infinitesimal deviation of a state variable from the critical point.  The definitions of some critical exponents relevant to the thermodynamics of the vdW fluid are given in Table~\ref{Tab:CritExps}.  Experimental data (see, e.g., Refs.~\onlinecite{Hocken1976}, \onlinecite{Sengers2009}) indicate that the liquid-gas transition belongs to the universality class of the three-dimensional Ising model, which is a three-dimensional (3D) model with short-range interactions and a scalar order parameter.\cite{Sengers2009}  The theoretical values for the critical exponents $\alpha,\ \beta$, $\gamma$ and $\delta$ for this model are given in Table~\ref{Tab:CritExps},\cite{Sengers2009} where the value of $\delta$ is obtained from the scaling law $\beta\delta=\beta+\gamma$.\cite{Kadanoff1967}  Also shown in Table~\ref{Tab:CritExps} are well-known critical exponents for the mean-field vdW fluid.\cite{Kadanoff1967,Stanley1971}  The critical exponents $\gamma_{\rm p}$ and $\gamma_{\rm p}^\prime$ are not commonly quoted.  One sees that the vdW exponents are in poor agreement with the 3D Ising model predictions and therefore also in poor agreement with the experimental values.  In the following we derive the vdW exponents together with the corresponding amplitudes expressed in our dimensionless reduced forms that are needed for comparison with our numerical calculations for temperatures near~$T_{\rm c}$.

\subsection{Heat Capacity at Constant Volume}

The heat capacity at constant volume in Eq.~(\ref{Eq:CVvdW}) for the vdW fluid is independent of temperature.  Therefore according to the definitions in Table~\ref{Tab:CritExps} one obtains
\be
\alpha=\alpha^\prime = 0
\ee
and the corresponding amplitudes $a$ and $a^\prime$ in Table~\ref{Tab:CritExps} are hence both undefined.

\subsection{Critical $\hat{p}$ versus $\widehat{V}$ Isotherm at $\hat{\tau}=1$}

\begin{figure}[t]
\includegraphics[width=3.3in]{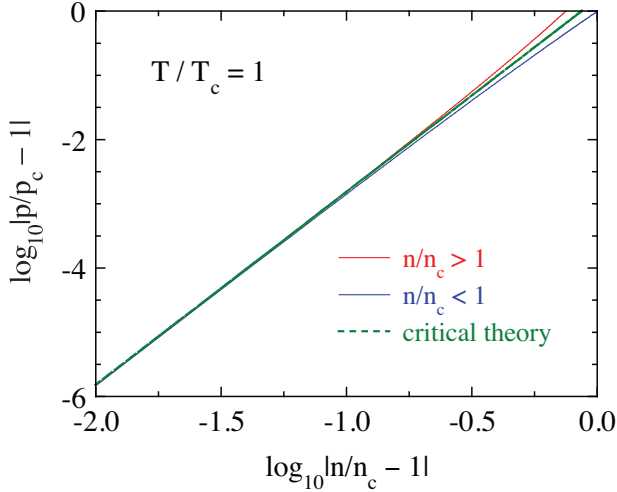}
\caption{(Color online) Log-log plot of $\hat{p}-1$ versus $\hat{n}-1$ for the critical $\hat{p}$ versus $\hat{n}$ isotherm at $\hat{\tau}=1$ in Fig.~\ref{Fig:vdW_p_vs_n} obtained from Eq.~(\ref{Eq:RedpVsRedn}).  On the far right, the top red solid curve is for $n/n_{\rm c}>1$ and the bottom blue solid curve is for $n/n_{\rm c}>1$.  The predicted asymptotic critical behavior $|\hat{p}-1| = d|\hat{n}-1|^\delta$ with $d=3/2$ and $\delta=3$ in Eq.~(\ref{Eq:pVCritVars}) is shown by the dashed green line.}
\label{Fig:vdW_p_vs_n_log}  
\end{figure}

For small deviations $p_0 = \hat{p}-1$ and $n_0 = \hat{n}-1$ of $\hat{p}$ and $\hat{n}$ from their critical values of unity and setting $\hat{\tau}=1$, a Taylor expansion of $p_0$  to lowest order in $n_0$ from Eq.~(\ref{Eq:RedpVsRedn}) gives
\be
p_0 = \frac{3}{2}n_0^3.
\label{Eq:pCrit}
\ee
A comparison of this result with the corresponding expression in Table~\ref{Tab:CritExps} yields the critical exponent $\delta$ and amplitude $d$ as
\be
\delta = 3,\qquad d = \frac{3}{2}.
\label{Eq:pVCritVars}
\ee

Thus the critical exponent and amplitude are the same on both sides of the critical point.  To determine the temperature region over which the critical behavior approximately describes the critical isotherm, shown in Fig.~\ref{Fig:vdW_p_vs_n_log} is a log-log plot of $\hat{p}-1$ versus $\hat{n}-1$.  The data are seen to follow the predicted asymptotic critical behavior $p_0 = n_0^\delta$ with amplitude $d=3/2$ and exponent $\delta=3$ for $0.9 \lesssim \hat{n} \lesssim 1.1$.  This region with $\hat{p} \sim 1\pm 0.001$ appears horizontal on the scale of Fig.~\ref{Fig:vdW_p_vs_n}.

\subsection{Critical Chemical Potential Isotherm versus $\hat{n}$}

From Eq.~(\ref{Eq:muNorm10}), there is no law of corresponding states for the behavior of the chemical potential of a vdW fluid near the critical point unless one only considers processes on the critical isotherm for which $\hat{\tau}=1$.  The value of the chemical potential at the critical point is
\be
\frac{\mu_{\rm c}}{\tau_{\rm c}} = -\left[\ln(2X) + \frac{7}{4}\right].
\ee
Expanding Eq.~(\ref{Eq:muNorm10}) with $\hat{\tau}=1$ in a Taylor series to the lowest three orders in $n_0\equiv \hat{n}-1$ gives
\be
\mu_0 \equiv \frac{\mu - \mu_{\rm c}}{\tau_{\rm c}} = \frac{9 n_0^3}{16} - \frac{9 n_0^4}{64} + \frac{81 n_0^5}{320}.
\ee
Comparing the first term of this expression with the critical behavior of the pressure in Eq.~(\ref{Eq:pCrit}), one obtains
\be
\mu_0 = \frac{3p_0}{8} = \frac{9 n_0^3}{16}.
\ee
Thus the critical exponent is the same as $\delta=3$ in Table~\ref{Tab:CritExps} for the critical $p$-$V$ isotherm but the amplitude is smaller than $d=\frac{3}{2}$ by a factor of 3/8.

\subsection{Liquid-Gas Transition Order Parameter}

We now determine the critical behavior of the difference in density between the liquid and gas phases on the coexistence line, which is the order parameter for the liquid-gas transition.  Equation~(\ref{Eq:DeltanX}) gives the asymptotic critical behavior as
\be
\Delta n_{\rm 0X} = n_{0l}-n_{0g} = 4\sqrt{-\tau_0}.
\label{Eq:dnl-dng}
\ee
Comparison of this expression with the definitions in Table~\ref{Tab:CritExps} gives the critical exponent and amplitude of the order parameter of the transition as
\be
\beta = \frac{1}{2},\qquad b = 4.
\label{Eq:DensityCritPars}
\ee
The exponent is typical of mean-field theories of second-order phase transitions.  The transition at the critical point is second order because the latent heat goes to zero at the critical point [see Fig.~\ref{Fig:Latent_Heat}(c) above].

\begin{figure}[t]
\includegraphics[width=3.3in]{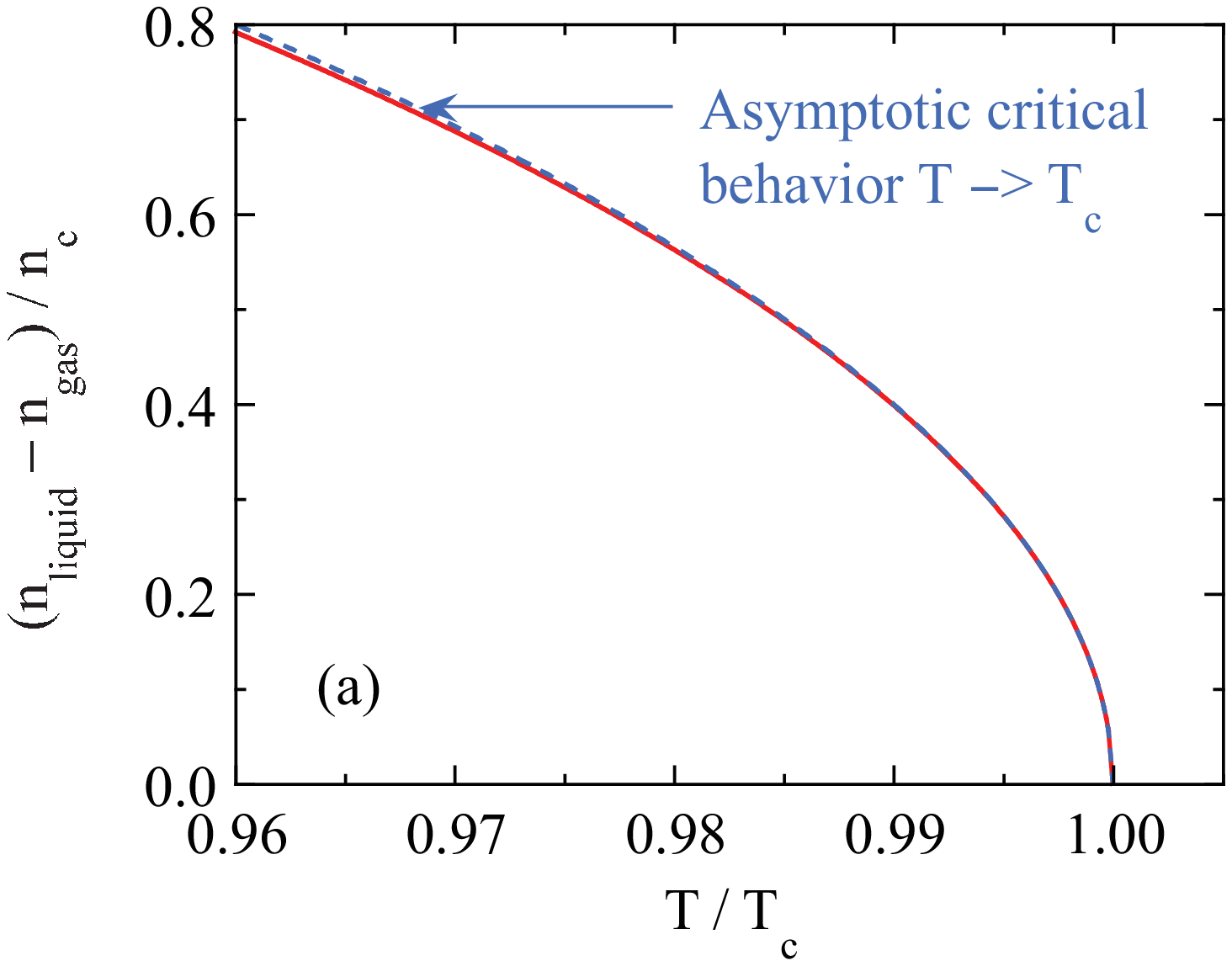}\vspace{-0.1in}
\includegraphics[width=3.3in]{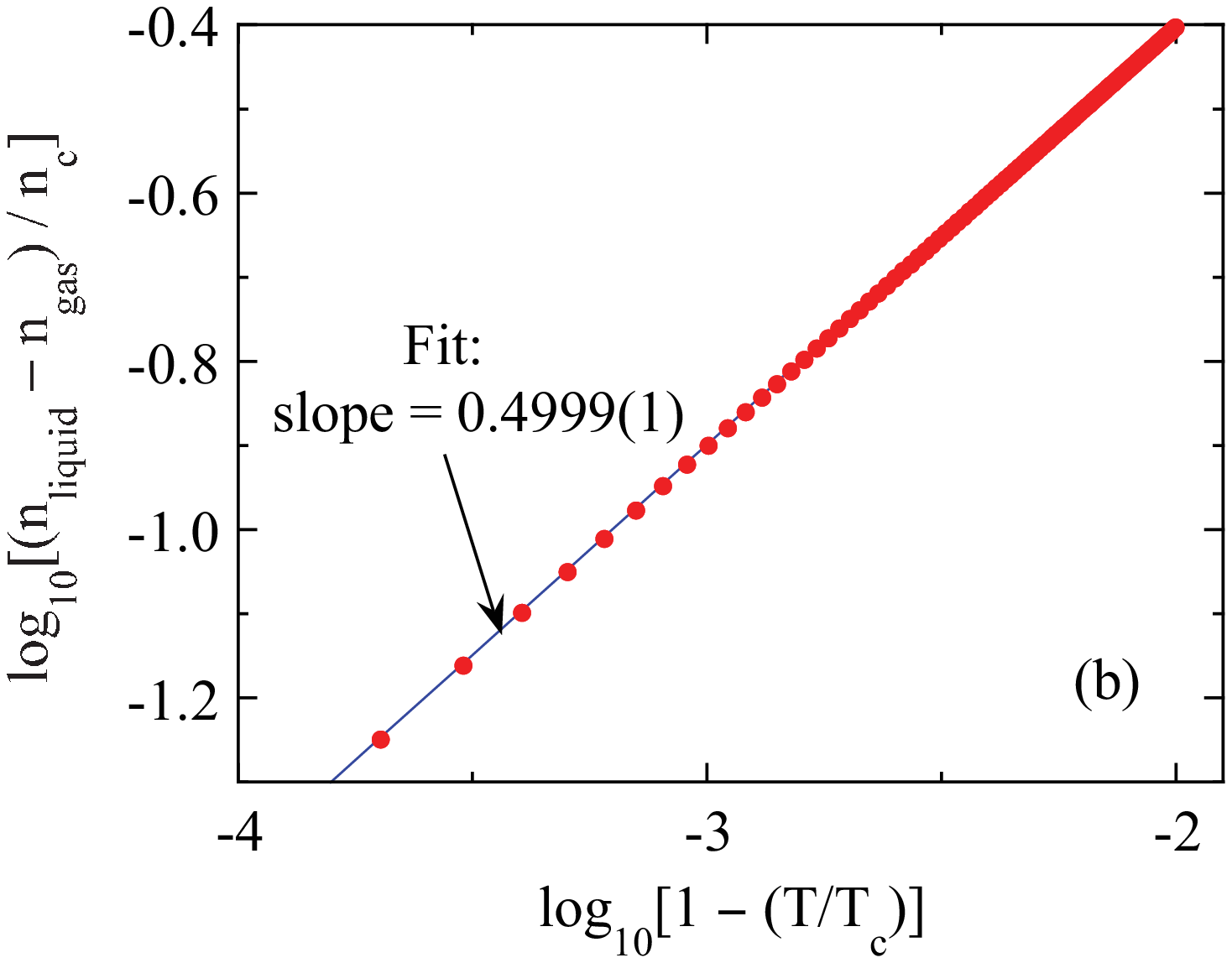}
\caption{(Color online) (a) Expanded plot from Fig.~\ref{Fig:DensityVsTemp}(b) of the difference $\Delta\hat{n}=\hat{n}_l-\hat{n}_g$ between the liquid and gas densities near the critical point $T/T_{\rm c}=1$ versus reduced temperature $T/T_{\rm c}$ (red curve).  $\Delta\hat{n}$ is the order parameter for the gas-liquid transition.  (b)~Logarithm to the base 10 of $\Delta\hat{n}$  versus the logarithm to the base 10 of the difference $1-T/T_{\rm c}$.  The fitted straight line for the data points on the lower left with $1-T/T_{\rm c}<10^{-3}$ is given by $\Delta n = b (1-T/T_{\rm c})^\beta$ with $b = 3.999$ and $\beta=0.4999$, consistent with the amplitude $b=4$ and exponent $\beta= 1/2$ predicted in Eq.~(\ref{Eq:DensityCritPars}).  This asymptotic critical behavior is shown as the dashed blue curve in panel~(a), where it is seen that this behavior is followed fairly accurately for $\hat{\tau}\gtrsim0.97$.}
\label{Fig:vdW_DeltaV_crit}
\end{figure}

Figure~\ref{Fig:vdW_DeltaV_crit}(a) shows an expanded plot of the data in Fig.~\ref{Fig:DensityVsTemp}(b) of the difference between the densities of the coexisting gas and liquid phases versus temperature.  One sees a sharp downturn as $T$ approaches $T_{\rm c}$.  In Fig.~\ref{Fig:vdW_DeltaV_crit}(b) is plotted  $\log_{10}\Delta n_{\rm 0X}$ versus $\log_{10}(1-\hat{\tau})$.  For $1-\hat{\tau} \lesssim 10^{-3}$, one obtains $\Delta n_{\rm 0X} = 3.999 (1-\hat{\tau})^{0.4999}$, consistent with the critical exponent and amplitude in Eq.~(\ref{Eq:DensityCritPars}).

\subsection{Isothermal Compressibility}

The critical behaviors of $\kappa_{\rm T}$ for $\hat{\tau}\to1^\pm$ are obtained using
\be
\frac{1}{\kappa_{\rm T}p_{\rm c}} = -\widehat{V}\left(\frac{\partial \hat{p}}{\partial \widehat{V}}\right)_{\hat{\tau}}\Bigg|_{\widehat{V}\to1,\ \hat{\tau}\to1} .
\label{Eq:1/kappa}
\ee
Differentiating the pressure in Eq.~(\ref{Eq:RedpVsRedV}) gives
\be
\left(\frac{\partial \hat{p}}{\partial \widehat{V}}\right)_{\hat{\tau}} = -\frac{24\hat{\tau}}{(3\widehat{V}-1)^2} + \frac{6}{\widehat{V}^3}.
\ee
Writing this expression in terms of the expansion parameters in Eqs.~(\ref{Eqs:deltaPars}) and Taylor expanding to lowest orders gives
\bse
\label{Eqs:dpdV}
\bea
\left(\frac{\partial \hat{p}}{\partial \widehat{V}}\right)_{\hat{\tau}} &=& -6\tau_0 -\frac{9}{2}v_0^2\quad (\hat{\tau}>1), \label{Eq:dpdVtau>1}\\*
\left(\frac{\partial \hat{p}}{\partial \widehat{V}}\right)_{\hat{\tau}} &=& \ \ 6t_0 -\frac{9}{2}v_0^2\quad (\hat{\tau}<1). \label{Eq:dpdVtau<1}
\eea
\ese

\subsubsection{Approach to the Critical Point along the Isochore $\widehat{V}=1$ with $\hat{\tau}>1$}

Setting $v_0=0$, Eqs.~(\ref{Eq:1/kappa}) and~(\ref{Eq:dpdVtau>1}) immediately give 
\be
\kappa_{\rm T}p_{\rm c} = \frac{1}{6\,\tau_0}.
\ee
Then the definition of the critical behavior of $\kappa_{\rm T}$ in Table~\ref{Tab:CritExps} gives the critical exponent $\gamma$ and amplitude $g$ as
\be
\gamma = 1,\qquad g = \frac{1}{6}.
\ee

\subsubsection{Approach to the Critical Point along Either Boundary of the Gas-Liquid Coexistence Curve on a $p$-$V$ Diagram with $\hat{\tau}<0$}

Defining the isothermal compressibility at either the pure gas or pure liquid coexistence points~G or~C on the $p$-$V$ isotherm in Fig.~\ref{Fig:Cubic_Eqn} has been used to define the critical behavior of $\kappa_{\rm T}$ for $\hat{\tau}<0$.  The value of $(\partial\hat{p}/\partial\widehat{V})_{\hat{\tau}}$ is the slope of a $p$-$V$ isotherm at either of those points since these become the same for $\hat{\tau}\to1$.  Referring to Fig.~\ref{Fig:Cubic_Eqn}, the reduced value of the volume $V_{\rm G}$ is what we called $\widehat{V}_l$ for the coexisting liquid phase above and $V_{\rm C}$ corresponds to  $\widehat{V}_g$ for the coexising gas phase. For either the liquid or gas phases, to lowest order in $t_0$ Eqs.~(\ref{Eqs:PropsHighT}) give
\be
v_{0l,g}^2 = 4\,t_0.
\ee 
Substituting this value into Eq.~(\ref{Eq:dpdVtau<1}) gives
\be
\left(\frac{\partial \hat{p}}{\partial \widehat{V}}\right)_{\hat{\tau}} = -12 t_0.
\label{Eq:dpdVtau<1B}
\ee
Then Eq.~(\ref{Eq:1/kappa}) becomes 
\be
\kappa_{\rm T}p_{\rm c} = \frac{1}{12\,t_0},
\ee
so the critical exponent $\gamma^\prime$ and amplitude $g^\prime$ are 
\be
\gamma^\prime = 1,\qquad g = \frac{1}{12}.
\ee
Thus the critical exponents are the same for $\hat{\tau}>1$ and~$\hat{\tau}<1$ but the amplitudes are a factor of two different.\cite{Stanley1971}  In the following section the critical exponents and amplitudes of $\alpha$ and~$\kappa_{\rm T}$ are found to be different from the above values when the critical point is approached along the critical isobar.

\subsection{Approach to the Critical Point along the Critical Isobar}

In this section we consider the critical exponents and amplitudes of $\kappa_{\rm T},\ \alpha$ and~$C_{\rm p}$ on approaching the critical point along the critical isobar, i.e. $\hat{p} = 1$.  We need these to compare with corresponding numerical calculations in Sec.~\ref{Sec:ConstPcalcs} below.  Setting $\hat{p}=1$, the equation of state~(\ref{Eq:vdWRed2}) becomes
\be
\hat{\tau} = \left(1+\frac{3}{\widehat{V}^2}\right)\frac{3\widehat{V}-1}{8}.
\ee
The lowest-order Taylor series expansion of this equation in the variables $\tau_0$ and~$v_0$ in Eqs.~(\ref{Eqs:deltaPars}) gives
\bea
\tau_0 &=& \frac{3}{8}v_0^3,\nonumber\\*
{\rm so}\quad  v_0 &=& \frac{2\tau_0^{1/3}}{3^{1/3}}\qquad (\hat{p}=1).
\label{Eq:p=1v0}
\eea

\subsubsection{Isothermal Compressibility}

For $\hat{\tau}>1$, substituting $v_0$ in Eq.~(\ref{Eq:p=1v0}) into~(\ref{Eqs:dpdV}) and using the definition in Eq.~(\ref{Eq:1/kappa}) gives, after a Taylor series expansion,
\bse
\label{Eqs:KappaCrit}
\be
\kappa_T p_{\rm c} = \frac{1}{3^{1/3}6}|\tau_0|^{-2/3} \qquad (\hat{p}=1).
\label{Eq:kappTp1}
\ee
Thus the critical exponents and amplitudes are
\be
\gamma_{\rm p} = \gamma_{\rm p}^\prime = \frac{2}{3},\qquad g_{\rm p} = g_{\rm p}^\prime = \frac{1}{3^{1/3}6}\approx 0.1156,
\label{Eq:kappaCritPars}
\ee
\ese
which are the same for $\hat{\tau} > 1$ and $\hat{\tau} < 1$ (see the critical isobar in the $V$-$T$ plane in Fig.~\ref{Fig:vdW_V_T_isotherms}).

\subsubsection{Volume Thermal Expansion Coefficient}

From Eq.~(\ref{Eq:alphaOVERkappaT}), near the critical point and on a path with $\hat{p}=1$ one has
\be
\frac{\alpha\tau_{\rm c}}{k_{\rm B}}\bigg|_{\hat{p}=1} = \left(\frac{\partial \hat{p}}{\partial \hat{\tau}}\right)_{\widehat{V}}\bigg|_{\widehat{V}=1}\kappa_{\rm T}p_{\rm c}.
\label{Eq:alphaCrit}
\ee
From Eq.~(\ref{Eq:RedpVsRedV}), the derivative is
\be
\left(\frac{\partial \hat{p}}{\partial \hat{\tau}}\right)_{\widehat{V}}\bigg|_{\widehat{V}=1} = 4.
\ee
Inserting this result and Eqs.~(\ref{Eqs:KappaCrit}) into~(\ref{Eq:alphaCrit}) gives
\be
\frac{\alpha\tau_{\rm c}}{k_{\rm B}} = 4 \kappa_{\rm T}p_{\rm c} = \frac{2}{3^{4/3}}|\tau_0|^{-2/3}\approx 0.4622 |\tau_0|^{-2/3}\quad (\hat{p}=1).
\label{Eq:alphaCritPars}
\ee
Thus $\alpha\tau_{\rm c}/k_{\rm B}$ has the same critical exponents as $\kappa_{\rm T} p_{\rm c}$ but with amplitudes four times larger than for $\kappa_{\rm T} p_{\rm c}$.

\subsubsection{Heat Capacity at Constant Pressure}

Inserting the above expressions for $\alpha\tau_{\rm c}/k_{\rm B}$ and $\kappa_{\rm T} p_{\rm c}$ near the critical point into the expression~(\ref{Eq:CpvdW}) for $C_{\rm p}$ gives
\be
\frac{C_{\rm p}}{Nk_{\rm B}} = \frac{3}{2} + \frac{1}{3^{1/3}}|\tau_0|^{-2/3} \approx \frac{3}{2} + 0.6934|\tau_0|^{-2/3}\quad (\hat{p}=1).
\label{Eq:CpCritp1}
\ee
When examining the critical part of $C_{\rm p}$, one would remove the noncritical part 3/2 due to $C_{\rm V}$ from the right-hand side.

The above critical exponents and amplitudes of the vdW fluid are listed in Table~\ref{Tab:CritExps}.

\section{\label{Sec:Hysteresis} Superheating and Supercooling}

\begin{figure}[t]
\includegraphics[width=3.3in]{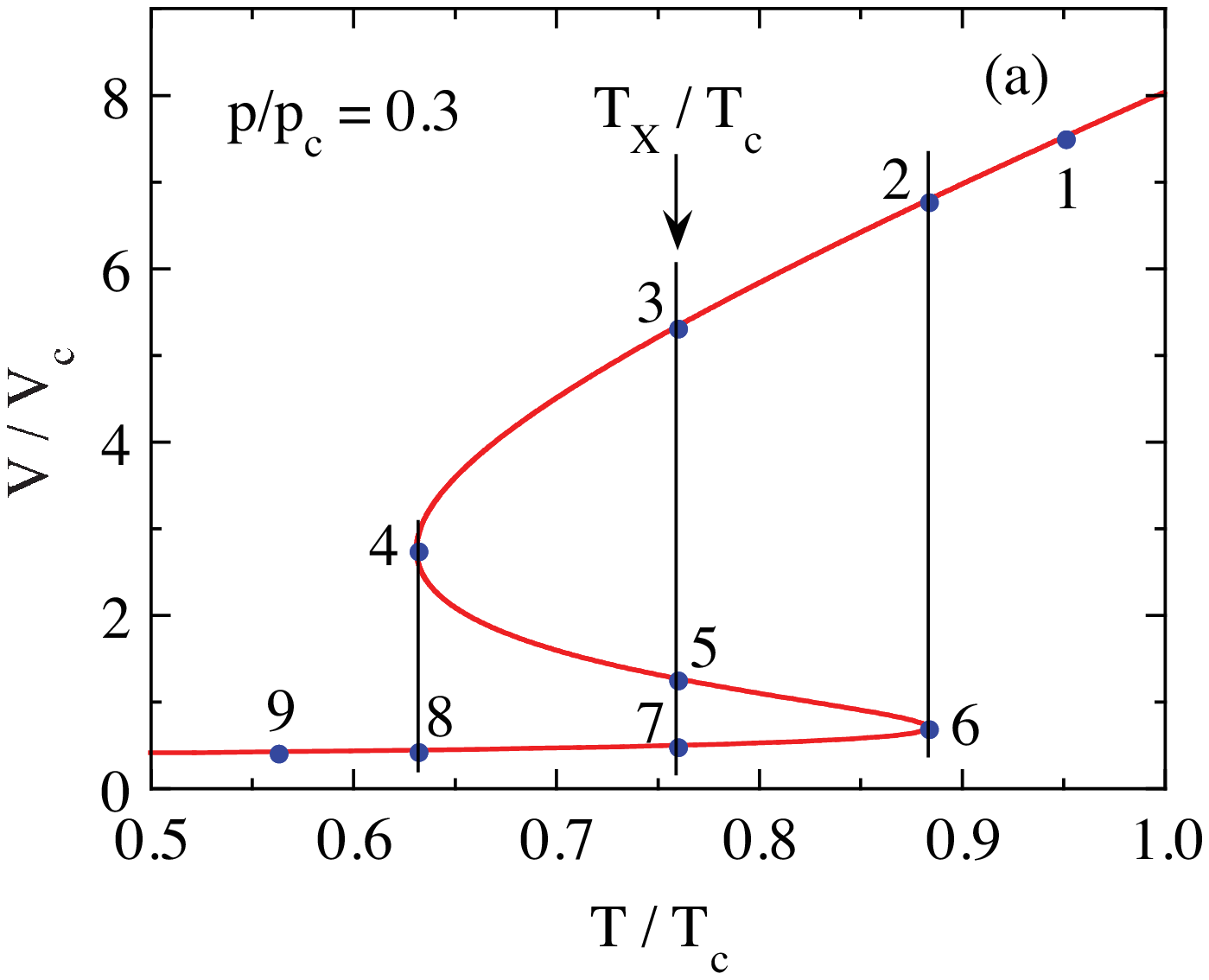}\vspace{-0.1in}
\includegraphics[width=3.3in]{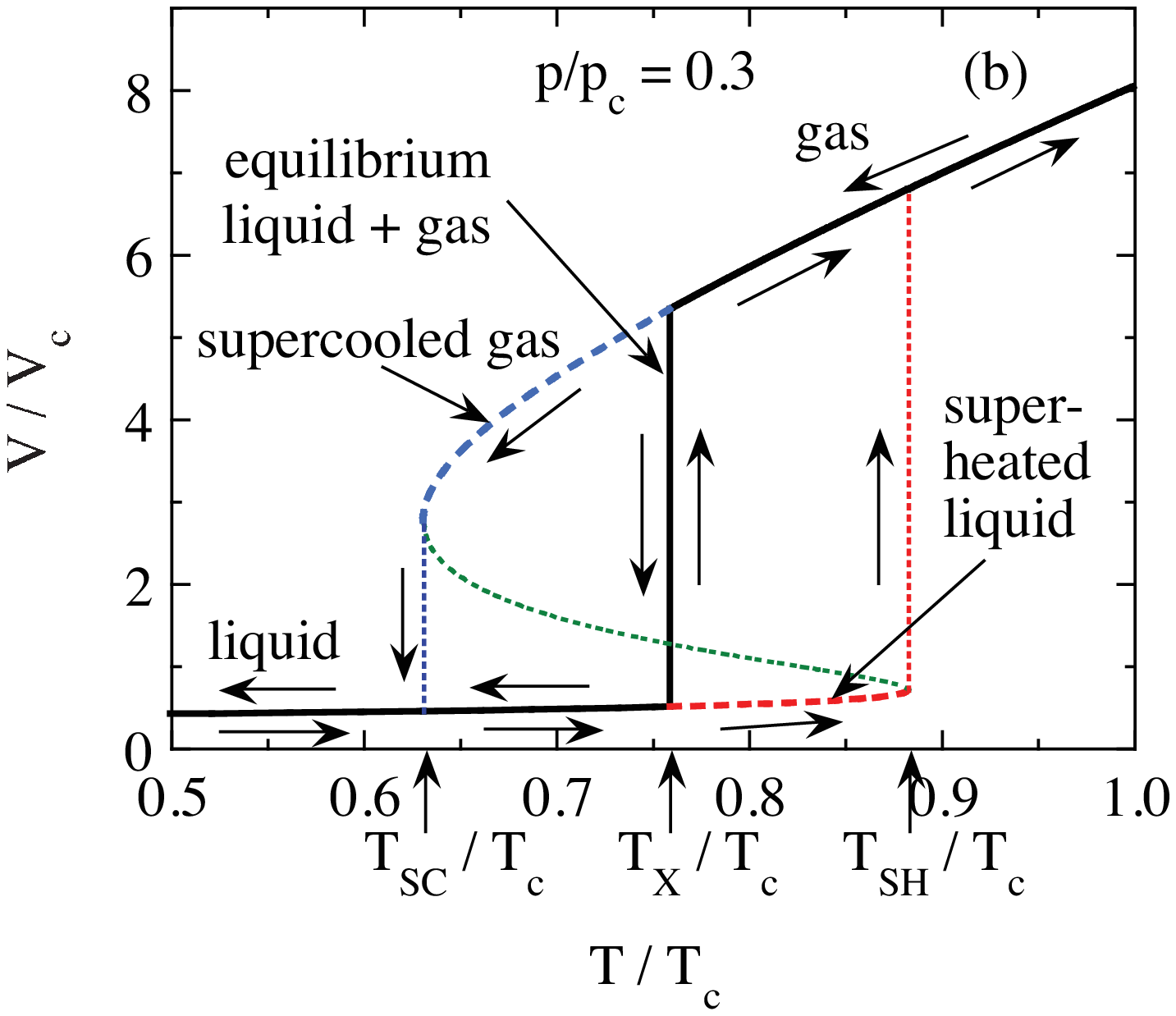}\vspace{-0.1in}
\caption{(Color online) (a) Reduced volume $\widehat{V} = V/V_{\rm c}$ versus reduced temperature $\hat{\tau} = T/T_{\rm c}$ at constant pressure $\hat{p} = p/p_{\rm c} = 0.3$.  Significant points on the curve are designated by numbers as discussed in the text.  The temperature $T_{\rm X}/T_{\rm c}$ is the equilibrium first-order phase transition temperature. (b) The equilibrium behavior of $\widehat{V}$ versus~$\hat{\tau}$ is shown by the heavy black line.  The vertical black line is the region of liquid-gas coexistence as in Fig.~\ref{Fig:vdW_V_T_isotherms}.  The regions of supercooled gas with minimum temperature $T_{\rm SC}/T_{\rm c}$ and superheated liquid with maximum temperature $T_{\rm SH}/T_{\rm c}$ are shown by dashed blue and red lines, respectively.  The dotted green curve with negative slope is not accessible by the vdW fluid.  The vertical blue and red dotted lines are not equilibrium mixtures of gas and liquid; these lines represent an irreversible decrease and increase in volume, respectively.}
\label{Fig:Volume_Hysteresis_p_0.3}  
\end{figure}

It is well known that systems exhibiting first-order phase transitions can exhibit hysteresis in the transition temperature and therefore in other physical properties upon cooling and warming, where the transition temperature is lower on cooling (supercooling) and higher on warming (superheating) than the equilibrium transition temperature $T_{\rm X}$.  The van der Waals fluid can also exhibit these properties. 

Shown in Fig.~\ref{Fig:Volume_Hysteresis_p_0.3}(a) is a plot of reduced volume $\widehat{V}$ versus reduced temperature $\hat{\tau}$ at fixed pressure $\hat{p}=0.3$ from Fig.~\ref{Fig:vdW_V_T_isobars} that is predicted from the vdW equation of state~(\ref{Eq:vdWRed2}).  Important points on the curve are labeled by numbers.  Points~1 and~9 correspond to pure gas and liquid phases, respectively, and are in the same regions as points~A and~I in the $p$ versus~$V$ isotherm in the top panel of  Fig.~\ref{Fig:vdW_V_T_isotherms}.  Points~3, 5 and~7 are at the gas-liquid coexistence temperature $T_{\rm X}/T_{\rm c}$ as in Fig.~\ref{Fig:vdW_V_T_isotherms}.  Points~3 and~7 thus correspond to points~C and~G in Fig.~\ref{Fig:Cubic_Eqn}.  Points~4 and~6 are points of infinite slope of $\widehat{V}$ versus~$\hat{\tau}$.  The volumes of points~4 and~6 do not correspond precisely with those points~D and~F in Fig.~\ref{Fig:Cubic_Eqn} at the same pressure, contrary to what might have been expected.  The curve 4-5-6 is not physically accessible by the vdW fluid because the thermal expansion coefficient is negative along this curve.

There is no physical constraint that prevents the system from following the path 1-2-3-4 on decreasing the temperature, where point~4 overshoots the equilibrium phase transition temperature.  When a liquid first nucleates as small droplets on cooling, the surface to volume ratio is large, and the surface tension (surface free energy) tends to prevent the liquid droplets from forming.  This free energy is not included in the treatment of the bulk van der Waals fluid, and represents a potential energy barrier that must be overcome by density fluctuations (homogeneous nucleation) or by interactions of the fluid with a surface or impurities (heterogeneous nucleation) before a bulk phase transition can occur.\cite{Kittel1980}  These mechanisms take time to nucleate sufficiently large liquid droplets, and therefore rapid cooling promotes this so-called supercooling.  The minimum possible supercooling temperature $\hat{\tau}_{\rm SC} = T_{\rm SC}/T_{\rm c}$ occurs at point~4 in Fig.~\ref{Fig:Volume_Hysteresis_p_0.3}(a), resulting in a supercooling curve given by the dashed blue curve in Fig.~\ref{Fig:Volume_Hysteresis_p_0.3}(b).  Similarly, superheating can occur with a maximum reduced temperature $\hat{\tau}_{\rm SH} = T_{\rm SH}/T_{\rm c}$ at point~6 in Fig.~\ref{Fig:Volume_Hysteresis_p_0.3}(a), resulting in a superheating curve given by the dashed red curve in Fig.~\ref{Fig:Volume_Hysteresis_p_0.3}(b).  The vertical dotted blue and red lines in Fig.~\ref{Fig:Volume_Hysteresis_p_0.3}(b) represent nonequilibrium irreversible transitions from supercooled gas to liquid and from superheated liquid to gas, respectively.  The latter can be dangerous because this transition can occur rapidly, resulting in explosive spattering of the liquid as it transforms into gas with a much higher volume.  The dashed supercooling and superheating curves in Fig.~\ref{Fig:Volume_Hysteresis_p_0.3}(b) are included in the $T$ versus~$V$ phase diagram in Fig.~\ref{Fig:p_T_vs_V_phase_diags}(b).

\begin{figure}[t]
\includegraphics[width=3.3in]{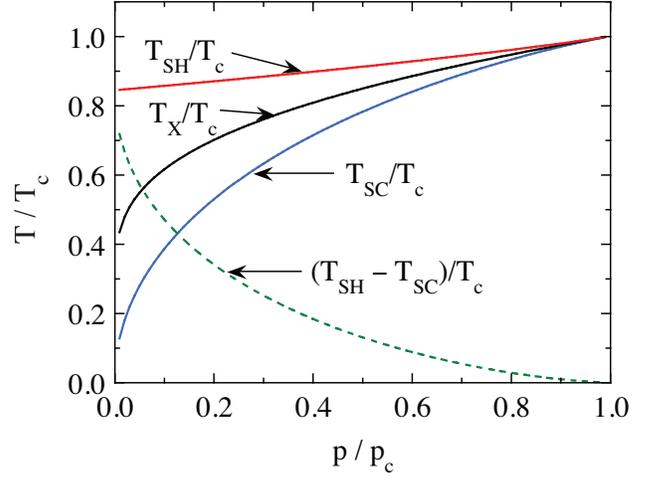}
\caption{(Color online) Reduced equilibrium phase transition temperature $\hat{\tau}_{\rm X} = T_{\rm X}/T_{\rm c}$, maximum superheating temperature $\hat{\tau}_{\rm SH} = T_{\rm SH}/T_{\rm c}$, minimum supercooling temperature $\hat{\tau}_{\rm SC} = T_{\rm SC}/T_{\rm c}$ and the difference $\hat{\tau}_{\rm SH} - \hat{\tau}_{\rm SC}$ versus reduced pressure~$\hat{p} = p/p_{\rm c}$ from $\hat{p}=0.01$ to~1.}
\label{Fig:SuperHtgCooling}  
\end{figure}

The reduced volumes~$\widehat{V}_4$ and~$\widehat{V}_6$ in Fig.~\ref{Fig:Volume_Hysteresis_p_0.3}(a) are calculated for a given pressure~$\hat{p}$ from the equation of state~(\ref{Eq:vdWRed2}) as the volumes at which $d\hat{\tau}/d\widehat{V}=0$ (and $d\widehat{V}/d\hat{\tau}=\infty$).  Then the reduced temperatures $\hat{\tau}_{\rm SC} = \hat{\tau}_4$ and~$\hat{\tau}_{\rm SH} = \hat{\tau}_6$ are determined from these volumes and the given $\hat{p}$ using Eq.~(\ref{Eq:vdWRed2}).  The equilibrium first-order transition temperature $\hat{\tau}_{\rm X}$ is calculated by first finding the volumes $\widehat{V}_3$ and~$\widehat{V}_7$ at which the chemical potentials in Eq.~(\ref{Eq:muNorm2}) are equal, where one also requires that $\hat{\tau}_3 = \hat{\tau}_6$ without explicitly calculating their values.  Once these volumes are determined, the value of $\hat{\tau}_{\rm X} =\hat{\tau}_3 = \hat{\tau}_6$ is determined from Eq.~(\ref{Eq:vdWRed2}).  Plots of $\hat{\tau}_{\rm X}$, $\hat{\tau}_{\rm SH}$, $\hat{\tau}_{\rm SC}$ and~$\hat{\tau}_{\rm SH} - \hat{\tau}_{\rm SC}$ are shown versus $\hat{p}$ from $\hat{p} = 0.01$ to~1 in Fig.~\ref{Fig:SuperHtgCooling}.  One sees that $\hat{\tau}_{\rm X}$ is roughly midway between  $\hat{\tau}_{\rm SH}$ and $\hat{\tau}_{\rm SC}$ over the whole pressure range, with $\hat{\tau}_{\rm SH} - \hat{\tau}_{\rm SC}$ decreasing monotonically with increasing temperature and going to zero at $\hat{p}=1$ as expected.

\section{\label{Sec:ConstPcalcs} Numerical Calculations at Constant Pressure of the Entropy, Internal Energy, Enthalpy, Thermal Expansion Coefficient, Isothermal Compressibility and Heat Capacity at Constant Pressure versus Temperature}

\subsection{Results for $p \geq p_{\rm c}$}

\begin{figure}[t]
\includegraphics[width=3.3in]{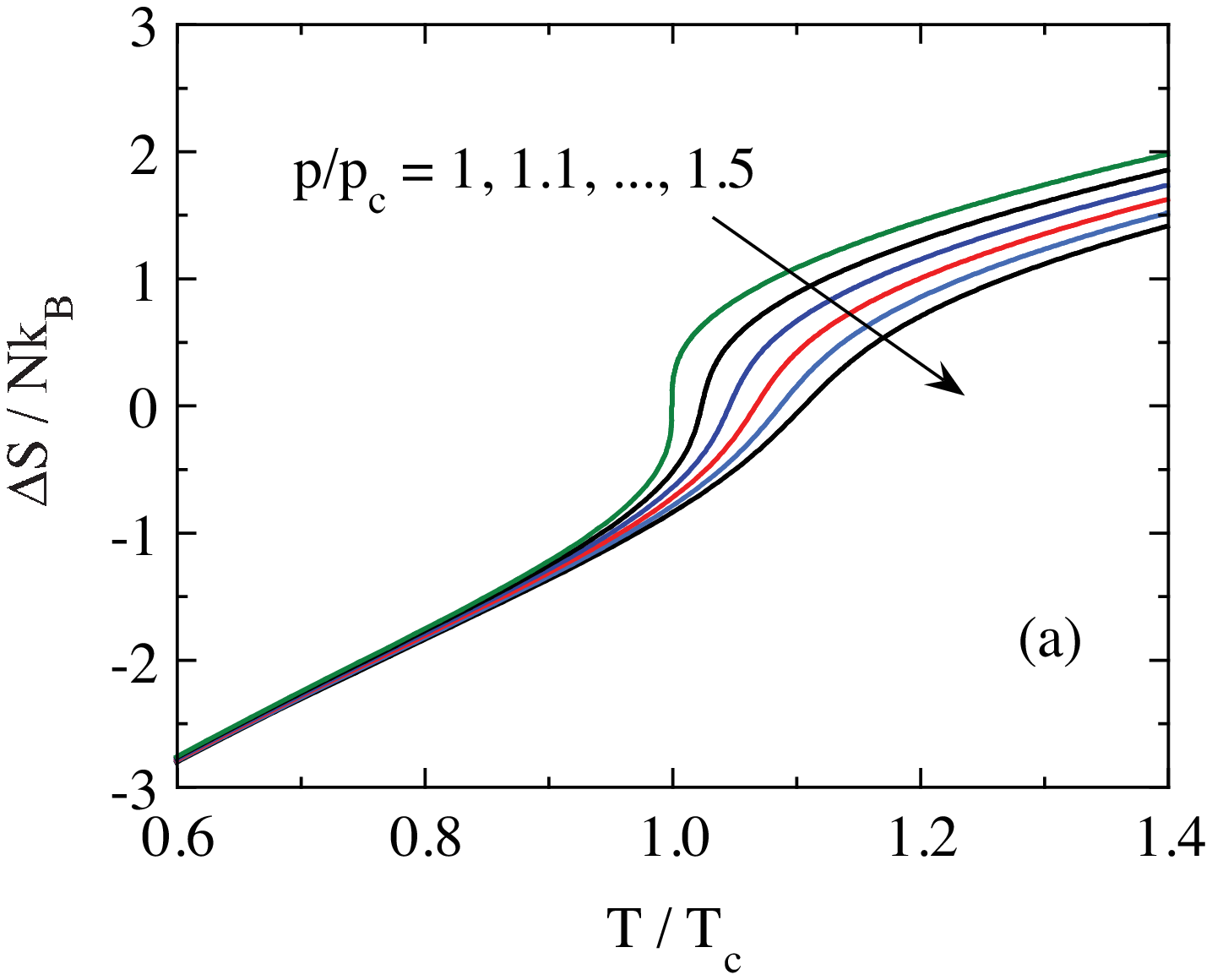}
\includegraphics[width=3.3in]{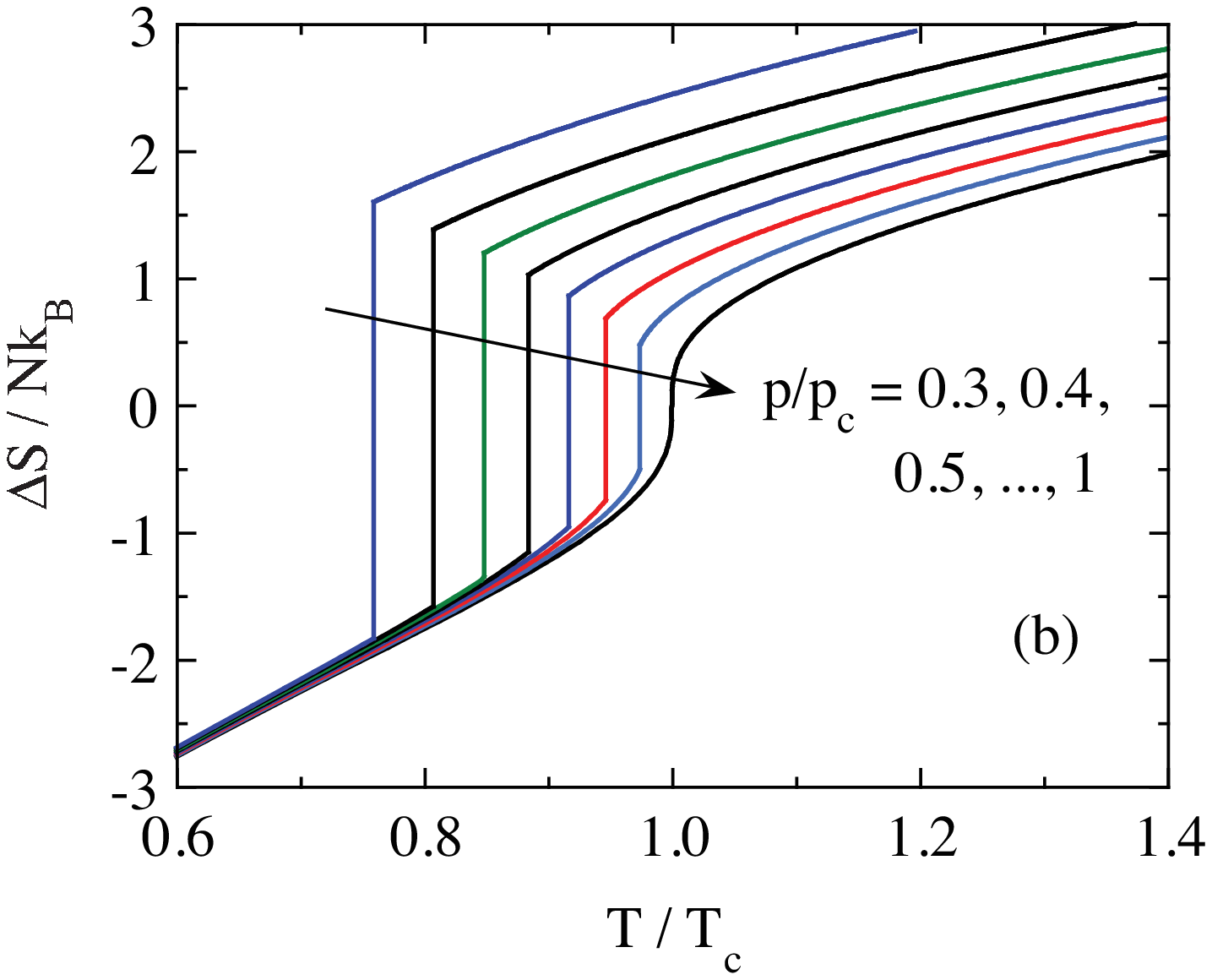}
\caption{(Color online) Equilibrium isobars of the difference in entropy  from that the critical point $\Delta S/Nk_{\rm B}$ versus reduced temperature $\hat{\tau} = T/T_{\rm c}$ calculated using Eq.~(\ref{Eq:DeltasigmaRed}) for (a) $\hat{p} = p/p_{\rm c} \geq 1$ and~(b) $\hat{p} \leq 1$.  The difference in entropy is calculated with respect to the entropy at the critical point $\hat{\tau}=\widehat{V}=\hat{p} = 1$.}
\label{Fig:vdW_const_pressure_sigma}  
\end{figure}

\begin{figure}[t]
\includegraphics[width=3.3in]{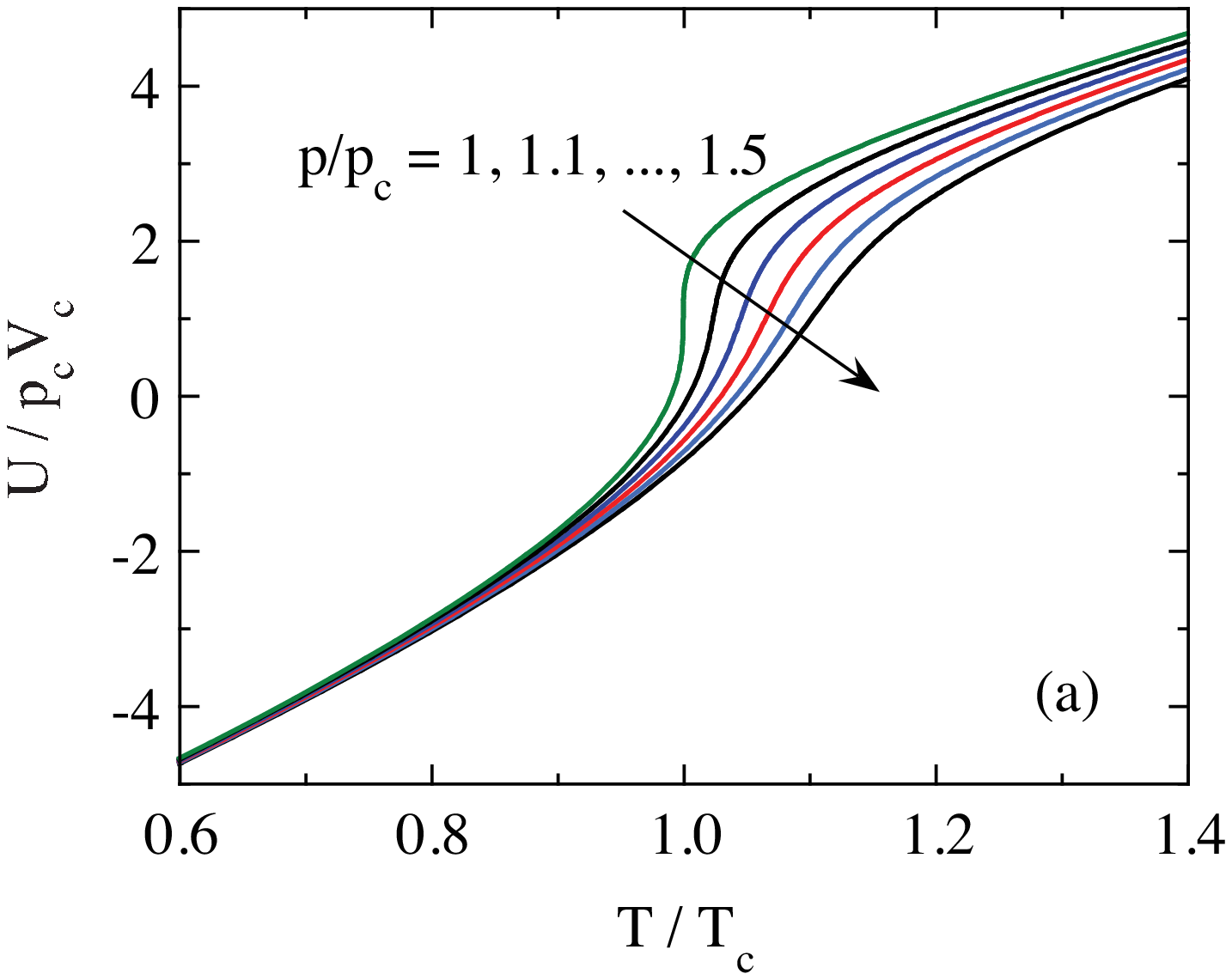}
\includegraphics[width=3.3in]{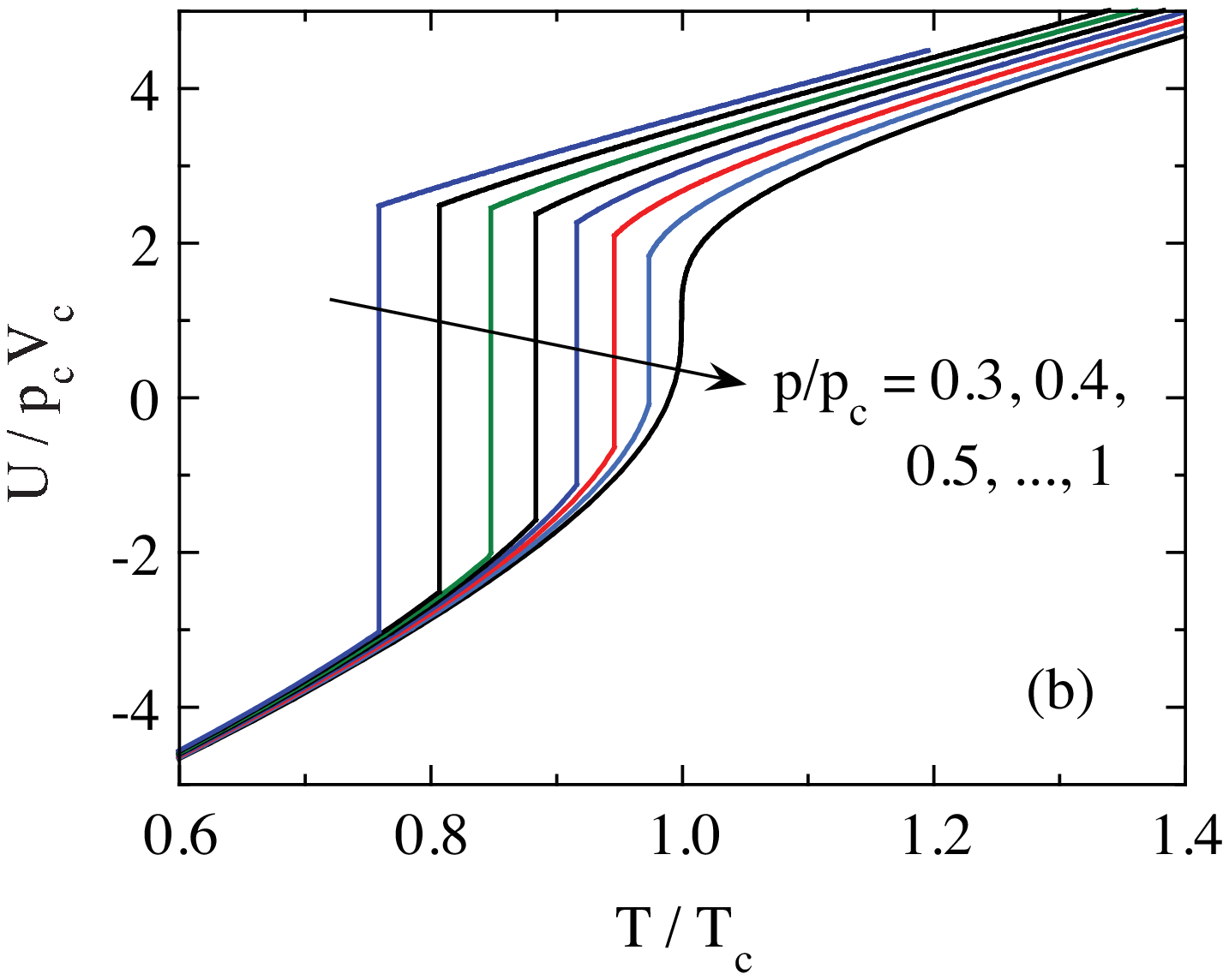}
\caption{(Color online) Equilibrium isobars of reduced internal energy $U/(p_{\rm c}V_{\rm c})$ versus reduced temperature $\hat{\tau} = T/T_{\rm c}$ calculated using Eq.~(\ref{Eq:URed}) for (a) $\hat{p} = p/p_{\rm c} \geq 1$ and~(b) $\hat{p} \leq 1$.}
\label{Fig:vdW_const_pressure_U}  
\end{figure}

\begin{figure}[t]
\includegraphics[width=3.3in]{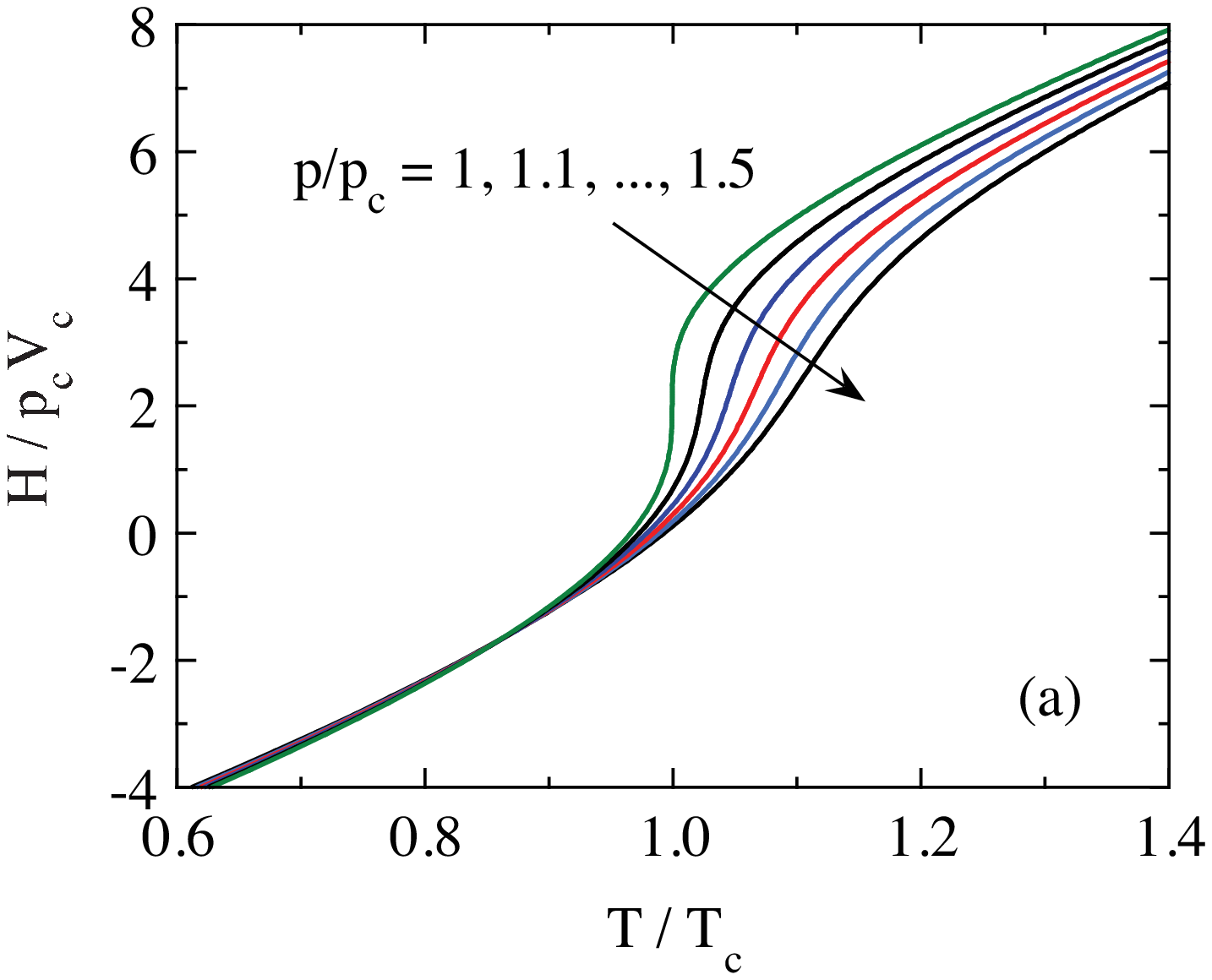}
\includegraphics[width=3.3in]{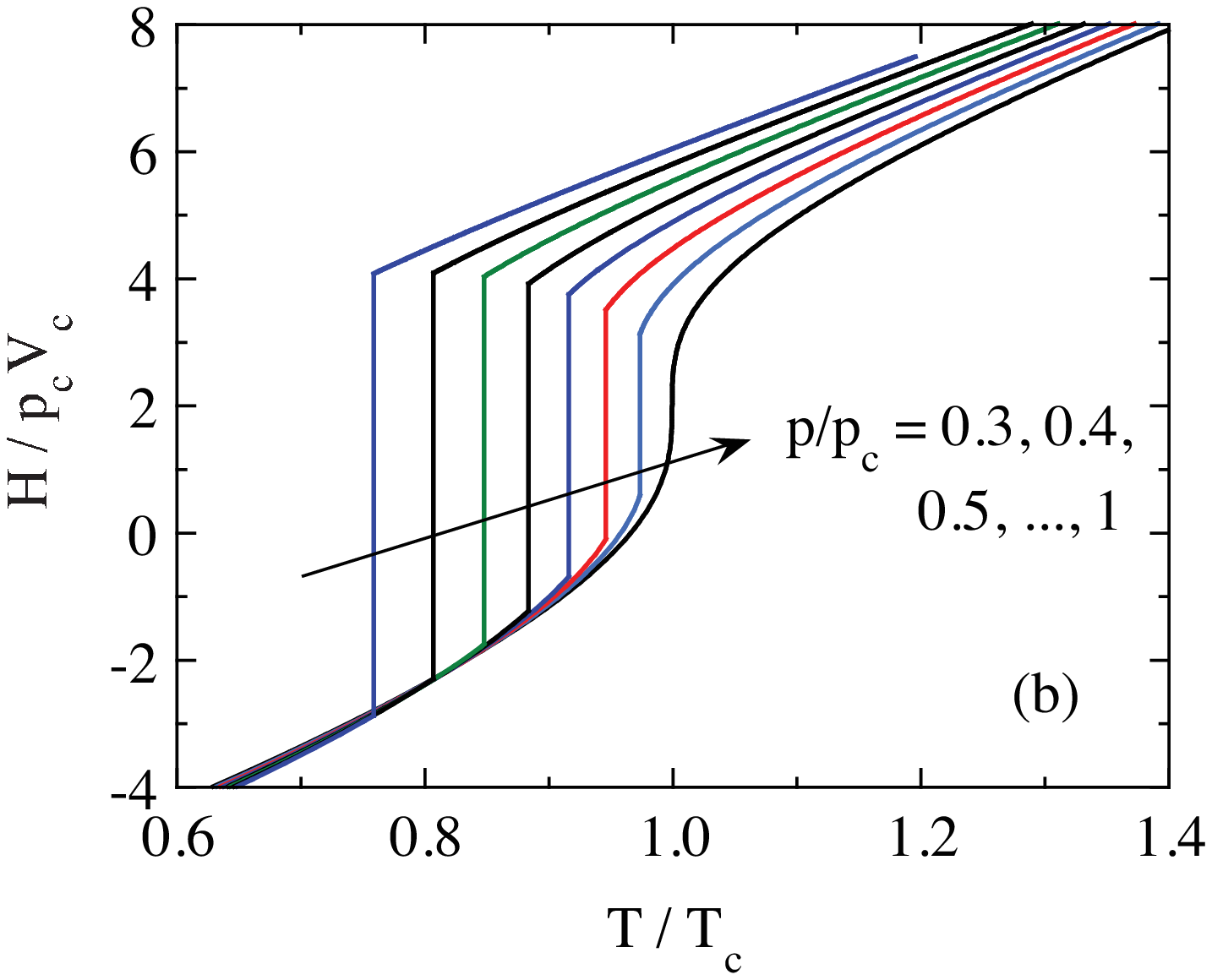}
\caption{(Color online) Equilibrium isobars of reduced enthalpy $H/(p_{\rm c}V_{\rm c})$ versus reduced temperature $\hat{\tau} = T/T_{\rm c}$ calculated using Eq.~(\ref{Eq:EnthalpyRed}) for (a) $\hat{p} = p/p_{\rm c} \geq 1$ and~(b) $\hat{p} \leq 1$.}
\label{Fig:vdW_const_pressure_H}  
\end{figure}

\begin{figure*}[t]
\includegraphics[width=3.3in]{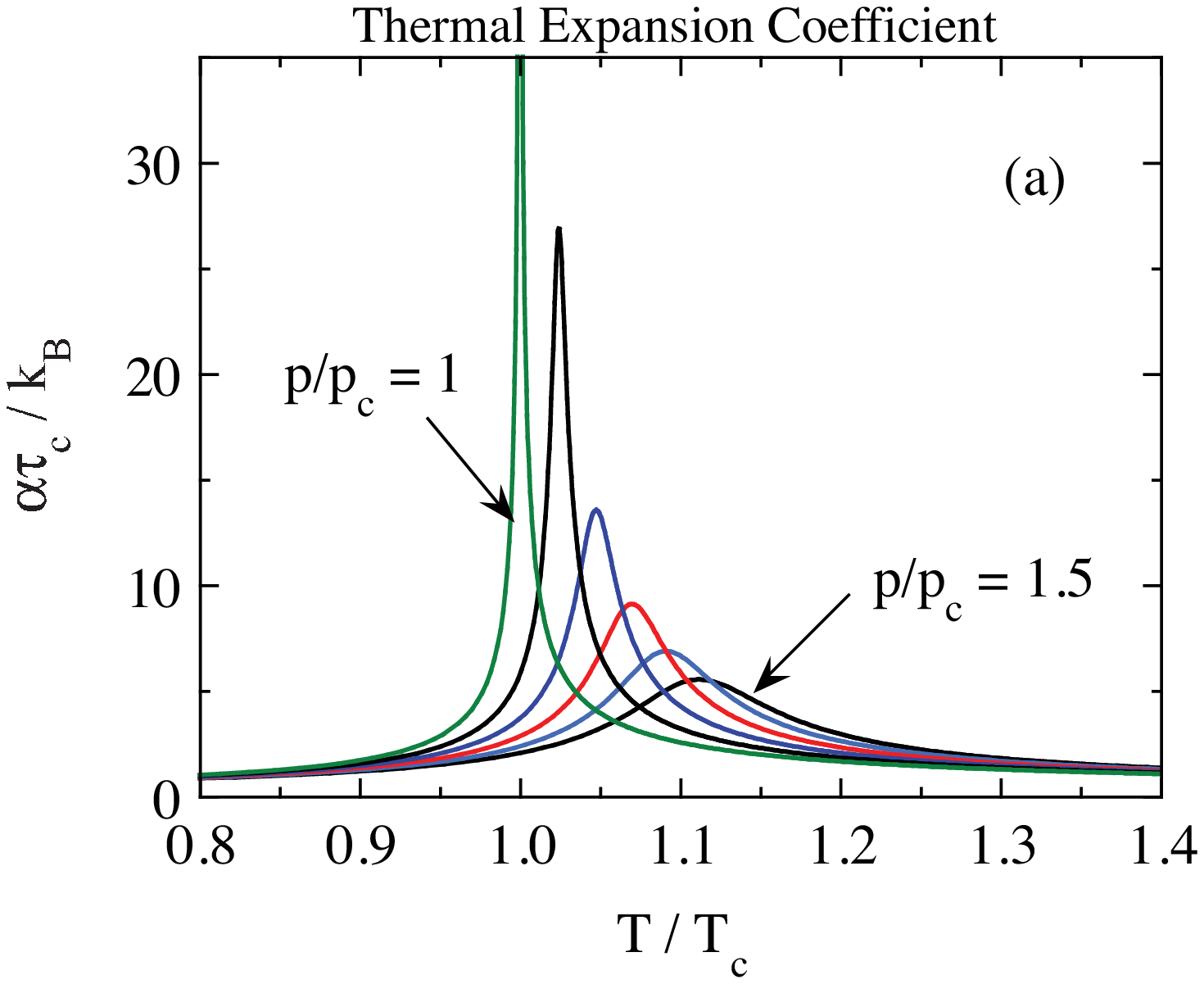}\vspace{-0.1in}
\includegraphics[width=3.3in]{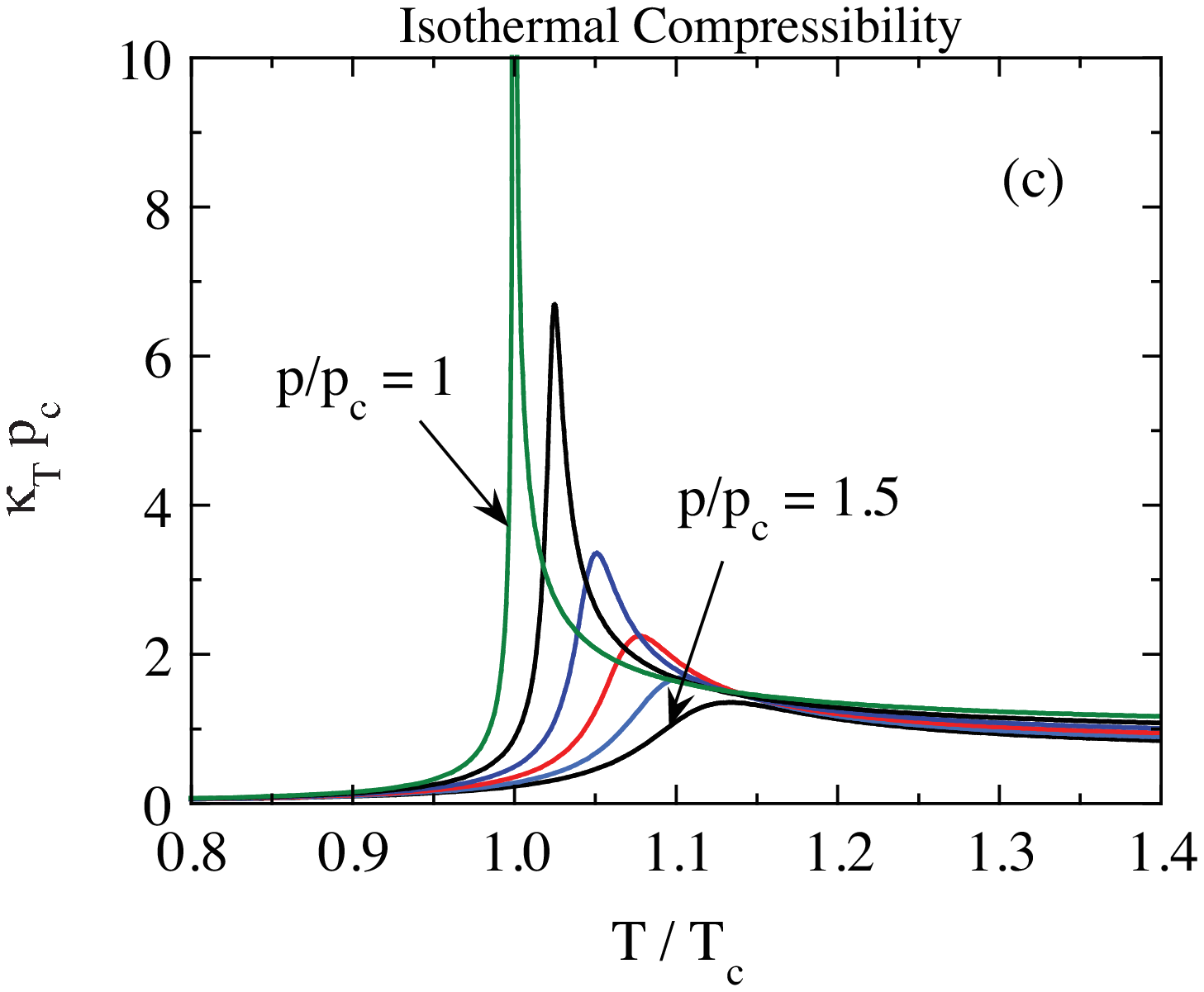}
\includegraphics[width=3.3in]{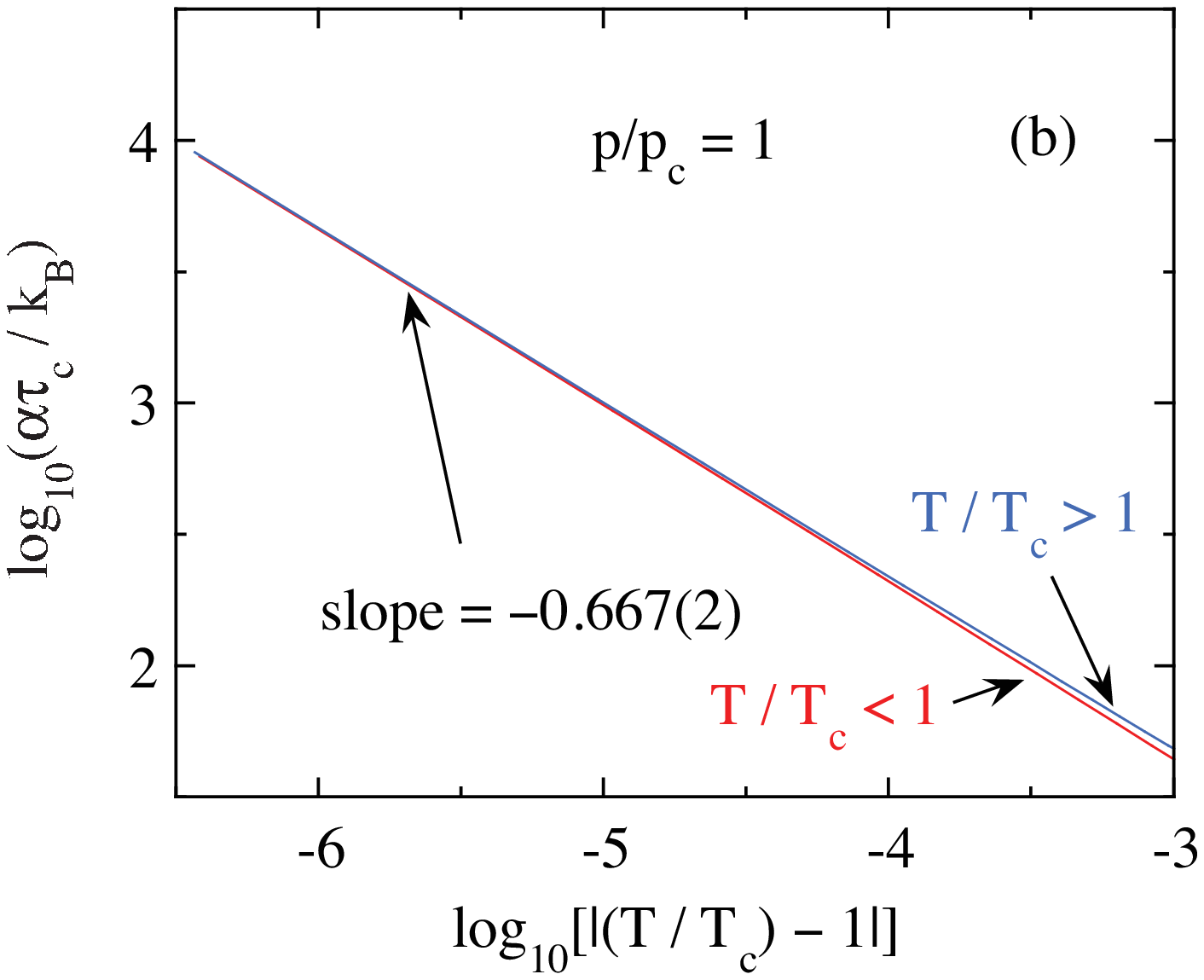}\vspace{-0.1in}
\includegraphics[width=3.3in]{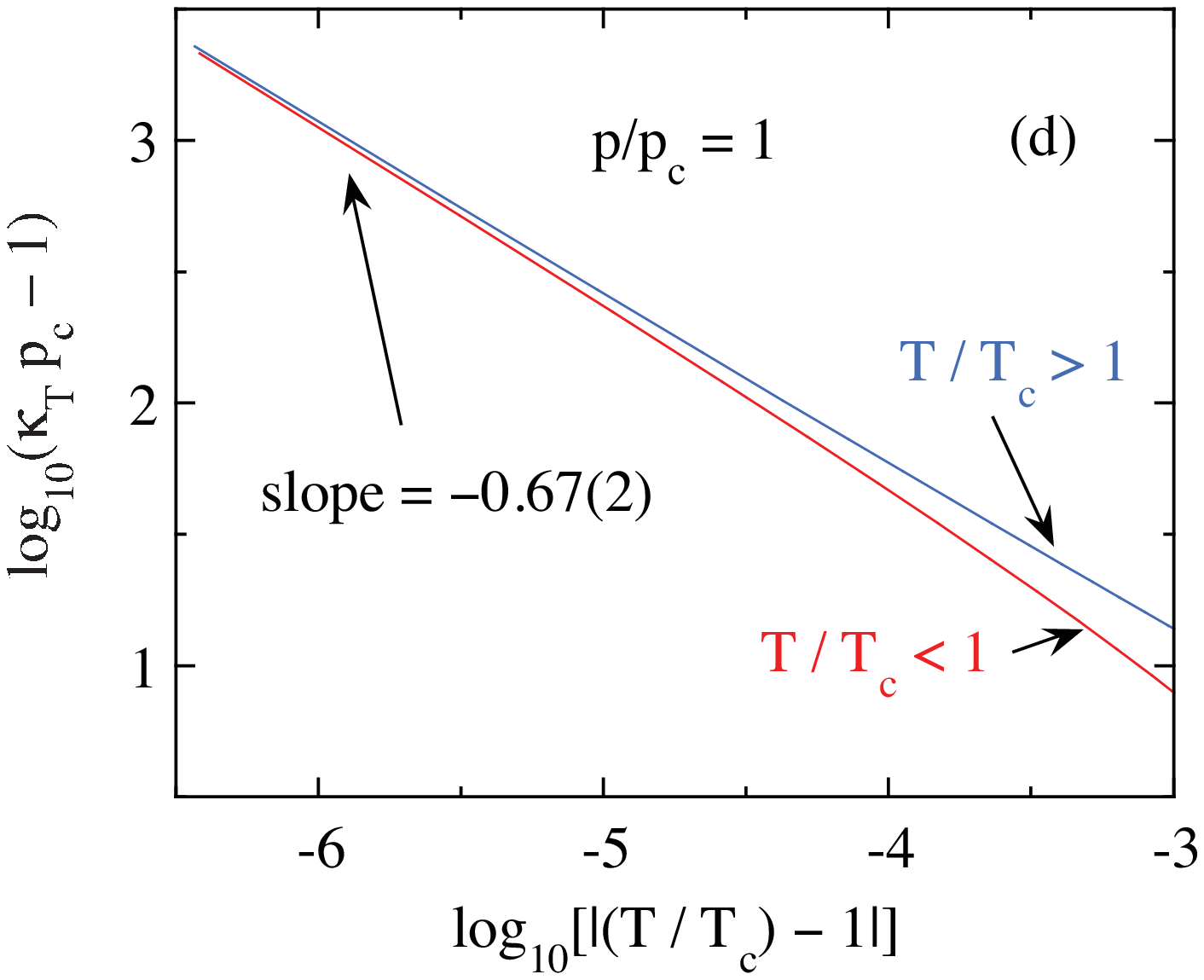}
\caption{(Color online) (a) Dimensionless reduced volume thermal expansion coefficient $\alpha\tau_{\rm c}/k_{\rm B}$ versus reduced temperature $T/T_{\rm c}$ for reduced pressures $p/p_{\rm c} = 1$ to 1.5 in 0.1 increments obtained from Eq.~(\ref{Eq:alphavdW}).  The divergence of $\alpha$ for $p/p_{\rm c} = 1$ is illustrated in (b) which shows a log-log plot of $\alpha\tau_{\rm c}/k_{\rm B}$ versus $\left|\frac{T}{T_{\rm c}} - 1\right|$ for both $T/T_{\rm c} < 1$ (red) and $T/T_{\rm c} > 1$ (blue).  Both data sets show the same divergent behavior $\alpha\tau_{\rm c}/k_{\rm B} \propto \left|\frac{T}{T_{\rm c}} - 1\right|^{-2/3}$ for $T\to T_{\rm c}$. Panels (c) and~(d) show the same plots for the reduced isothermal compressibility $\kappa p_{\rm c}$ versus $T/T_{\rm c}$ obtained using Eq.~(\ref{Eq:kappaTRedvdW}) at the same pressures.  The critical exponent in~(d) is seen to be the same as for the thermal expansion coefficient in~(b) to within the respective error bars.}
\label{Fig:vdW_const_pressure_alpha}  
\end{figure*}

\begin{figure}[t]
\includegraphics[width=3.3in]{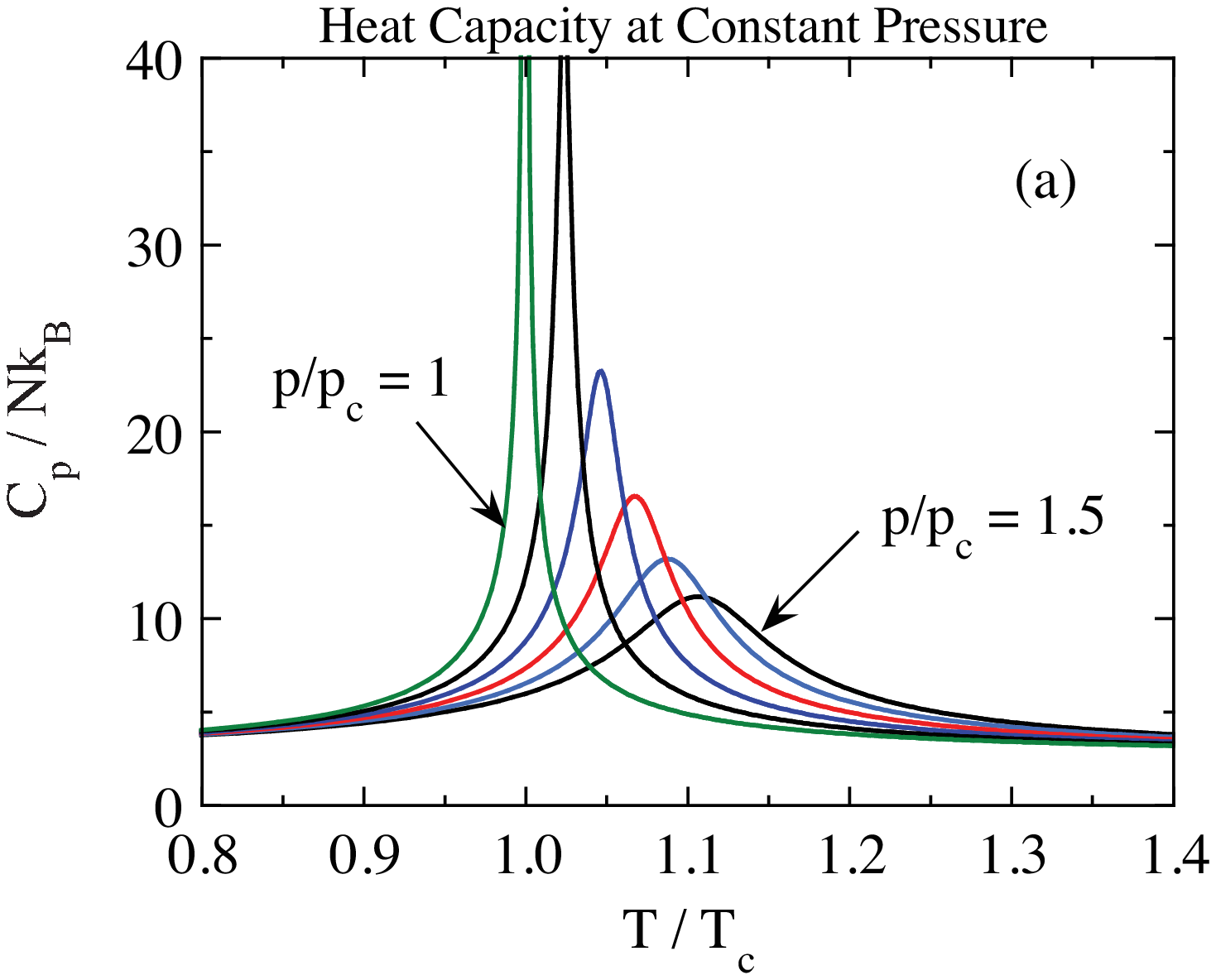}\vspace{-0.1in}
\includegraphics[width=3.2in]{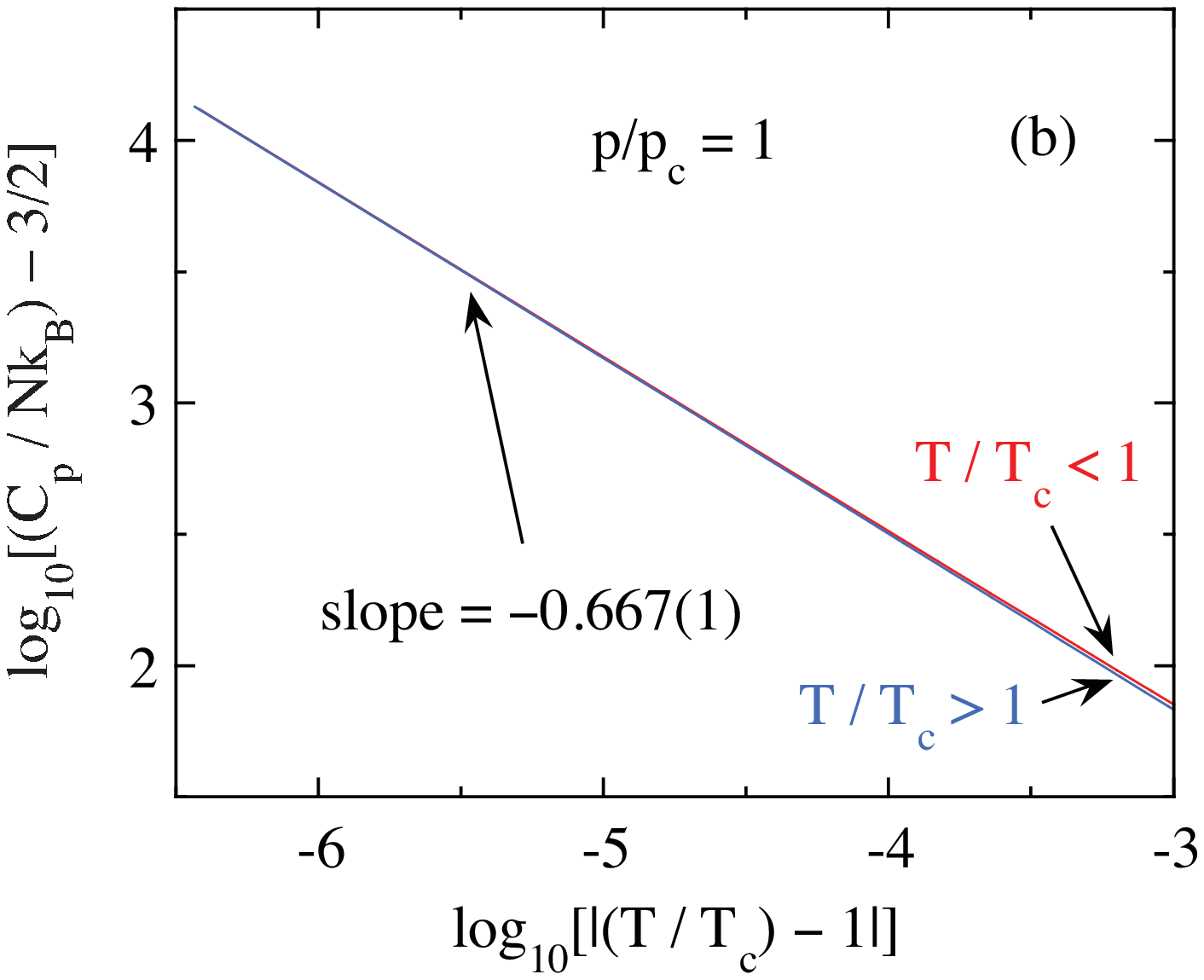}
\caption{(Color online) (a) Heat capacity at constant pressure $C_{\rm p}/Nk_{\rm B}$ versus temperature $T/T_{\rm c}$ for reduced pressures $p/p_{\rm c} = 1$ to 1.5 in 0.1 increments.  The divergence of $C_{\rm p}$ for $p/p_{\rm c} = 1$ is illustrated in (b) which shows a log-log plot of the critical part $\frac{C_{\rm p}}{Nk_{\rm B}} - \frac{3}{2}$ versus $\left|\frac{T}{T_{\rm c}} - 1\right|$ for both $T/T_{\rm c} < 1$ (red) and $T/T_{\rm c} > 1$ (blue).  Both data sets show the same divergent behavior $\frac{C_{\rm p}}{Nk_{\rm B}} - \frac{3}{2} \propto \left|\frac{T}{T_{\rm c}} - 1\right|^{-2/3}$.}
\label{Fig:vdW_const_pressure_calcs}  
\end{figure}

The entropy relative to that at the critical point $\Delta S/Nk_{\rm B}$ versus reduced temperature $\hat{\tau} = T/T_{\rm c}$ calculated using Eq.~(\ref{Eq:DeltasigmaRed}) at constant pressure for $\hat{p} = p/p_{\rm c} \geq 1$ is shown in Fig.~\ref{Fig:vdW_const_pressure_sigma}(a). As the pressure decreases towards the critical point $\hat{p}=1$, an inflection point develops in $\Delta S$ versus~$T$ with a slope that increases to~$\infty$ at $\hat{\tau}=1$, signaling entrance into a phase-separated temperature range with decreasing pressure.  The development of an infinite slope in $\Delta S$ versus~$T$ with decreasing pressure results in the onset of a divergence in the heat capacity at constant pressure at the critical point discussed below.  Similar behaviors are found for the internal energy and enthalpy using Eqs.~(\ref{Eq:URed}) and~(\ref{Eq:EnthalpyRed}), respectively, as shown in Figs.~\ref{Fig:vdW_const_pressure_U}(a) and~\ref{Fig:vdW_const_pressure_H}(a), respectively.

The thermal expansion coefficient $\alpha\tau_{\rm c}/k_{\rm B}$ versus $T/T_{\rm c}$ calculated from Eq.~(\ref{Eq:alphavdW}) is plotted in Fig.~\ref{Fig:vdW_const_pressure_alpha}(a) for $p/p_{\rm c} = 1$ to 1.5 in 0.1 increments.  It is interesting that that the molecular interactions have a large influence on $\alpha$ (and $\kappa_{\rm T}$ and $C_{\rm p}$, see below) even when $p$ is significantly larger than $p_{\rm c}$.  The data show divergent behavior for $\hat{p}=1$ at $T\to T_{\rm c}$ which is found in Fig.~\ref{Fig:vdW_const_pressure_alpha}(b) to be given by $\alpha\tau_{\rm c}/k_{\rm B} = 0.462(16)\left|\frac{T}{T_{\rm c}} - 1\right|^{-0.667(2)}$ for both $T\to T_{\rm c}^\pm$, where the exponent and amplitude are equal to the analytical values of $-2/3$ and $2/3^{4/3}$ in Eqs.~(\ref{Eq:alphaCritPars}) to within the error bars.

The isothermal compressibility $\kappa p_{\rm c}$ versus $T/T_{\rm c}$ calculated from $\alpha\tau_{\rm c}/k_{\rm B}$ and Eq.~(\ref{Eq:alphakapparat}) is plotted in Fig.~\ref{Fig:vdW_const_pressure_alpha}(c) for $p/p_{\rm c} = 1$ to 1.5 in 0.1 increments.  The data again show divergent behavior for $\hat{p}=1$ at $T\to T_{\rm c}$ which is found in Fig.~\ref{Fig:vdW_const_pressure_alpha}(d) to be given by $\kappa p_{\rm c} - 1 = 0.114(22) \left|\frac{T}{T_{\rm c}} - 1\right|^{-0.668(11)}$ for both $T\to T_{\rm c}^\pm$, where the exponent and amplitude are equal to the analytical values of $-2/3$ and $1/(3^{1/3}6)$ in Eqs.~(\ref{Eq:kappaCritPars}) to within the error bars.  The noncritical background compressibility of the ideal gas $\kappa_{\rm T}p_{\rm c} = p_{\rm c}/p=1$ in Eq.~(\ref{Eq:kappTIG}) from the calculated $\kappa p_{\rm c}$ versus $T/T_{\rm c}$ data before making the plot in Fig.~\ref{Fig:vdW_const_pressure_alpha}(d).

The $C_{\rm p}(T)$ predicted by Eq.~(\ref{Eq:CpvdW2}) is plotted for $p/p_{\rm c} = 1.0$--1.5 in Fig.~\ref{Fig:vdW_const_pressure_calcs}(a).  One sees that as $p$ decreases towards~$p_{\rm c}$ from above, a peak occurs at a temperature somewhat above $T_{\rm c}$ that develops into a divergent behavior at $T=T_{\rm c}$ when $p=p_{\rm c}$.  The critical part of the divergent behavior $\frac{C_{\rm p}}{Nk_{\rm B}} - \frac{3}{2}$ for $p = p_{\rm c}$ is plotted versus $\left|\frac{T}{T_{\rm c}} - 1\right|$ for both $T/T_{\rm c} < 1$ and $T/T_{\rm c} > 1$ in a log-log plot in Fig.~\ref{Fig:vdW_const_pressure_calcs}(b).  The same critical behavior $\frac{C_{\rm p}}{Nk_{\rm B}} - \frac{3}{2} = 0.693(12)\left|\frac{T}{T_{\rm c}} - 1\right|^{-0.667(1)}$ is observed at the critical point for both $T\to T_{\rm c}^+$ and $T\to T_{\rm c}^-$, as shown, where the exponent and amplitude are equal to the analytical values of $-2/3$ and $1/3^{1/3}$ in Eqs.~(\ref{Eq:CpCritp1}) to within the error bars.  From Eq.~(\ref{Eq:CpCVvdW}), this critical exponent is consistent with the critical exponents of~$-2/3$ determined above for both $\alpha$ and $\kappa_{\rm T}$ obtained on approaching the critical point at constant pressure versus temperature from either side of the critical point.

If instead of approaching the critical point in Fig.~\ref{Fig:vdW_p_vs_T_phase_diag} horizontally at constant pressure $\hat{p}=1$ versus temperature as above, one approaches it vertically at constant temperature $\hat{\tau}=1$ versus pressure, we find that the critical behavior of $[C_{\rm p}/(Nk_{\rm B}) - 3/2]$, $\alpha\tau_{\rm c}/k_{\rm B}$ and $(\kappa p_{\rm c} - 1)$ all still follow the same behavior $ \propto \left|\frac{T}{T_{\rm c}} - 1\right|^{-2/3}$ to within the error bars of 0.001 to 0.01 on the respective exponents.

\subsection{\label{Sec:ConstP<1} Results for $p\leq p_{\rm c}$}

\begin{figure}[t]
\includegraphics[width=2.75in]{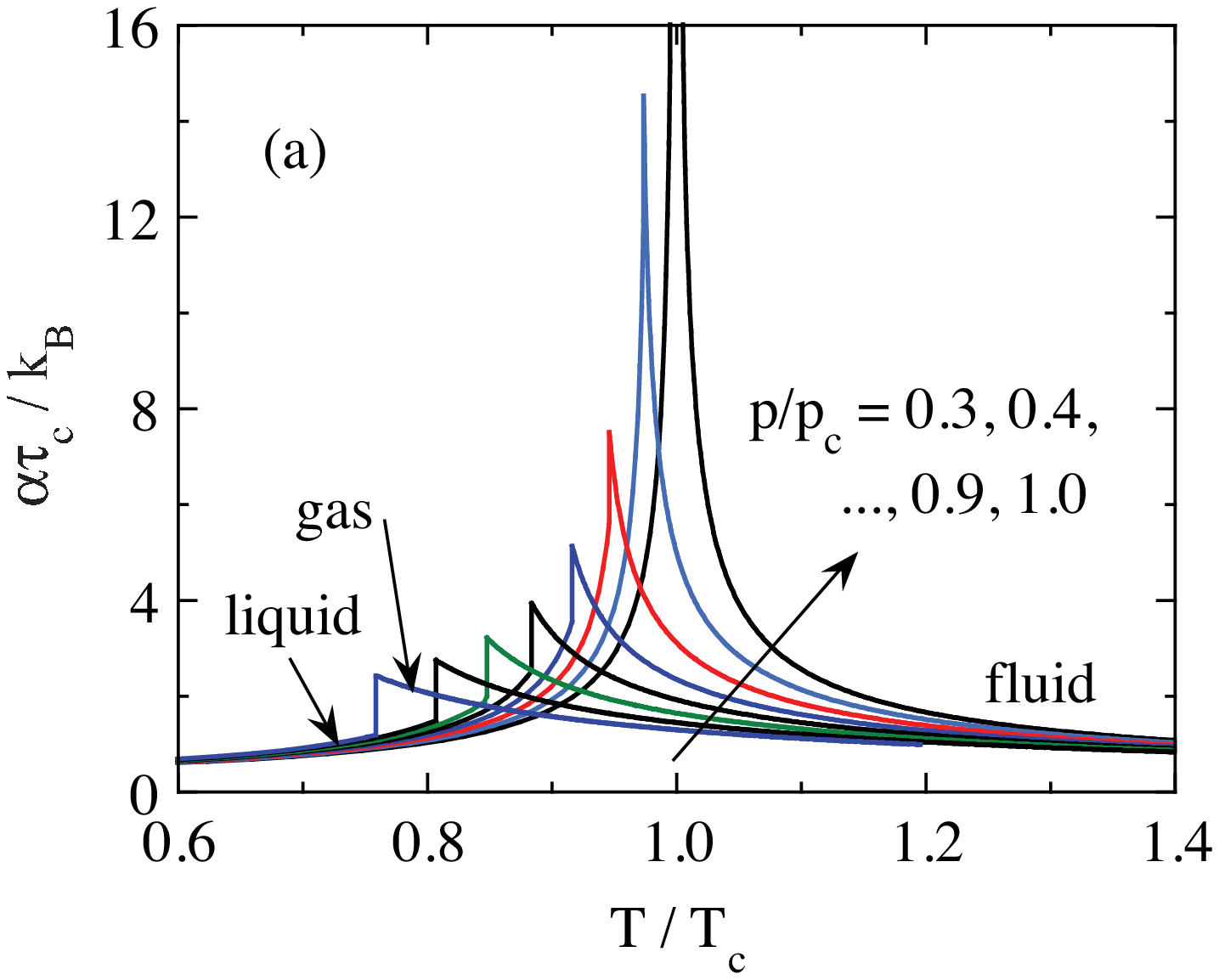}\vspace{-0.1in}
\includegraphics[width=2.75in]{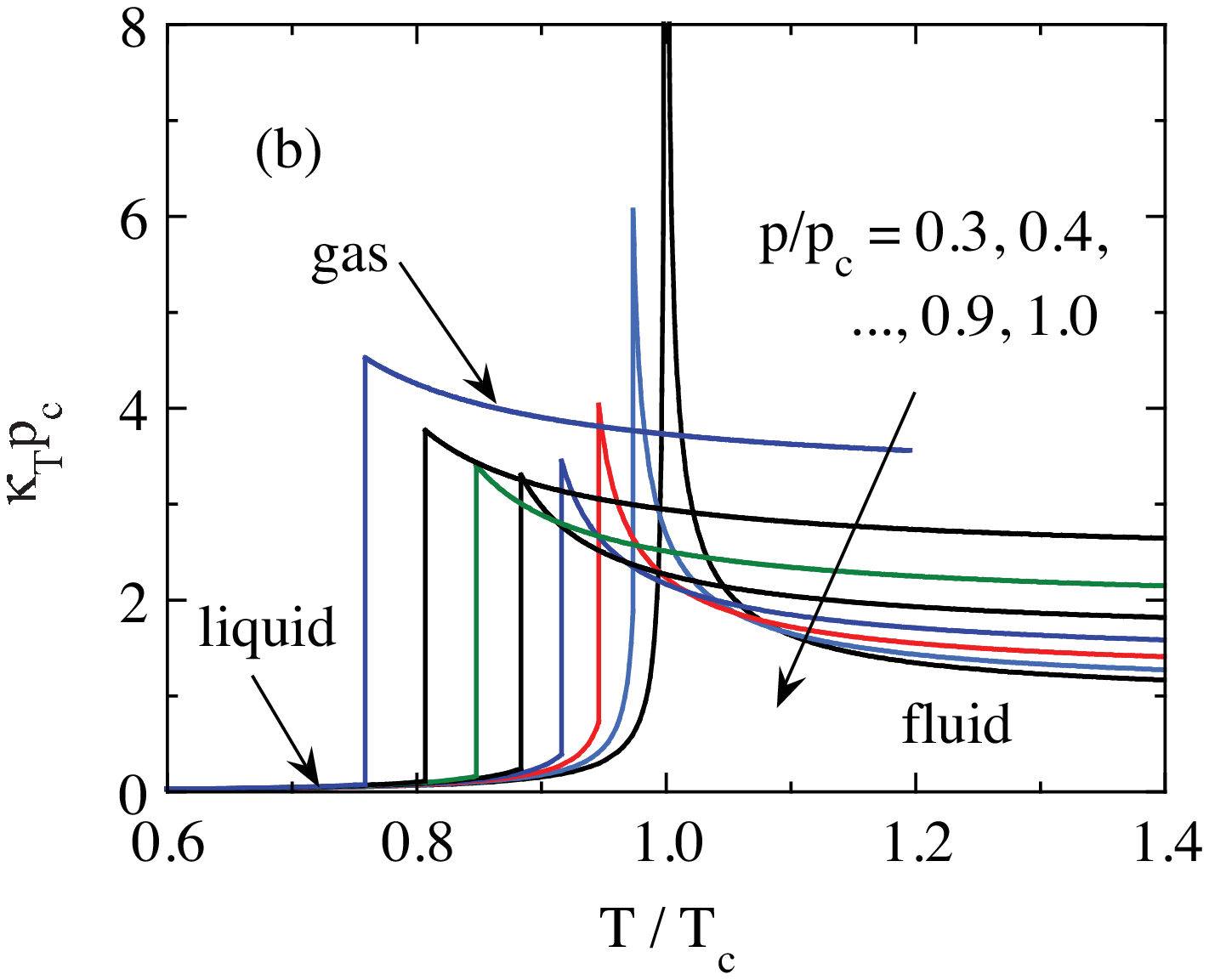}\vspace{-0.1in}
\includegraphics[width=2.8in]{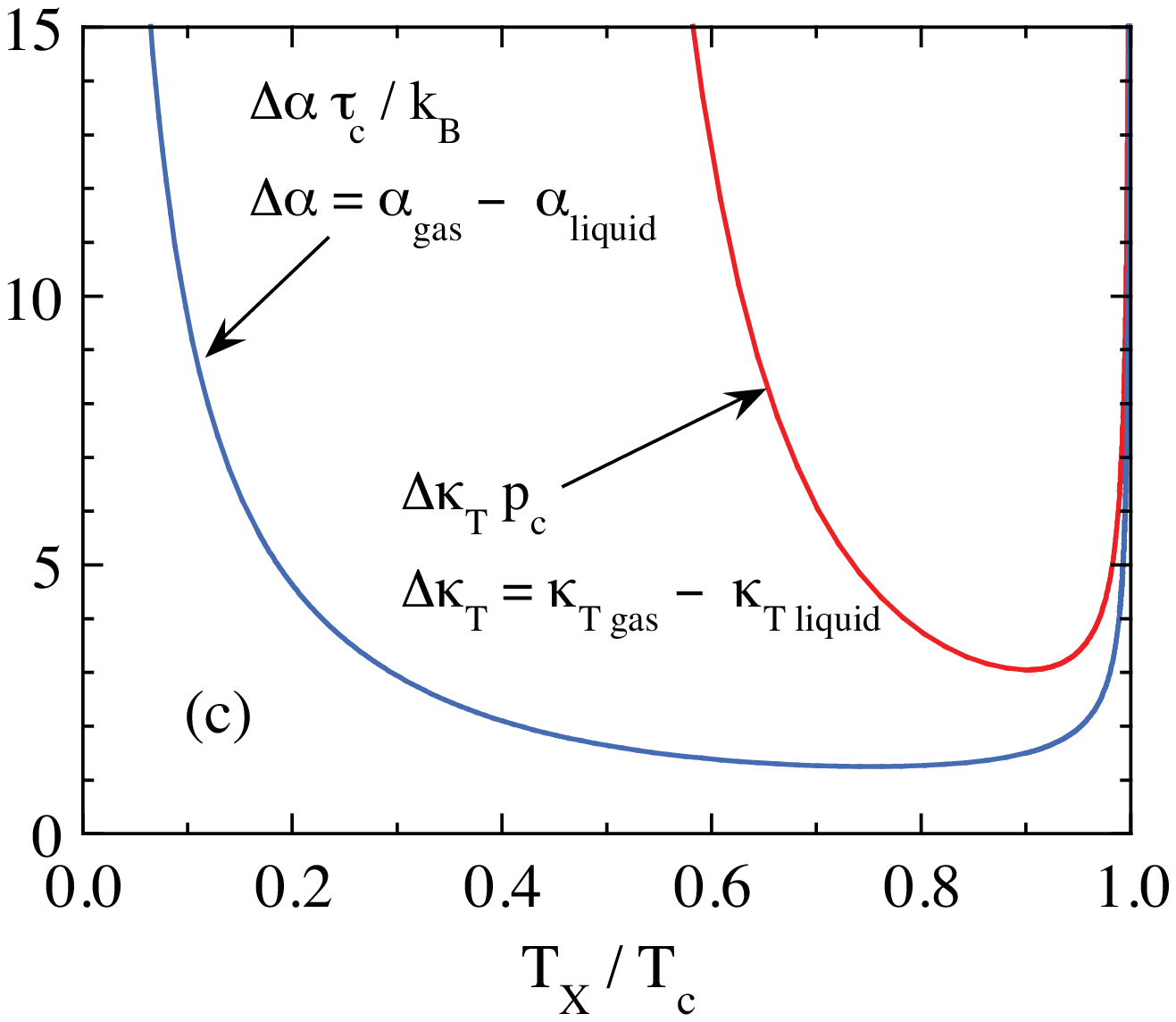}
\caption{(Color online) Isobars of the (a) thermal expansion coefficient $\alpha\tau_{\rm c}$ and (b) isothermal compressibility $\kappa_{\rm T}p_{\rm c}$ vs temperature $\hat{\tau} = T/T_{\rm c}$.  The respective discontinuities vs $\hat{\tau}_{\rm X}$ predicted by Eqs.~(\ref{Eq:DeltaAlpha}) and~(\ref{Eq:DeltaKappa}) are shown in~(c).}
\label{Fig:vdW_const_p_alpha_equilib}  
\end{figure}

\begin{figure}[t]
\includegraphics[width=2.9in]{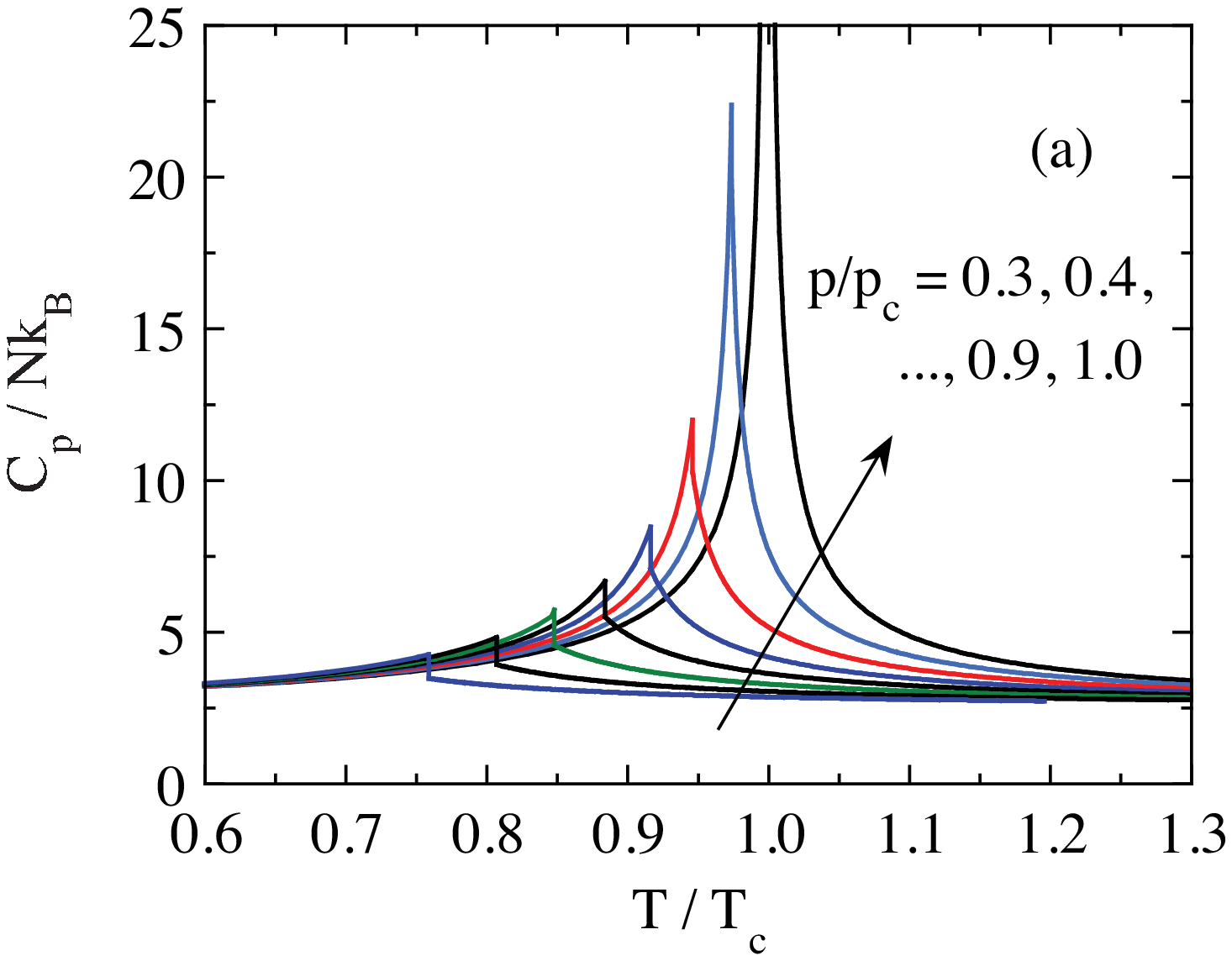}\vspace{-0.1in}
\includegraphics[width=2.9in]{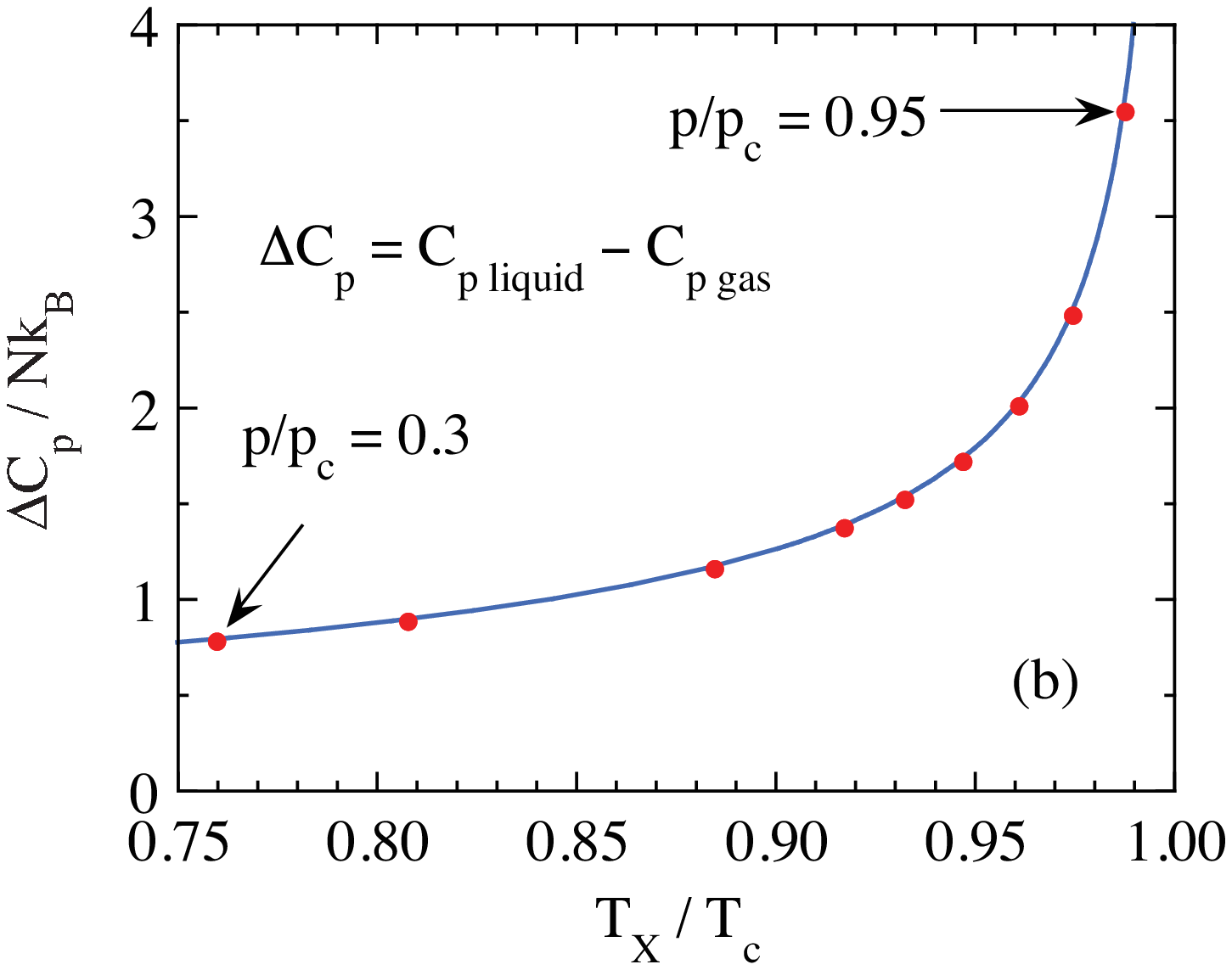}\vspace{-0.1in}
\includegraphics[width=2.9in]{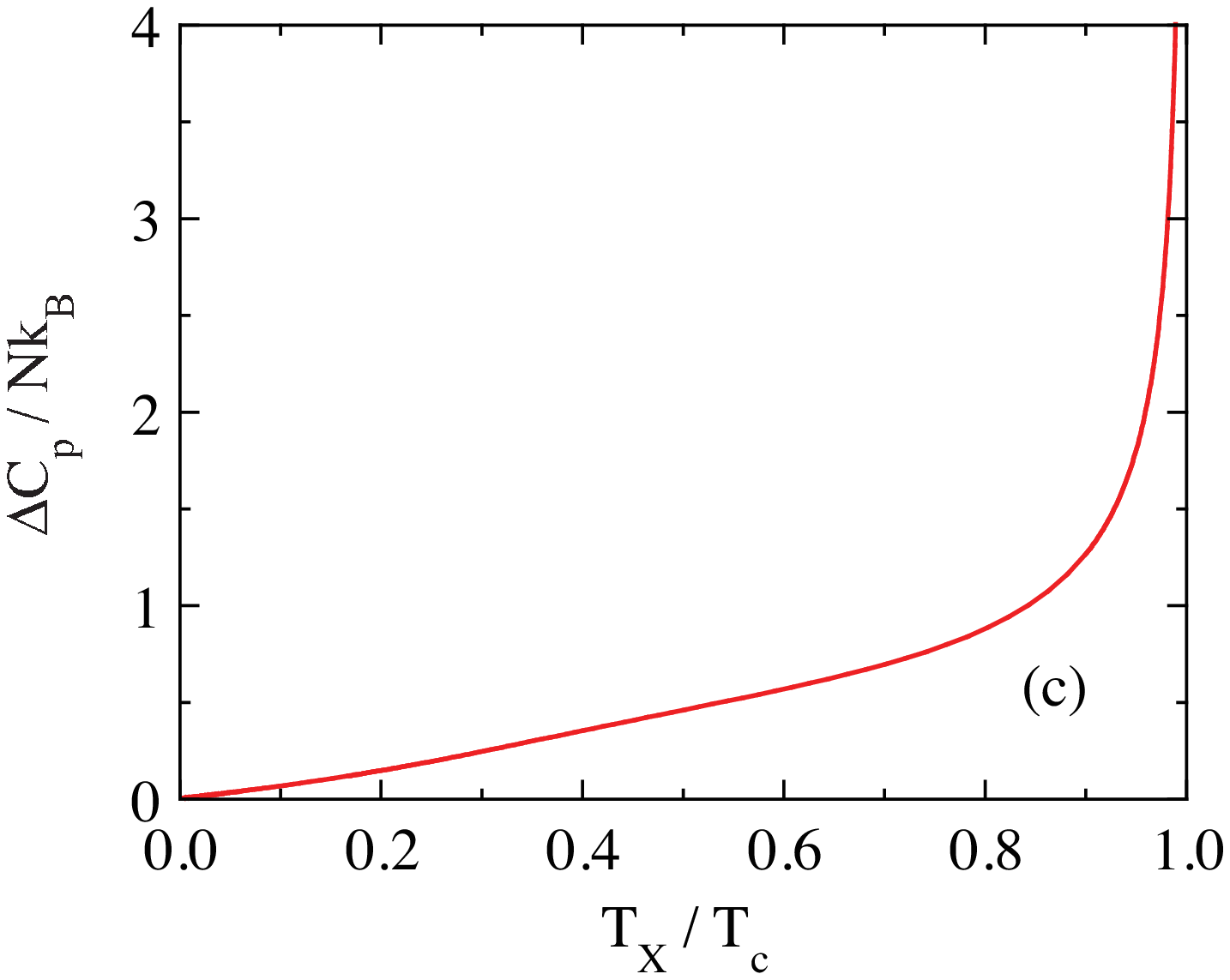}
\caption{(Color online) (a) $C_{\rm p}/(Nk_{\rm B})$ versus $\hat{\tau} = T/T_{\rm c}$ isobars.  (b) Heat capacity jump $\Delta C_{\rm p}/Nk_{\rm B}$ versus transition temperature $\hat{\tau}_{\rm X} = T_{\rm X}/T_{\rm c}$.  Two values of $\hat{p}$ are indicated.  The data points (filled red circles) are measured from isobars such as in panel~(a).  The solid curve is the prediction in Eq.~(\ref{Eq:DeltaCpThy}). (c) The prediction of Eq.~(\ref{Eq:DeltaCpThy}) for $\Delta C_{\rm p}/Nk_{\rm B}$ over the full temperature range.}
\label{Fig:vdW_const_p_Cp_p_leq_1}  
\end{figure}

The equilibrium $\Delta S/Nk_{\rm B}$ versus $\hat{\tau}$ calculated using Eq.~(\ref{Eq:DeltasigmaRed}) at constant pressure for $\hat{p} \leq 1$, augmented by the above calculations of the gas-liquid coexistence region, is shown in Fig.~\ref{Fig:vdW_const_pressure_sigma}(b).  For $\hat{p} < 1$ a discontinuity in the entropy occurs at a temperature-dependent transition temperature $T_{\rm X}$ that decreases with decreasing~$\hat{p}$ according to the pressure versus temperature phase diagram in Fig.~\ref{Fig:vdW_p_vs_T_phase_diag}.  The change in entropy at the transition $\Delta S_{\rm X}/Nk_{\rm B}$ versus the reduced transition temperature $T_{\rm X}/T_{\rm c}$ is plotted above in Fig.~\ref{Fig:vdW_transition_entropy}.  Similar behaviors are found for the internal energy and enthalpy as shown in Figs.~\ref{Fig:vdW_const_pressure_U}(b) and~\ref{Fig:vdW_const_pressure_H}(b), respectively.

The reduced thermal expansion coefficient $\alpha\tau_{\rm c}/k_{\rm B}$ and reduced isothermal compressibility $\kappa_{\rm T}p_{\rm c}$ versus reduced temperature $\hat{\tau} = T/T_{\rm c}$ for several values of reduced pressure $\hat{p} =p/p_{\rm c}$ are plotted in Figs.~\ref{Fig:vdW_const_p_alpha_equilib}(a) and~\ref{Fig:vdW_const_p_alpha_equilib}(b), respectively. Both quantities show discontinuous increases (jumps) at the first-order transition temperature $\hat{\tau}_{\rm X}< 1$ from liquid to gas phases with increasing temperature.  These data are for the pure gas and liquid phases on either side of the coexistence curve in Fig.~\ref{Fig:vdW_p_vs_T_phase_diag}. Remarkably, the jumps vary nonmonically with temperature for both quantitities.  This is confirmed in Fig.~\ref{Fig:vdW_const_p_alpha_equilib}(c) where the jumps calculated from the parametric solutions to them in Eqs.~(\ref{Eq:DeltaAlpha}) and~(\ref{Eq:DeltaKappa}) are plotted.

The reduced heat capacity at constant pressure $C_{\rm p}/(Nk_{\rm B})$ versus reduced temperature $\hat{\tau}$ is shown in Fig.~\ref{Fig:vdW_const_p_Cp_p_leq_1}(a) for $\hat{p} = 0.3,\ 0.4, \ldots,\  1$.  The transition from pure gas to pure liquid on cooling below the reduced transition temperature $\hat{\tau}_{\rm X}= T_{\rm X}/T_{\rm c}$ results in a peak in the heat capacity and a jump $\Delta C_{\rm p}/Nk_{\rm B}$ at $\hat{\tau}_{\rm X}$.  In addition, there is a latent heat at the transition that is not considered here.  The heat capacity jump is plotted versus $\hat{\tau}_{\rm X}$ as filled circles in Fig.~\ref{Fig:vdW_const_p_Cp_p_leq_1}(b), where it is seen to initially strongly decrease with decreasing~$\hat{\tau}_{\rm X}$ and then become much less dependent on~$\hat{\tau}_{\rm X}$.  The exact parametric solution for $\Delta C_{\rm p}/Nk_{\rm B}$ versus $\hat{\tau}_{\rm X}$ obtained from Eq.~(\ref{Eq:DeltaCpThy}) is plotted as the solid red curve in Fig.~\ref{Fig:vdW_const_p_Cp_p_leq_1}(c), where, in contrast to the jumps in $\kappa_{\rm T}$ and~$\alpha$ in Fig.~\ref{Fig:vdW_const_p_alpha_equilib}, $\Delta C_{\rm p}/Nk_{\rm B}$ decreases monotonically with decreasing $\hat{\tau}_{\rm X}$ and goes linearly to zero at $T=0$.

Representative values of the jumps versus $\hat{\tau}_{\rm X}$ in $\kappa_{\rm T}$, $\alpha$ and~$C_{\rm p}$ on crossing the coexistence curve in Fig.~\ref{Fig:vdW_p_vs_T_phase_diag}, calculated from Eqs.~(\ref{Eq:DeltaKappa}), (\ref{Eq:DeltaAlpha}) and~(\ref{Eq:DeltaCpThy}), are listed in Table~\ref{Tab:LeknerTableB} in Appendix~\ref{App:Tables}.

\section{\label{Sec:Expansions} Adiabatic Free Expansion and Joule-Thomson Expansion}

\subsection{Adiabatic Free Expansion}

\begin{figure}[t]
\includegraphics[width=3.3in]{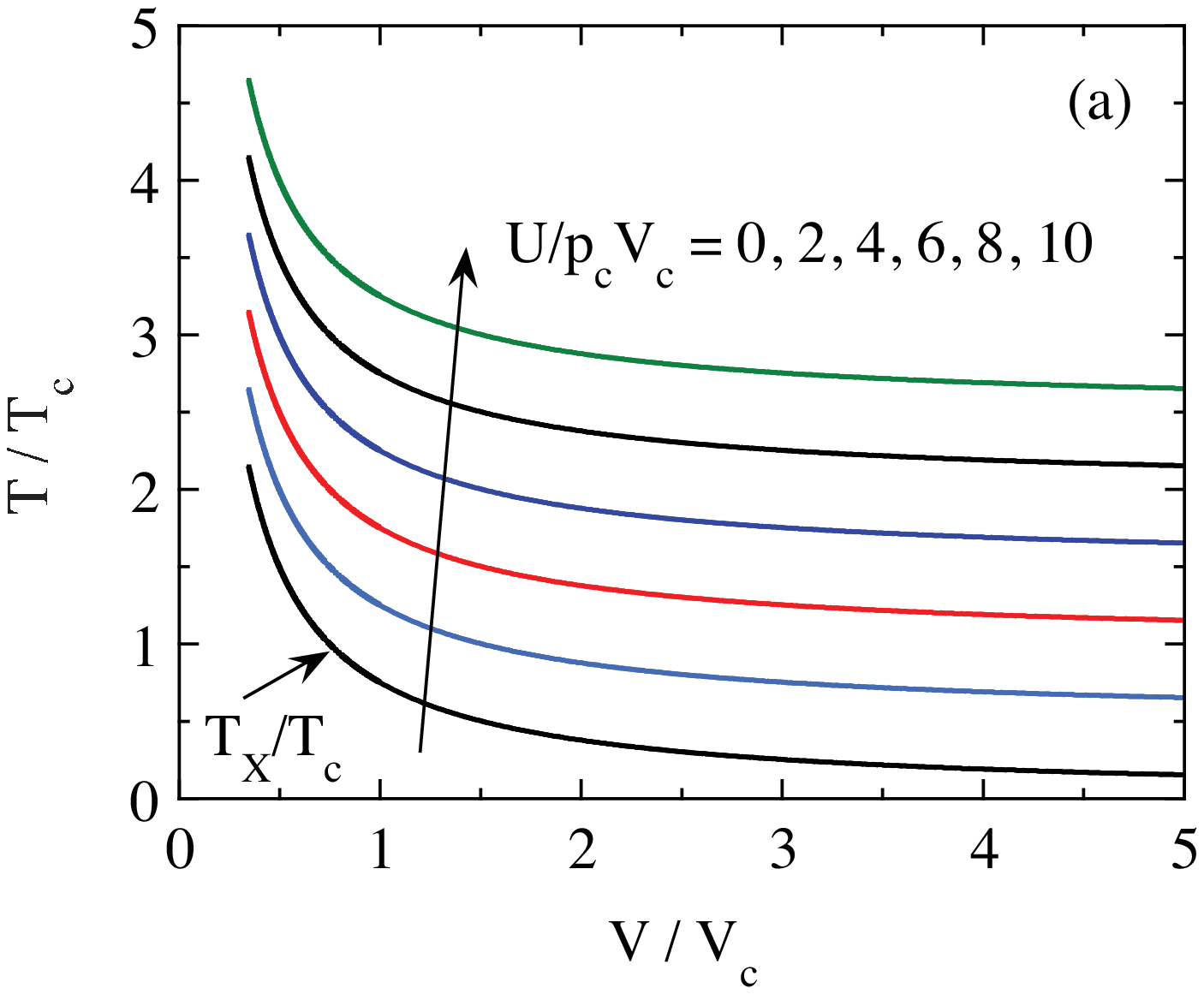}
\includegraphics[width=3.2in]{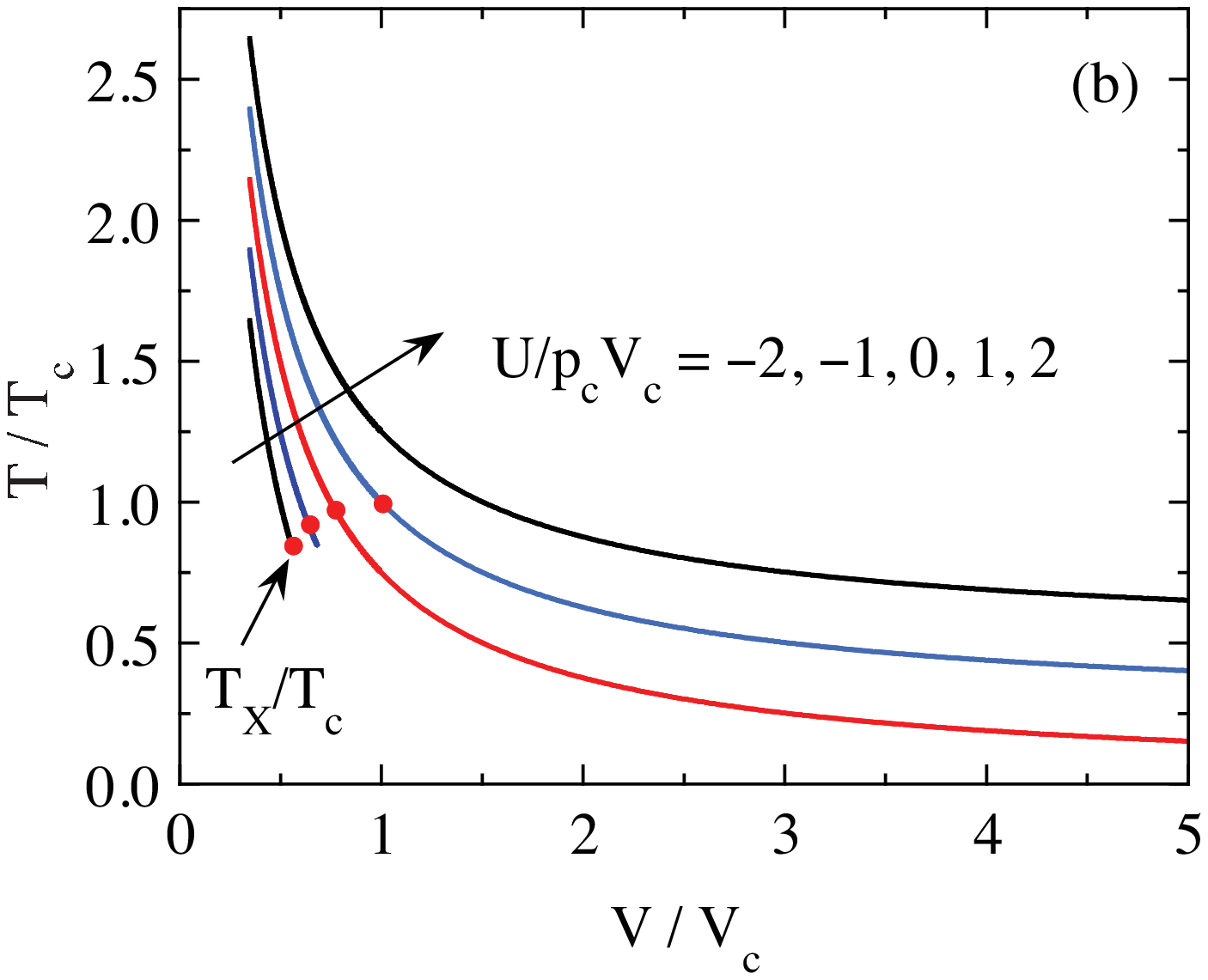}
\caption{(Color online) (a) Reduced temperature $\hat{\tau} = T/T_{\rm c}$ versus reduced volume $\widehat{V} = V/V_{\rm c}$ at fixed values of reduced internal energy $U/(p_{\rm c}V_{\rm c})$ as indicated for the ranges (a) $U/(p_{\rm c}V_{\rm c}) = 0$ to 10 and (b) $-2$ to 2.  The filled red circles in~(b) denote the phase transtion from pure gas to pure liquid due to the cooling associated with the free expansion.  In~(b), the curves with $U/(p_{\rm c}V_{\rm c}) = -2$ and~$-1$ terminate because the calculated pressure becomes negative at the ends of these plotted curves.  For $U/(p_{\rm c}V_{\rm c}) = 0$ and~1, a wide range of possible initial and final volumes for the adiabatic expansion result in the liquification of the expanding gas.}
\label{Fig:vdW_const_U}  
\end{figure}

In an adiabatic free expansion of a gas from an initial volume $V_1$ to a final volume~$V_2$, the heat absorbed by the fluid $Q$ and the work done by the fluid~$W$ during the expansion are both zero, so the change in the internal energy $U$ of the fluid obtained from the first law of thermodynamics is
\be
\Delta U \equiv U_2-U_1 = Q-W=0.
\ee
From the expression for the reduced internal energy of the vdW fluid in Eq.~(\ref{Eq:URed}), one has
\be
\frac{U_2-U_1}{p_{\rm c}V_{\rm c}} = 4(\hat{\tau}_2-\hat{\tau}_1) - 3\left(\frac{1}{\widehat{V}_2} - \frac{1}{\widehat{V}_1}\right).
\ee
Setting this equal to zero gives
\be
\hat{\tau}_2-\hat{\tau}_1 = \frac{3}{4}\left(\frac{1}{\widehat{V}_2} - \frac{1}{\widehat{V}_1}\right).
\ee
By definition of an expansion one has $\widehat{V}_2 > \widehat{V}_1$, yielding
\be
\hat{\tau}_2 < \hat{\tau}_1,
\ee
so the adiabatic free expansion of a vdW fluid cools it.  This contasts with an ideal gas where $\hat{\tau}_2 = \hat{\tau}_1$ because according to Eq.~(\ref{Eq:UIG}), $U$ does not depend on volume for an ideal gas.

The above considerations are valid if there is no gas to liquid phase transition.  To clarify this issue, in Fig.~\ref{Fig:vdW_const_U} are plotted $\hat{\tau}$ versus $\widehat{V}$ obtained using Eq.~(\ref{Eq:URed}) at the fixed values of $U/(p_{\rm c}V_{\rm c})$ indicated.  We have also calculated the pressure of the gas along each curve using Eq.~(\ref{Eq:RedpVsRedV}) (not shown) and compared it with the liquifaction pressure $\hat{p}_{\rm X}(\hat{\tau})$ in Fig.~\ref{Fig:vdW_p_vs_T_phase_diag}.  For a range of $U/(p_{\rm c}V_{\rm c})$ approximately between 0 and~1, we find that the gas liquifies as it expands and cools.   Thus under limited circumstances, adiabatic free expansion of the van der Waals gas can liquify it.

We note the caveat discussed by Reif\cite{Reif1965} that an adiabatic free expansion is an irreversible ``one-shot'' expansion that necessarily has to cool the solid container that the gas is confined in.  The container would likely have a substantial heat capacity compared to that of the gas.  Therefore the actual amount of cooling of the gas is likely significantly smaller than calculated above.  This limitation is eliminated in the steady-state expansion of a gas through a ``throttle'' in a tube from high to low pressure as discussed in the next section, where in the steady state the walls of the tube on either side of the throttle have reached a steady temperature.

\subsection{Joule-Thomson Expansion}

\begin{figure}[t]
\includegraphics[width=3.3in]{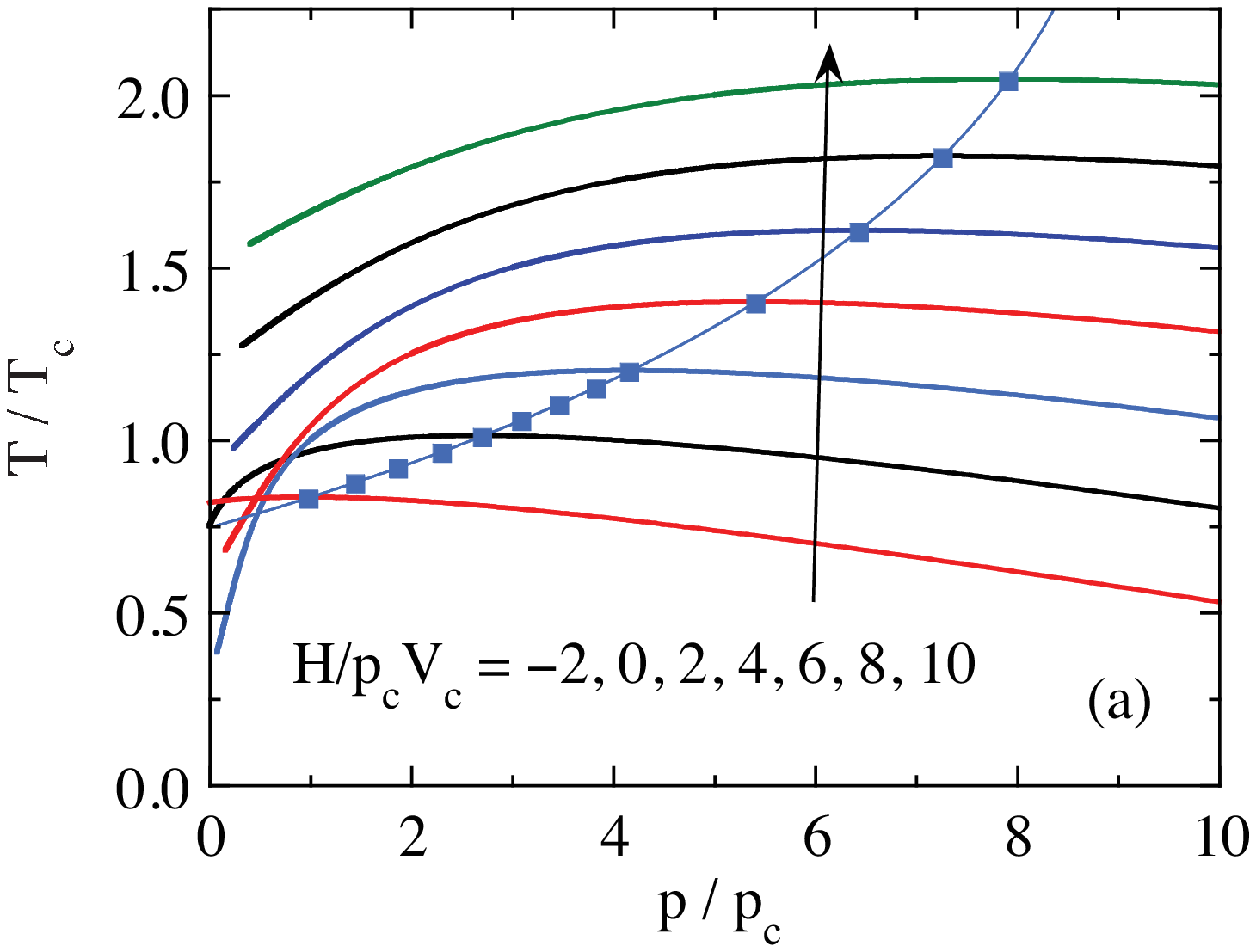}
\includegraphics[width=3.2in]{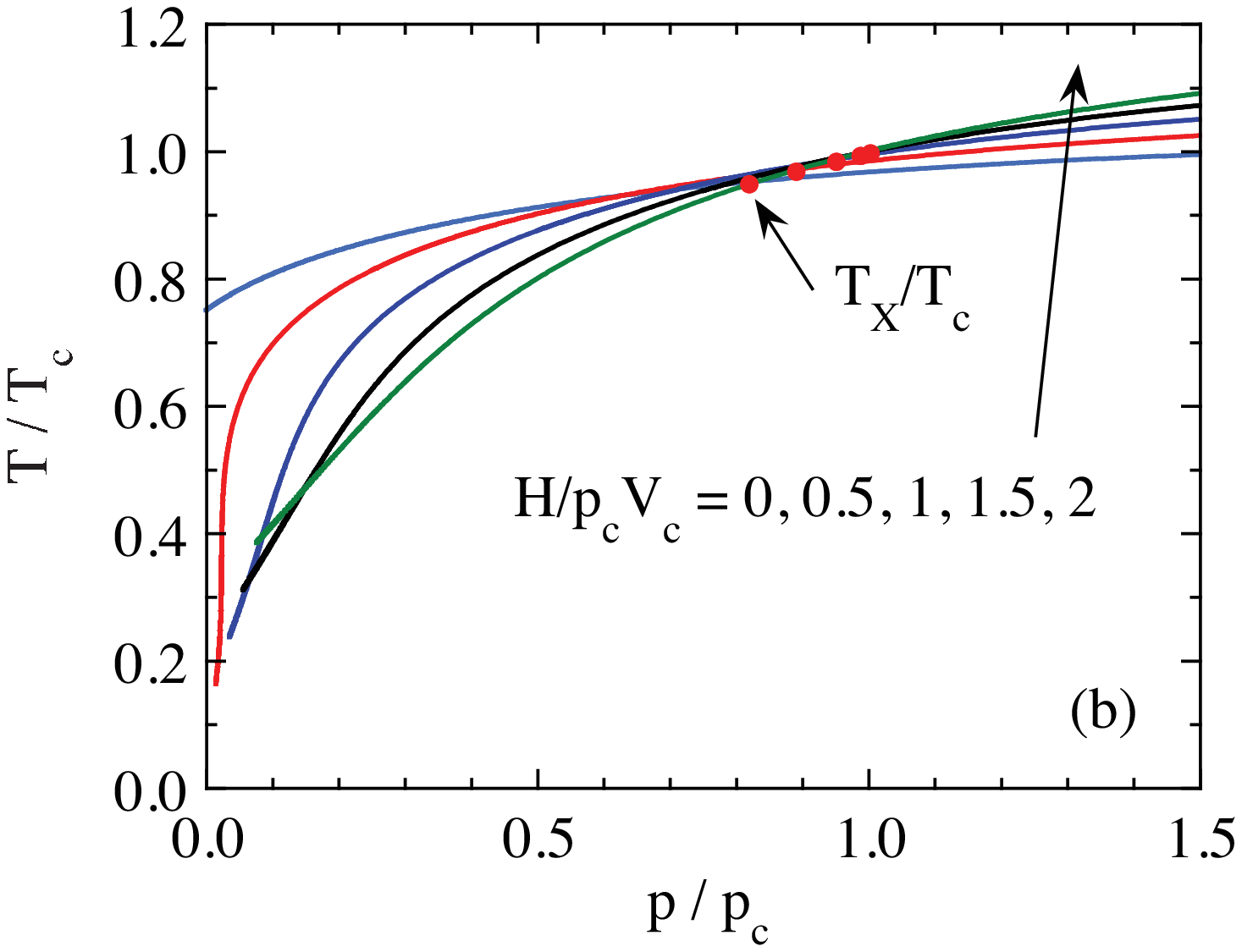}
\caption{(Color online) (a) Reduced temperature $\hat{\tau} = T/T_{\rm c}$ versus reduced pressure $\hat{p} = p/p_{\rm c}$ at fixed values of reduced enthalpy $H/(p_{\rm c}V_{\rm c})$ as indicated for the range (a) $H/(p_{\rm c}V_{\rm c}) = -2$ to 10.  The filled blue squares are the inversion points with zero slope where the gas is first warmed instead of cooled if the starting pressure it to the right of the squares.  The thin blue line through the blue squares is the analytic prediction for the inversion point in Eq.~(\ref{InvMinus}).  (b) Explanded plots of $\hat{\tau}$ versus $\hat{p}$ for $H/(p_{\rm c}V_{\rm c}) = 0$ to 2.  The filled red circle for each curve is $\hat{p}_{\rm X}(\hat{\tau})$ for the given value of $H/(p_{\rm c}V_{\rm c})$ that separates the region of pure gas to the right of the circle and pure liquid to the left.}
\label{Fig:vdW_const_H}  
\end{figure}

In Joule-Thomson (or Joule-Kelvin) expansion, gas at a high pressure $p_1$ passes through a constriction (throttle)  that might be a porous plug or small valve to a region with low pressure~$p_2$ in a thermally insulated tube.\cite{Reif1965}  In such an expansion, the enthalpy~$H$ instead of the internal energy~$U$ is found to be constant during the expansion:
\be
\frac{H_2(\hat{\tau}_2,\hat{p}_2)}{p_{\rm c}V_{\rm c}} = \frac{H_1(\hat{\tau}_1,\hat{p}_1)}{p_{\rm c}V_{\rm c}}.
\ee
Whether heating or cooling of the gas occurs due to the expansion depends on how $\hat{p}$ and~$\hat{\tau}$ vary at constant~$H$\@.  Therefore it is useful to plot $\hat{\tau}$ versus $\hat{p}$ at fixed $H$ to characterize how the fluid temperature changes on passing from the high to the low pressure side of the throttle.

From Eq.~(\ref{Eq:EnthalpyRed}) one can express the temperature in terms of the enthalpy and volume as
\be
\hat{\tau} = \frac{3\widehat{V}-1}{4(5\widehat{V}-1)}\left(\frac{H}{p_{\rm c}V_{\rm c}} + \frac{6}{\widehat{V}}\right).
\label{Eq:tauVsV}
\ee
However, one needs to plot $\hat{\tau}$ versus~$\hat{p}$ instead of versus~$\widehat{V}$ at fixed~$H$\@.  Therefore in a parametric solution one calculates $\hat{\tau}$ versus~$\widehat{V}$ using Eq.~(\ref{Eq:tauVsV}) and $\hat{p}$ versus~$\widehat{V}$ using Eq.~(\ref{Eq:RedpVsRedV}) and then obtains $\hat{\tau}$ versus~$\hat{p}$ with~$\widehat{V}$ as an implicit parameter.  We note that according to Eq.~(\ref{Eq:Hcrit}), a value of $H/(p_{\rm c}V_{\rm c}) = 2$ gives rise to a plot of $\hat{\tau}$ versus~$\hat{p}$ that passes through the critical point $\hat{\tau}=\hat{p}=\widehat{V}=1$.   Thus for $H/(p_{\rm c}V_{\rm c}) < 2$ one might expect that expansion of the vdW gas through a throttle could liquify the gas in a continuous steady-state process.  This is confirmed below.

Shown in Fig.~\ref{Fig:vdW_const_H}(a) are plots of reduced temperature $\hat{\tau}$ versus reduced pressure $\hat{p}$ at fixed values of reduced enthalpy $H/(p_{\rm c}V_{\rm c})$ for $H/(p_{\rm c}V_{\rm c}) = -2$ to 10.  Each curve has a smooth maximum.  The point at the maximum of a curve is called an inversion point (labeled by a filled blue square) and the locus of these points versus $H$ is known as the ``inversion curve''.  If the high pressure~$p_2$ is greater than the pressure of the inversion point, the gas would initially warm on expanding instead of cooling, whereas if $p_2$ is at a lower pressure than this, then the gas only cools as it expands through the throttle.  Thus in using the Joule-Thomson expansion to cool a gas, one normally takes the high pressure $p_2$ to be at a lower pressure than the pressure of the inversion point.  The low pressure $p_1$ can be adjusted according to the application.

The slope of $T$ versus $p$ at fixed~$H$ is\cite{Reif1965}
\be
\left(\frac{\partial T}{\partial p}\right)_H = \frac{V}{C_{\rm p}}(T\alpha_{\rm T} - 1).
\label{Eq:dTdpH}
\ee
In reduced variables~(\ref{Eq:RedVar}), this becomes
\be
\left(\frac{\partial\hat{\tau}}{\partial\hat{p}}\right)_H = \frac{3\widehat{V}}{8[C_{\rm p}/(Nk_{\rm B})]}\left[\hat{\tau}\left(\frac{\alpha_{\rm T}\tau_{\rm c}}{k_{\rm B}}\right) - 1\right].
\label{Eq:dTdpHRed}
\ee
Thus the inversion~(I) point for a particular plot where the slope $\partial\hat{\tau}/\partial\hat{p}$ at fixed~$H$ changes from positive at low pressures to negative at high pressures is given by setting the right side of Eq.~(\ref{Eq:dTdpHRed}) to zero, yielding
\be
\hat{\tau}_{\rm I}(\hat{\tau},\widehat{V}) = \frac{1}{\alpha_{\rm T}\tau_{\rm c}/k_{\rm B}} = \frac{3[4\hat{\tau}\widehat{V}^3 - (3\widehat{V}-1)^2]}{4(3\widehat{V}-1)\widehat{V}^2},
\label{Eq:hattauI}
\ee
where the second equality was obtained using the expression for $\alpha_{\rm T}$ in Eq.~(\ref{Eq:alphavdW}).  Equation~(\ref{Eq:hattauI}) allows an accurate determination of the inversion point by locating the value of $\widehat{V}$ at which the calculated $\hat{\tau}(\widehat{V})$ crosses $\hat{\tau}_{\rm I}(\widehat{V})$ for the particular value of~$H$\@.

\begin{figure}[t]
\includegraphics[width=3.3in]{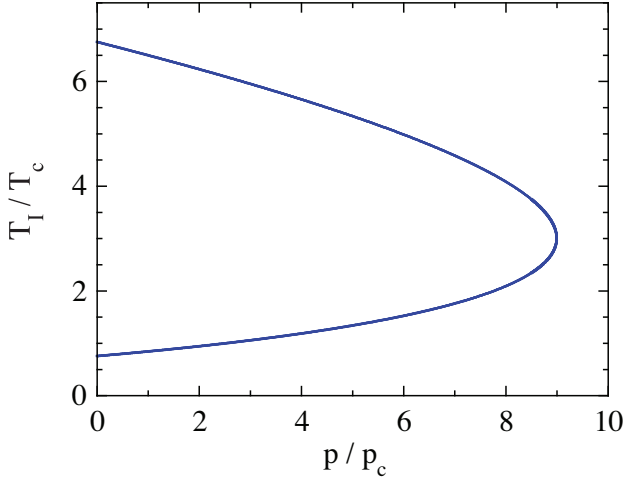}
\caption{(Color online) Reduced inversion temperature $\hat{\tau}_{\rm I} = T_{\rm I}/T_{\rm c}$ versus reduced pressure $\hat{p} = p/p_{\rm c}$ for Joule-Thomson expansion at constant enthalpy obtained using Eqs.~(\ref{Eqs:InvPlusMinus}).}
\label{Fig:vdW_const_H_Inversion}  
\end{figure}

One can also determine an analytic equation for the inversion curve of ${\tau}_{\rm I}$ versus $\hat{p}$ and important points along it.  By equating the temperatures $\hat{\tau}(H,\widehat{V})$ in Eq.~(\ref{Eq:tauVsV}) and $\hat{\tau}_{\rm I}$ in Eq.~(\ref{Eq:hattauI}) one obtains the reduced volume versus enthalpy at the inversion point as
\be
\widehat{V}_{\rm I} = \frac{15 + \sqrt{3(30 + h)}}{45 - h},
\label{Eq:VI}
\ee
where we have introduced the abbreviation
\be
h \equiv \frac{H}{p_{\rm c}V_{\rm c}}.
\ee
Then $\hat{\tau}_{\rm I}$ is given in terms of $h$ by inserting Eq.~(\ref{Eq:VI}) into~(\ref{Eq:tauVsV}), yielding
\be
\hat{\tau}_{\rm I} = \frac{1}{4}\left[42 + h -4\sqrt{3(30+h)}\right].
\label{Eq:tauIh}
\ee
The reduced inversion pressure versus~$h$ is obtained by inserting $\widehat{V}_{\rm I}(h)$ in Eq.~(\ref{Eq:VI}) and  $\hat{\tau}_{\rm I}(h)$ in Eq.~(\ref{Eq:tauIh}) into the equation of state~(\ref{Eq:RedpVsRedV}), yielding 
\be
\hat{p} = 3\left[-75-h+8\sqrt{3(30+h)}\right].
\ee
Solving this expression for $h$ gives the two solutions
\be
h_\pm = 18 + \frac{9-\hat{p}}{3} \pm 8(9-\hat{p}).
\label{Eq:hpm}
\ee
Finally, inserting these two enthalpies into Eq.~(\ref{Eq:tauIh}) and simplifying gives the two-branch solution for the inversion curve of $\hat{\tau}_{\rm I}(h)$ versus $\hat{p}$ as
\bse
\label{Eqs:InvPlusMinus}
\bea
\hat{\tau}_{\rm I} &=& 3 + \frac{9 - \hat{p}}{12} + \sqrt{9 - \hat{p}}\qquad(\hat{\tau}_{\rm I} \geq 3),\label{InvPlus}\\*
\hat{\tau}_{\rm I} &=& 3 + \frac{9 - \hat{p}}{12} - \sqrt{9 - \hat{p}}\qquad(\hat{\tau}_{\rm I} \leq 3).\label{InvMinus}
\eea
\ese
The inverse relation was obtained in Ref.~\onlinecite{LeVent2001} as
\be
\hat{p}_{\rm I} = 9\left(3-2\sqrt{\frac{\hat{\tau}}{3}}\right)\left(2\sqrt{\frac{\hat{\tau}}{3}}-1\right),
\ee
which yields Eqs.~(\ref{Eqs:InvPlusMinus}) on solving for $\hat{\tau}_{\rm I}(\hat{p})$.

Important points along the inversion curve are\cite{Kwok1979}
\bea
\hat{p}_{\rm max} &=& 9,\nonumber\\*
\hat{\tau}_{\rm I}(\hat{p}=0) &=& \frac{3}{4}\ {\rm and}\ \frac{27}{4},\quad h(\hat{p}=0) = -3 \ {\rm and}\ 45,\nonumber\\*
\hat{\tau}_{\rm I}(\hat{p}=\hat{p}_{\rm max}) &=& 3,\quad h(\hat{p}=\hat{p}_{\rm max}) = 18.
\eea
These points are consistent with the inversion point data in Fig.~\ref{Fig:vdW_const_H}(a).  A plot of $\hat{\tau}_{\rm I}$ versus $\hat{p}$ obtained using Eq.~(\ref{InvMinus}) is shown in Fig.~\ref{Fig:vdW_const_H}(a) and a plot using both of Eqs.~(\ref{Eqs:InvPlusMinus}) is shown in Fig.~\ref{Fig:vdW_const_H_Inversion}.  One notes from Fig.~\ref{Fig:vdW_const_H_Inversion} that the curve is asymmetric with respect to a horizontal line through the apex of the curve.

Expanded plots of $\hat{\tau}$ versus $\hat{p}$ for $H/(p_{\rm c}V_{\rm c}) = 0$ to~2 are shown in Fig.~\ref{Fig:vdW_const_H}(b) to emphasize the low-temperature and low-pressure region.  On each curve is appended the data point ($\hat{\tau},\ \hat{p}_{\rm X}$) (filled red circle) which is the corresponding point on the coexistence curve~$\hat{p}_{\rm X}(\tau)$ in Fig.~\ref{Fig:vdW_p_vs_T_phase_diag}.  If the final pressure is to the left of the red circle for the curve, the fluid on the low-pressure side of the throttle is in the liquid phase, whereas if the final pressure is to the right of the red circle, the fluid is in the gas phase.  Thus using Joule-Thomson expansion, one can convert gas into liquid as the fluid cools within the throttle if one appropriately chooses the operating conditions.

\section{\label{Eq:Summary} Summary} 

The van der Waals theory of fluids is a mean-field theory in which attractive molecular interactions give rise to a first-order phase transition between gas and liquid phases.  The theory can be solved exactly analytically or to numerical accuracy for all thermodynamic properties versus temperature, pressure and volume.  Here new understandings of these properties are provided, which also necessitated review of  important results about the vdW fluid already known.

The main contributions of this work include resolving the long-standing contentious question about the influence of the vdW interaction parameters~$a$ and~$b$ on the pressure of the vdW gas with respect to that of an ideal gas at the same temperature and volume, and resolving a common misconception about the meaning of the Boyle temperature.   The calculation of the coexistence region between gas and liquid using the conventional parametric solution with volume as the implicit parameter is described in detail.  Lekner's elegant parametric solution\cite{Lekner1982} of gas-liquid coexistence of the vdW fluid using the entropy difference between the gas and liquid phases as the implicit parameter is developed in detail, including determining the limiting behaviors of thermodynamic properties as the temperature approaches zero and the critical temperature to augment the corresponding results of Refs.~\onlinecite{Lekner1982} and~\onlinecite{Berberan-Santos2008}.  Using Lekner's formulation, analytic solutions  are presented in Appendix~\ref{Sec:Discont} for the discontinuities on crossing the $p$-$T$ gas-liquid coexistence curve of the isothermal compressibilty~$\kappa_{\rm T}$, thermal expansion coefficient~$\alpha$ and heat capacity at constant pressure~$C_{\rm p}$.

Although it is well known that hysteresis in the transition temperature can occur for first-order phase transitions, this aspect of the vdW fluid has been little discussed.  Quantitative numerical calculations are given of the maximum superheating and minimum supercooling temperatures on heating and cooling through the equilibrium transition temperature, respectively.  

Numerical values of thermodyanic  quantities versus temperature, volume, or entropy from the above studies are given in tables in Appendix~\ref{App:Tables}.

The critical exponents of the vdW fluid are calculated for several thermodynamic properties on approach to the critical point and within the region of overlap they agree with previous well-known results.\cite{Stanley1971}  The critical amplitudes are also calculated in terms of our reduced parameters.  The critical exponent and/or amplitude for a given property depend, in general, on the path of approach to the critical point.

The temperature dependences of $\kappa_{\rm T}$, $\alpha$ and $C_{\rm p}$ of the vdW fluid not associated with the critical point have been little studied previously.  A systematic numerical investigation at constant pressure is presented of $\kappa_{\rm T}$, $\alpha$ and $C_{\rm p}$ versus temperature where the pressure is either greater than, equal to, or less than the critical pressure.  For pressures above the critical pressure, these properties are strongly influenced by proximity to the critical point even at pressures significantly above the critical pressure.   At the critical pressure, the numerical critical behaviors near the critical point agree with the predictions of analytic theory discussed above.  Below the critical pressure, the numerical values of the discontinuities of $\kappa_{\rm T}$, $\alpha$ and $C_{\rm p}$ on crossing the coexistence curve are in agreement with the above analytic results.  The latent heat and entropy change on crossing the liquid-gas coexistence curve in the pressure-temperature plane were also thoroughly discussed.  

Expansion of a continuous flow of a non-ideal gas at high pressure through a throttle to a lower-pressure region, called Joule-Thomson expansion, can be used to cool or liquify the gas.  This process occurs at constant enthalpy when the temperature on the high-pressure side is below the so-called inversion temperature.  Isenthalps in the temperature-pressure plane for the vdW gas were generated numerically and the inversion points determined.  Formulas for the two branches of the inversion temperature versus pressure were derived analytically and found to be consistent with an earlier calculation by Le~Vent.\cite{LeVent2001}  Conditions under which liquid can be produced by Joule-Thomson expansion of a vdW gas are described.


\appendix

\section{\label{App:Tables} Tables of Values}

\begin{table*}
\caption{\label{Tab:CoexistData} Phase coexistence points in the $p$-$V$-$T$ phase space of the vdW fluid.  The data columns versus temperature are labeled by the pressure and volume notations in Fig.~\ref{Fig:Cubic_Eqn}.  }
\begin{ruledtabular}
\begin{tabular}{cccccccccccc}
$T/T_{\rm c}$ & $p_{\rm X}/p_{\rm c}$   &  $V_{\rm G}/V_{\rm c}$ &  $V_{\rm F}/V_{\rm c}$ &  $V_{\rm D}/V_{\rm c}$ &  $V_{\rm C}/V_{\rm c}~\Big|$ & $T/T_{\rm c}$ & $p_{\rm X}/p_{\rm c}$   &  $V_{\rm G}/V_{\rm c}$&  $V_{\rm F}/V_{\rm c}$ &  $V_{\rm D}/V_{\rm c}$ &  $V_{\rm C}/V_{\rm c}$\\	
\hline
1 & 1 & 1 & 1 & 1 & 1 & 0.57 & 6.419e-02 & 0.4241 & 0.5246 & 3.158 & 2.191e+01 \\
0.99 & 9.605e-01 & 0.8309 & 0.8946 & 1.128 & 1.243e+00 & 0.56 & 5.764e-02 & 0.4214 & 0.5209 & 3.232 & 2.411e+01 \\
0.98 & 9.219e-01 & 0.7755 & 0.8561 & 1.189 & 1.376e+00 & 0.55 & 5.158e-02 & 0.4188 & 0.5173 & 3.308 & 2.661e+01 \\
0.97 & 8.843e-01 & 0.7376 & 0.8283 & 1.240 & 1.496e+00 & 0.54 & 4.598e-02 & 0.4163 & 0.5137 & 3.387 & 2.947e+01 \\
0.96 & 8.476e-01 & 0.7082 & 0.8059 & 1.287 & 1.612e+00 & 0.53 & 4.081e-02 & 0.4138 & 0.5102 & 3.469 & 3.274e+01 \\
0.95 & 8.119e-01 & 0.6841 & 0.7870 & 1.330 & 1.727e+00 & 0.52 & 3.607e-02 & 0.4114 & 0.5068 & 3.553 & 3.652e+01 \\
0.94 & 7.771e-01 & 0.6637 & 0.7704 & 1.372 & 1.844e+00 & 0.51 & 3.174e-02 & 0.4091 & 0.5034 & 3.641 & 4.089e+01 \\
0.93 & 7.432e-01 & 0.6459 & 0.7556 & 1.412 & 1.963e+00 & 0.50 & 2.779e-02 & 0.4068 & 0.5000 & 3.732 & 4.598e+01 \\
0.92 & 7.102e-01 & 0.6302 & 0.7422 & 1.451 & 2.087e+00 & 0.49 & 2.420e-02 & 0.4045 & 0.4967 & 3.827 & 5.195e+01 \\
0.91 & 6.782e-01 & 0.6161 & 0.7299 & 1.490 & 2.215e+00 & 0.48 & 2.097e-02 & 0.4023 & 0.4934 & 3.925 & 5.897e+01 \\
0.90 & 6.470e-01 & 0.6034 & 0.7186 & 1.529 & 2.349e+00 & 0.47 & 1.806e-02 & 0.4002 & 0.4902 & 4.028 & 6.729e+01 \\
0.89 & 6.167e-01 & 0.5918 & 0.7080 & 1.567 & 2.489e+00 & 0.46 & 1.545e-02 & 0.3981 & 0.4870 & 4.134 & 7.722e+01 \\
0.88 & 5.874e-01 & 0.5811 & 0.6981 & 1.605 & 2.636e+00 & 0.45 & 1.313e-02 & 0.3960 & 0.4838 & 4.246 & 8.915e+01 \\
0.87 & 5.589e-01 & 0.5712 & 0.6888 & 1.644 & 2.791e+00 & 0.44 & 1.108e-02 & 0.3940 & 0.4807 & 4.362 & 1.036e+02 \\
0.86 & 5.312e-01 & 0.5620 & 0.6800 & 1.682 & 2.955e+00 & 0.43 & 9.283e-03 & 0.3921 & 0.4776 & 4.483 & 1.212e+02 \\
0.85 & 5.045e-01 & 0.5534 & 0.6717 & 1.721 & 3.128e+00 & 0.42 & 7.710e-03 & 0.3901 & 0.4746 & 4.611 & 1.429e+02 \\
0.84 & 4.786e-01 & 0.5453 & 0.6637 & 1.760 & 3.311e+00 & 0.41 & 6.347e-03 & 0.3883 & 0.4716 & 4.744 & 1.698e+02 \\
0.83 & 4.535e-01 & 0.5377 & 0.6561 & 1.800 & 3.506e+00 & 0.40 & 5.175e-03 & 0.3864 & 0.4686 & 4.883 & 2.036e+02 \\
0.82 & 4.293e-01 & 0.5306 & 0.6489 & 1.840 & 3.714e+00 & 0.39 & 4.175e-03 & 0.3846 & 0.4656 & 5.030 & 2.465e+02 \\
0.81 & 4.059e-01 & 0.5238 & 0.6419 & 1.880 & 3.936e+00 & 0.38 & 3.332e-03 & 0.3828 & 0.4627 & 5.184 & 3.015e+02 \\
0.80 & 3.834e-01 & 0.5174 & 0.6352 & 1.921 & 4.172e+00 & 0.37 & 2.627e-03 & 0.3811 & 0.4597 & 5.346 & 3.729e+02 \\
0.79 & 3.616e-01 & 0.5113 & 0.6288 & 1.963 & 4.426e+00 & 0.36 & 2.044e-03 & 0.3794 & 0.4568 & 5.518 & 4.670e+02 \\
0.78 & 3.406e-01 & 0.5055 & 0.6225 & 2.005 & 4.698e+00 & 0.35 & 1.567e-03 & 0.3777 & 0.4540 & 5.699 & 5.926e+02 \\
0.77 & 3.205e-01 & 0.5000 & 0.6165 & 2.049 & 4.990e+00 & 0.34 & 1.183e-03 & 0.3761 & 0.4511 & 5.890 & 7.631e+02 \\
0.76 & 3.011e-01 & 0.4947 & 0.6107 & 2.092 & 5.304e+00 & 0.33 & 8.785e-04 & 0.3745 & 0.4483 & 6.093 & 9.986e+02 \\
0.75 & 2.825e-01 & 0.4896 & 0.6051 & 2.137 & 5.643e+00 & 0.32 & 6.401e-04 & 0.3729 & 0.4455 & 6.308 & 1.330e+03 \\
0.74 & 2.646e-01 & 0.4848 & 0.5996 & 2.183 & 6.009e+00 & 0.31 & 4.569e-04 & 0.3713 & 0.4426 & 6.537 & 1.806e+03 \\
0.73 & 2.475e-01 & 0.4801 & 0.5943 & 2.229 & 6.406e+00 & 0.30 & 3.188e-04 & 0.3698 & 0.4399 & 6.781 & 2.506e+03 \\
0.72 & 2.311e-01 & 0.4756 & 0.5891 & 2.277 & 6.835e+00 & 0.29 & 2.170e-04 & 0.3683 & 0.4371 & 7.041 & 3.560e+03 \\
0.71 & 2.154e-01 & 0.4713 & 0.5841 & 2.326 & 7.302e+00 & 0.28 & 1.437e-04 & 0.3668 & 0.4343 & 7.321 & 5.193e+03 \\
0.70 & 2.005e-01 & 0.4672 & 0.5792 & 2.376 & 7.811e+00 & 0.27 & 9.225e-05 & 0.3654 & 0.4315 & 7.620 & 7.801e+03 \\
0.69 & 1.862e-01 & 0.4632 & 0.5744 & 2.427 & 8.366e+00 & 0.26 & 5.723e-05 & 0.3639 & 0.4288 & 7.943 & 1.211e+04 \\
0.68 & 1.726e-01 & 0.4593 & 0.5698 & 2.479 & 8.973e+00 & 0.25 & 3.417e-05 & 0.3625 & 0.4260 & 8.291 & 1.951e+04 \\
0.67 & 1.597e-01 & 0.4556 & 0.5652 & 2.532 & 9.639e+00 & 0.24 & 1.953e-05 & 0.3612 & 0.4233 & 8.668 & 3.276e+04 \\
0.66 & 1.475e-01 & 0.4520 & 0.5608 & 2.587 & 1.037e+01 & 0.23 & 1.063e-05 & 0.3598 & 0.4205 & 9.077 & 5.767e+04 \\
0.65 & 1.358e-01 & 0.4485 & 0.5564 & 2.644 & 1.118e+01 & 0.22 & 5.477e-06 & 0.3585 & 0.4178 & 9.524 & 1.071e+05 \\
0.64 & 1.249e-01 & 0.4451 & 0.5522 & 2.702 & 1.207e+01 & 0.21 & 2.647e-06 & 0.3571 & 0.4150 & 10.01 & 2.116e+05 \\
0.63 & 1.145e-01 & 0.4419 & 0.5480 & 2.761 & 1.305e+01 & 0.20 & 1.189e-06 & 0.3558 & 0.4122 & 10.55 & 4.485e+05 \\
0.62 & 1.047e-01 & 0.4387 & 0.5439 & 2.823 & 1.415e+01 & 0.19 & 4.909e-07 & 0.3546 & 0.4095 & 11.14 & 1.032e+06 \\
0.61 & 9.550e-02 & 0.4356 & 0.5399 & 2.886 & 1.537e+01 & 0.18 & 1.836e-07 & 0.3533 & 0.4067 & 11.80 & 2.615e+06 \\
0.60 & 8.687e-02 & 0.4326 & 0.5359 & 2.951 & 1.673e+01 & 0.17 & 6.113e-08 & 0.3521 & 0.4039 & 12.54 & 7.416e+06 \\
0.59 & 7.878e-02 & 0.4297 & 0.5321 & 3.018 & 1.826e+01 & 0.16 & 1.773e-08 & 0.3508 & 0.4011 & 13.37 & 2.406e+07 \\
0.58 & 7.123e-02 & 0.4269 & 0.5283 & 3.087 & 1.997e+01 & 0.15 & 4.360e-09 & 0.3496 & 0.3982 & 14.31 & 9.174e+07 \\
\end{tabular}
\end{ruledtabular}
\end{table*}

\begin{table*}
\caption{\label{Tab:CoexistData_pVT} Representative phase coexistence points in the $p$-$V$-$T$ phase space of the vdW fluid.  The data columns versus reduced pressure $p/p_{\rm c}$ are labeled by the temperature and volume notations in Fig.~\ref{Fig:Volume_Hysteresis_p_0.3}.  }
\begin{ruledtabular}
\begin{tabular}{ccccccccccc}
$p/p_{\rm c}$ & $T_{\rm SC}/T_{\rm c}$   &  $T_{\rm X}/T_{\rm c}$ &  $T_{\rm SH}/T_{\rm c}$ &  $V_7/V_{\rm c}$ &  $V_3/V_{\rm c}$ & $V_5/V_{\rm c}$ &  $V_8/V_{\rm c}$ &  $V_4/V_{\rm c}$ &  $V_6/V_{\rm c}$ &  $V_2/V_{\rm c}$ \\	
\hline
0.0002 & 0.0183 & 0.2880 & 0.8438 & 0.3680 & 3.836e+03 & 3.542 & 0.3352 & 1.221e+02 & 0.6667 & 1.125e+04 \\
0.0004 & 0.0259 & 0.3062 & 0.8438 & 0.3707 & 2.038e+03 & 3.308 & 0.3359 & 8.627e+01 & 0.6667 & 5.624e+03 \\
0.0006 & 0.0317 & 0.3180 & 0.8438 & 0.3726 & 1.410e+03 & 3.172 & 0.3365 & 7.037e+01 & 0.6667 & 3.749e+03 \\
0.0008 & 0.0365 & 0.3270 & 0.8439 & 0.3740 & 1.087e+03 & 3.075 & 0.3370 & 6.090e+01 & 0.6667 & 2.812e+03 \\
0.0100 & 0.1274 & 0.4341 & 0.8450 & 0.3929 & 1.135e+02 & 2.243 & 0.3470 & 1.698e+01 & 0.6677 & 2.243e+02 \\
0.0120 & 0.1393 & 0.4446 & 0.8453 & 0.3949 & 9.656e+01 & 2.185 & 0.3483 & 1.547e+01 & 0.6679 & 1.868e+02 \\
0.0140 & 0.1502 & 0.4539 & 0.8455 & 0.3968 & 8.425e+01 & 2.136 & 0.3496 & 1.429e+01 & 0.6681 & 1.600e+02 \\
0.0160 & 0.1603 & 0.4622 & 0.8458 & 0.3985 & 7.487e+01 & 2.095 & 0.3509 & 1.335e+01 & 0.6683 & 1.400e+02 \\
0.0180 & 0.1697 & 0.4698 & 0.8460 & 0.4001 & 6.748e+01 & 2.058 & 0.3520 & 1.256e+01 & 0.6685 & 1.243e+02 \\
0.0200 & 0.1786 & 0.4768 & 0.8463 & 0.4016 & 6.148e+01 & 2.025 & 0.3531 & 1.190e+01 & 0.6687 & 1.118e+02 \\
0.0300 & 0.2174 & 0.5057 & 0.8475 & 0.4081 & 4.298e+01 & 1.901 & 0.3581 & 9.648e+00 & 0.6697 & 7.433e+01 \\
0.0400 & 0.2496 & 0.5283 & 0.8488 & 0.4134 & 3.333e+01 & 1.814 & 0.3624 & 8.305e+00 & 0.6707 & 5.558e+01 \\
0.0500 & 0.2777 & 0.5473 & 0.8500 & 0.4181 & 2.735e+01 & 1.749 & 0.3664 & 7.388e+00 & 0.6717 & 4.433e+01 \\
0.0600 & 0.3028 & 0.5637 & 0.8513 & 0.4224 & 2.327e+01 & 1.696 & 0.3701 & 6.711e+00 & 0.6728 & 3.682e+01 \\
0.0700 & 0.3257 & 0.5783 & 0.8526 & 0.4264 & 2.029e+01 & 1.652 & 0.3736 & 6.184e+00 & 0.6738 & 3.147e+01 \\
0.0800 & 0.3468 & 0.5915 & 0.8539 & 0.4301 & 1.801e+01 & 1.614 & 0.3770 & 5.758e+00 & 0.6749 & 2.745e+01 \\
0.0900 & 0.3665 & 0.6037 & 0.8552 & 0.4337 & 1.621e+01 & 1.581 & 0.3802 & 5.406e+00 & 0.6759 & 2.432e+01 \\
0.1000 & 0.3849 & 0.6150 & 0.8564 & 0.4371 & 1.474e+01 & 1.552 & 0.3834 & 5.107e+00 & 0.6770 & 2.182e+01 \\
0.1200 & 0.4188 & 0.6354 & 0.8590 & 0.4436 & 1.251e+01 & 1.502 & 0.3895 & 4.626e+00 & 0.6792 & 1.806e+01 \\
0.1400 & 0.4496 & 0.6536 & 0.8616 & 0.4498 & 1.087e+01 & 1.461 & 0.3953 & 4.251e+00 & 0.6814 & 1.538e+01 \\
0.1800 & 0.5039 & 0.6855 & 0.8669 & 0.4614 & 8.633e+00 & 1.395 & 0.4067 & 3.696e+00 & 0.6860 & 1.180e+01 \\
0.2000 & 0.5283 & 0.6997 & 0.8695 & 0.4671 & 7.828e+00 & 1.368 & 0.4122 & 3.483e+00 & 0.6884 & 1.055e+01 \\
0.2200 & 0.5512 & 0.7130 & 0.8722 & 0.4726 & 7.159e+00 & 1.343 & 0.4178 & 3.299e+00 & 0.6908 & 9.524e+00 \\
0.2400 & 0.5728 & 0.7255 & 0.8749 & 0.4781 & 6.595e+00 & 1.322 & 0.4233 & 3.138e+00 & 0.6933 & 8.668e+00 \\
0.2600 & 0.5933 & 0.7374 & 0.8776 & 0.4835 & 6.111e+00 & 1.302 & 0.4288 & 2.995e+00 & 0.6959 & 7.943e+00 \\
0.2800 & 0.6128 & 0.7486 & 0.8803 & 0.4890 & 5.691e+00 & 1.283 & 0.4343 & 2.868e+00 & 0.6985 & 7.321e+00 \\
0.3000 & 0.6314 & 0.7594 & 0.8831 & 0.4944 & 5.323e+00 & 1.267 & 0.4399 & 2.753e+00 & 0.7011 & 6.781e+00 \\
0.3200 & 0.6491 & 0.7698 & 0.8859 & 0.4999 & 4.997e+00 & 1.251 & 0.4455 & 2.649e+00 & 0.7039 & 6.308e+00 \\
0.3400 & 0.6661 & 0.7797 & 0.8886 & 0.5053 & 4.707e+00 & 1.237 & 0.4511 & 2.553e+00 & 0.7067 & 5.890e+00 \\
0.3600 & 0.6824 & 0.7892 & 0.8915 & 0.5109 & 4.446e+00 & 1.223 & 0.4568 & 2.466e+00 & 0.7095 & 5.518e+00 \\
0.3800 & 0.6981 & 0.7985 & 0.8943 & 0.5165 & 4.210e+00 & 1.210 & 0.4627 & 2.385e+00 & 0.7125 & 5.184e+00 \\
0.4000 & 0.7132 & 0.8074 & 0.8971 & 0.5221 & 3.996e+00 & 1.198 & 0.4686 & 2.310e+00 & 0.7155 & 4.883e+00 \\
0.4400 & 0.7418 & 0.8245 & 0.9029 & 0.5337 & 3.620e+00 & 1.176 & 0.4807 & 2.174e+00 & 0.7218 & 4.362e+00 \\
0.4800 & 0.7685 & 0.8406 & 0.9088 & 0.5457 & 3.301e+00 & 1.157 & 0.4934 & 2.055e+00 & 0.7285 & 3.925e+00 \\
0.5200 & 0.7936 & 0.8558 & 0.9148 & 0.5583 & 3.025e+00 & 1.139 & 0.5068 & 1.948e+00 & 0.7357 & 3.553e+00 \\
0.5600 & 0.8170 & 0.8704 & 0.9209 & 0.5715 & 2.785e+00 & 1.122 & 0.5209 & 1.852e+00 & 0.7433 & 3.232e+00 \\
0.6000 & 0.8391 & 0.8843 & 0.9271 & 0.5856 & 2.571e+00 & 1.107 & 0.5359 & 1.763e+00 & 0.7516 & 2.951e+00 \\
0.6400 & 0.8600 & 0.8977 & 0.9334 & 0.6006 & 2.380e+00 & 1.093 & 0.5522 & 1.682e+00 & 0.7605 & 2.702e+00 \\
0.6800 & 0.8796 & 0.9106 & 0.9399 & 0.6169 & 2.207e+00 & 1.080 & 0.5698 & 1.607e+00 & 0.7702 & 2.479e+00 \\
0.7200 & 0.8982 & 0.9230 & 0.9465 & 0.6347 & 2.049e+00 & 1.068 & 0.5891 & 1.535e+00 & 0.7810 & 2.277e+00 \\
0.7600 & 0.9157 & 0.9350 & 0.9533 & 0.6545 & 1.903e+00 & 1.056 & 0.6107 & 1.468e+00 & 0.7930 & 2.092e+00 \\
0.8000 & 0.9322 & 0.9466 & 0.9603 & 0.6769 & 1.766e+00 & 1.046 & 0.6352 & 1.403e+00 & 0.8066 & 1.921e+00 \\
0.8400 & 0.9478 & 0.9579 & 0.9676 & 0.7027 & 1.636e+00 & 1.035 & 0.6637 & 1.339e+00 & 0.8224 & 1.760e+00 \\
0.8800 & 0.9624 & 0.9688 & 0.9750 & 0.7338 & 1.510e+00 & 1.026 & 0.6981 & 1.276e+00 & 0.8414 & 1.605e+00 \\
0.9200 & 0.9761 & 0.9795 & 0.9828 & 0.7733 & 1.382e+00 & 1.017 & 0.7422 & 1.210e+00 & 0.8655 & 1.451e+00 \\
0.9600 & 0.9887 & 0.9899 & 0.9910 & 0.8300 & 1.245e+00 & 1.008 & 0.8059 & 1.137e+00 & 0.8998 & 1.287e+00 \\
1 & 1 & 1 & 1 & 1 & 1 & 1 & 1 & 1 & 1 & 1 \\
\end{tabular}
\end{ruledtabular}
\end{table*}


\begin{table*}
\caption{\label{Tab:LeknerTableA} Representative values for the quantities listed, calculated in terms of the implicit parameter~$y$ using Lekner's parametric solution.\cite{Lekner1982} The subscript~X refers to a property associated with the coexistence curve in Fig.~\ref{Fig:vdW_p_vs_T_phase_diag}.  $\hat{\tau}_{\rm X}$ and $\hat{p}_{\rm X}$ are the coordinates of the curve; $\Delta\widehat{V}_{\rm X}$ is the difference in volume of the coexisting gas and liquid phases;  $\hat{n}_g$  and $\hat{n}_l$ are the number densities of the two coexisting phases, $\Delta \hat{n}_{\rm X}$ is the difference in density between the coexisting phases, which is the order parameter for the gas-liquid transition; $\hat{n}_{\rm ave}$ is the average of $\hat{n}_g$  and $\hat{n}_l$; $\Delta S_{\rm X}/(Nk_{\rm B}) = (S_g-S_l)/(Nk_{\rm B})$ is the entropy difference between the coexisting phases; and $L/(p_{\rm c}V_{\rm c})$ is the latent heat (enthalpy) of vaporization.  These data are complementary to those in Table~\ref{Tab:CoexistData}.}
\begin{ruledtabular}
\begin{tabular}{ccccccccccc}
$y$ & $\hat{\tau}_{\rm X}$   &  $\hat{p}_{\rm X}$ &  $d\hat{p}_{\rm X}/d\hat{\tau}_{\rm X}$ &  $ \Delta \widehat{V}_{\rm X}$ &  $\hat{n}_g$ & $\hat{n}_l$ &  $\Delta \hat{n}_{\rm X}$ &  $\hat{n}_{\rm ave}$ &  $\underline{S_g-S_l}$ &  $L/(p_{\rm c}V_{\rm c})$ \\
    & $ = T_{\rm X}/T_{\rm c}$ & $ = p_{\rm X}/p_{\rm c}$ & & $ = (V_g-V_l)/V_{\rm c}$ & $=n_g/n_{\rm c}$ & $=n_l/n_{\rm c}$ & $ = \hat{n}_l-\hat{n}_g$ &$=\frac{\hat{n}_g+\hat{n}_l}{2}$ & $Nk_{\rm B}$ & \\
\hline
0       &     1   &     1      &      4     &     0      &     1      &   1    &    0    &    1   &    0    &     0   \\
0.10000 & 0.99889 & 9.9557e-01 & 3.9893e+00 & 1.3369e-01 & 9.3384e-01 & 1.0671 & 0.13321 & 1.0004 & 0.20000 & 0.53274 \\ 
0.11220 & 0.99860 & 9.9442e-01 & 3.9866e+00 & 1.5011e-01 & 9.2584e-01 & 1.0753 & 0.14944 & 1.0006 & 0.22440 & 0.59757 \\ 
0.12589 & 0.99824 & 9.9298e-01 & 3.9831e+00 & 1.6857e-01 & 9.1689e-01 & 1.0845 & 0.16762 & 1.0007 & 0.25179 & 0.67025 \\ 
0.14125 & 0.99779 & 9.9118e-01 & 3.9788e+00 & 1.8934e-01 & 9.0688e-01 & 1.0949 & 0.18801 & 1.0009 & 0.28251 & 0.75169 \\ 
0.15849 & 0.99722 & 9.8891e-01 & 3.9733e+00 & 2.1274e-01 & 8.9569e-01 & 1.1065 & 0.21085 & 1.0011 & 0.31698 & 0.84292 \\ 
0.17783 & 0.99650 & 9.8606e-01 & 3.9664e+00 & 2.3911e-01 & 8.8318e-01 & 1.1196 & 0.23644 & 1.0014 & 0.35566 & 0.94510 \\ 
0.19953 & 0.99560 & 9.8248e-01 & 3.9578e+00 & 2.6887e-01 & 8.6921e-01 & 1.1343 & 0.26510 & 1.0018 & 0.39905 & 1.0595 \\ 
0.22387 & 0.99446 & 9.7801e-01 & 3.9469e+00 & 3.0251e-01 & 8.5363e-01 & 1.1508 & 0.29717 & 1.0022 & 0.44774 & 1.1874 \\ 
0.25119 & 0.99304 & 9.7240e-01 & 3.9333e+00 & 3.4059e-01 & 8.3626e-01 & 1.1693 & 0.33305 & 1.0028 & 0.50238 & 1.3304 \\ 
0.28184 & 0.99126 & 9.6540e-01 & 3.9163e+00 & 3.8382e-01 & 8.1693e-01 & 1.1901 & 0.37315 & 1.0035 & 0.56368 & 1.4900 \\ 
0.31623 & 0.98902 & 9.5666e-01 & 3.8949e+00 & 4.3301e-01 & 7.9544e-01 & 1.2134 & 0.41793 & 1.0044 & 0.63246 & 1.6680 \\ 
0.35481 & 0.98622 & 9.4579e-01 & 3.8683e+00 & 4.8920e-01 & 7.7161e-01 & 1.2395 & 0.46786 & 1.0055 & 0.70963 & 1.8663 \\ 
0.39811 & 0.98272 & 9.3231e-01 & 3.8350e+00 & 5.5365e-01 & 7.4523e-01 & 1.2687 & 0.52346 & 1.0070 & 0.79621 & 2.0865 \\ 
0.44668 & 0.97835 & 9.1564e-01 & 3.7935e+00 & 6.2800e-01 & 7.1611e-01 & 1.3013 & 0.58524 & 1.0087 & 0.89337 & 2.3307 \\ 
0.50119 & 0.97291 & 8.9514e-01 & 3.7420e+00 & 7.1432e-01 & 6.8408e-01 & 1.3378 & 0.65372 & 1.0109 & 1.0024 & 2.6006 \\ 
0.56234 & 0.96615 & 8.7007e-01 & 3.6784e+00 & 8.1535e-01 & 6.4900e-01 & 1.3784 & 0.72941 & 1.0137 & 1.1247 & 2.8976 \\ 
0.63096 & 0.95779 & 8.3964e-01 & 3.6000e+00 & 9.3476e-01 & 6.1078e-01 & 1.4235 & 0.81273 & 1.0171 & 1.2619 & 3.2231 \\ 
0.70795 & 0.94748 & 8.0304e-01 & 3.5040e+00 & 1.0776e+00 & 5.6940e-01 & 1.4734 & 0.90404 & 1.0214 & 1.4159 & 3.5774 \\ 
0.79433 & 0.93485 & 7.5951e-01 & 3.3872e+00 & 1.2507e+00 & 5.2494e-01 & 1.5284 & 1.0035 & 1.0267 & 1.5887 & 3.9604 \\ 
0.89125 & 0.91947 & 7.0849e-01 & 3.2465e+00 & 1.4642e+00 & 4.7765e-01 & 1.5887 & 1.1111 & 1.0332 & 1.7825 & 4.3705 \\ 
1.0000 & 0.90088 & 6.4971e-01 & 3.0787e+00 & 1.7324e+00 & 4.2793e-01 & 1.6543 & 1.2264 & 1.0411 & 2.0000 & 4.8047 \\ 
1.1220 & 0.87864 & 5.8343e-01 & 2.8812e+00 & 2.0769e+00 & 3.7642e-01 & 1.7251 & 1.3487 & 1.0508 & 2.2440 & 5.2579 \\ 
1.2589 & 0.85233 & 5.1066e-01 & 2.6528e+00 & 2.5310e+00 & 3.2401e-01 & 1.8008 & 1.4768 & 1.0624 & 2.5179 & 5.7228 \\ 
1.4125 & 0.82165 & 4.3327e-01 & 2.3938e+00 & 3.1471e+00 & 2.7183e-01 & 1.8806 & 1.6088 & 1.0762 & 2.8251 & 6.1900 \\ 
1.5849 & 0.78646 & 3.5408e-01 & 2.1075e+00 & 4.0108e+00 & 2.2124e-01 & 1.9638 & 1.7425 & 1.0925 & 3.1698 & 6.6477 \\ 
1.7783 & 0.74684 & 2.7672e-01 & 1.8004e+00 & 5.2677e+00 & 1.7374e-01 & 2.0489 & 1.8751 & 1.1113 & 3.5566 & 7.0831 \\ 
1.9953 & 0.70320 & 2.0516e-01 & 1.4831e+00 & 7.1752e+00 & 1.3083e-01 & 2.1345 & 2.0036 & 1.1326 & 3.9905 & 7.4830 \\ 
2.2387 & 0.65627 & 1.4305e-01 & 1.1693e+00 & 1.0211e+01 & 9.3795e-02 & 2.2188 & 2.1250 & 1.1563 & 4.4774 & 7.8358 \\ 
2.5119 & 0.60711 & 9.2953e-02 & 8.7509e-01 & 1.5309e+01 & 6.3517e-02 & 2.3003 & 2.2368 & 1.1819 & 5.0238 & 8.1333 \\ 
2.8184 & 0.55698 & 5.5764e-02 & 6.1578e-01 & 2.4410e+01 & 4.0272e-02 & 2.3773 & 2.3370 & 1.2088 & 5.6368 & 8.3722 \\ 
3.1623 & 0.50721 & 3.0600e-02 & 4.0325e-01 & 4.1824e+01 & 2.3679e-02 & 2.4485 & 2.4249 & 1.2361 & 6.3246 & 8.5544 \\ 
3.5481 & 0.45904 & 1.5218e-02 & 2.4302e-01 & 7.7868e+01 & 1.2777e-02 & 2.5133 & 2.5006 & 1.2631 & 7.0963 & 8.6867 \\ 
3.9811 & 0.41346 & 6.7958e-03 & 1.3318e-01 & 1.5943e+02 & 6.2571e-03 & 2.5714 & 2.5651 & 1.2888 & 7.9621 & 8.7788 \\ 
4.4668 & 0.37112 & 2.6990e-03 & 6.5526e-02 & 3.6357e+02 & 2.7476e-03 & 2.6227 & 2.6200 & 1.3127 & 8.9337 & 8.8412 \\ 
5.0119 & 0.33233 & 9.4325e-04 & 2.8554e-02 & 9.3610e+02 & 1.0678e-03 & 2.6678 & 2.6668 & 1.3344 & 10.024 & 8.8833 \\ 
5.6234 & 0.29716 & 2.8653e-04 & 1.0860e-02 & 2.7617e+03 & 3.6204e-04 & 2.7073 & 2.7069 & 1.3538 & 11.247 & 8.9122 \\ 
6.3096 & 0.26545 & 7.4573e-05 & 3.5467e-03 & 9.4880e+03 & 1.0539e-04 & 2.7418 & 2.7417 & 1.3710 & 12.619 & 8.9327 \\ 
7.0795 & 0.23699 & 1.6350e-05 & 9.7698e-04 & 3.8647e+04 & 2.5875e-05 & 2.7720 & 2.7720 & 1.3860 & 14.159 & 8.9479 \\ 
7.9433 & 0.21148 & 2.9615e-06 & 2.2247e-04 & 1.9043e+05 & 5.2513e-06 & 2.7985 & 2.7985 & 1.3992 & 15.887 & 8.9594 \\ 
8.9125 & 0.18867 & 4.3334e-07 & 4.0941e-05 & 1.1610e+06 & 8.6131e-07 & 2.8217 & 2.8217 & 1.4108 & 17.825 & 8.9682 \\ 
10.000 & 0.16828 & 4.9947e-08 & 5.9361e-06 & 8.9845e+06 & 1.1130e-07 & 2.8421 & 2.8421 & 1.4211 & 20.000 & 8.9751 \\ 
11.220 & 0.15007 & 4.4071e-09 & 6.5899e-07 & 9.0807e+07 & 1.1012e-08 & 2.8601 & 2.8601 & 1.4300 & 22.440 & 8.9804 \\ 
12.589 & 0.13381 & 2.8826e-10 & 5.4240e-08 & 1.2379e+09 & 8.0783e-10 & 2.8759 & 2.8759 & 1.4380 & 25.179 & 8.9846 \\ 
14.125 & 0.11930 & 1.3482e-11 & 3.1924e-09 & 2.3598e+10 & 4.2376e-11 & 2.8899 & 2.8899 & 1.4450 & 28.251 & 8.9879 \\ 
15.849 & 0.10636 & 4.3290e-13 & 1.2901e-10 & 6.5518e+11 & 1.5263e-12 & 2.9023 & 2.9023 & 1.4511 & 31.698 & 8.9904 \\ 
\end{tabular}
\end{ruledtabular}
\end{table*}

\begin{table}
\caption{\label{Tab:LeknerTableB} Representative values of the jumps in three properties on crossing the coexistence curve in Fig.~\ref{Fig:vdW_p_vs_T_phase_diag}.  The reduced temperatures~$\hat{\tau}_{\rm X}$ (and $y$ values) are the same as in Table~\ref{Tab:LeknerTableA}. The quantities listed are the jumps in the reduced isothermal compressibility $\kappa_{\rm T}$, thermal expansion coefficient~$\alpha$ and heat capacity at constant pressure $C_{\rm p}$ calculated from Eqs.~(\ref{Eq:DeltaKappa}), (\ref{Eq:DeltaAlpha}) and~(\ref{Eq:DeltaCpThy}), as shown in Figs.~\ref{Fig:vdW_const_p_alpha_equilib} and~\ref{Fig:vdW_const_p_Cp_p_leq_1}.}
\begin{ruledtabular}
\begin{tabular}{cccc}
$\hat{\tau}_{\rm X}$   &  $\Delta\kappa_{\rm T}p_{\rm c}$ &  $\Delta\alpha\,\tau_{\rm c}/k_{\rm B}$ & $\Delta C_{\rm p}/(Nk_{\rm B})$\\
    & $ = (\kappa_{{\rm T}g} - \kappa_{{\rm T}l})p_{\rm c}$ & $ = \frac{(\alpha_g - \alpha_l)\tau_{\rm c}}{k_{\rm B}}$ & $=\frac{C_{{\rm p}l}-C_{{\rm p}g}}{Nk_{\rm B}}$\\
\hline
0.99889 & 1.8110e+01 & 12.035 & 1.2007e+01 \\ 
0.99860 & 1.6164e+01 & 10.734 & 1.0703e+01 \\ 
0.99824 & 1.4418e+01 & 9.5642 & 9.5408e+00 \\ 
0.99779 & 1.2882e+01 & 8.5343 & 8.5053e+00 \\ 
0.99722 & 1.1513e+01 & 7.6148 & 7.5827e+00 \\ 
0.99650 & 1.0297e+01 & 6.7967 & 6.7607e+00 \\ 
0.99560 & 9.2177e+00 & 6.0688 & 6.0283e+00 \\ 
0.99446 & 8.2611e+00 & 5.4214 & 5.3760e+00 \\ 
0.99304 & 7.4144e+00 & 4.8460 & 4.7950e+00 \\ 
0.99126 & 6.6665e+00 & 4.3350 & 4.2776e+00 \\ 
0.98902 & 6.0077e+00 & 3.8815 & 3.8169e+00 \\ 
0.98622 & 5.4293e+00 & 3.4795 & 3.4069e+00 \\ 
0.98272 & 4.9242e+00 & 3.1238 & 3.0420e+00 \\ 
0.97835 & 4.4860e+00 & 2.8096 & 2.7175e+00 \\ 
0.97291 & 4.1097e+00 & 2.5329 & 2.4290e+00 \\ 
0.96615 & 3.7913e+00 & 2.2899 & 2.1726e+00 \\ 
0.95779 & 3.5282e+00 & 2.0776 & 1.9449e+00 \\ 
0.94748 & 3.3190e+00 & 1.8933 & 1.7428e+00 \\ 
0.93485 & 3.1645e+00 & 1.7347 & 1.5636e+00 \\ 
0.91947 & 3.0676e+00 & 1.5998 & 1.4048e+00 \\ 
0.90088 & 3.0351e+00 & 1.4872 & 1.2642e+00 \\ 
0.87864 & 3.0789e+00 & 1.3958 & 1.1398e+00 \\ 
0.85233 & 3.2197e+00 & 1.3250 & 1.0295e+00 \\ 
0.82165 & 3.4926e+00 & 1.2747 & 9.3181e-01 \\ 
0.78646 & 3.9586e+00 & 1.2451 & 8.4488e-01 \\ 
0.74684 & 4.7256e+00 & 1.2374 & 7.6714e-01 \\ 
0.70320 & 5.9936e+00 & 1.2531 & 6.9701e-01 \\ 
0.65627 & 8.1520e+00 & 1.2947 & 6.3297e-01 \\ 
0.60711 & 1.2003e+01 & 1.3654 & 5.7362e-01 \\ 
0.55698 & 1.9313e+01 & 1.4688 & 5.1776e-01 \\ 
0.50721 & 3.4261e+01 & 1.6090 & 4.6451e-01 \\ 
0.45904 & 6.7577e+01 & 1.7899 & 4.1347e-01 \\ 
0.41346 & 1.4940e+02 & 2.0145 & 3.6480e-01 \\ 
0.37112 & 3.7328e+02 & 2.2849 & 3.1908e-01 \\ 
0.33233 & 1.0636e+03 & 2.6019 & 2.7714e-01 \\ 
0.29716 & 3.4944e+03 & 2.9662 & 2.3968e-01 \\ 
0.26545 & 1.3415e+04 & 3.3788 & 2.0700e-01 \\ 
0.23699 & 6.1168e+04 & 3.8425 & 1.7898e-01 \\ 
0.21148 & 3.3768e+05 & 4.3618 & 1.5514e-01 \\ 
0.18867 & 2.3077e+06 & 4.9427 & 1.3490e-01 \\ 
0.16828 & 2.0021e+07 & 5.5928 & 1.1765e-01 \\ 
0.15007 & 2.2691e+08 & 6.3207 & 1.0288e-01 \\ 
0.13381 & 3.4691e+09 & 7.1361 & 9.0177e-02 \\ 
0.11930 & 7.4175e+10 & 8.0499 & 7.9206e-02 \\ 
0.10636 & 2.3100e+12 & 9.0743 & 6.9692e-02 \\ 

\end{tabular}
\end{ruledtabular}
\end{table}

\clearpage

\section{\label{Sec:Discont} Discontinuities in Isothermal Compressibility, Thermal Expansion and Heat Capacity versus Temperature at Constant Pressure on Crossing the Liquid-Gas Coexistence Curve}

The following equations for the discontinuities in the isothermal compressibility $\kappa_{\rm T}$, volume thermal expansion coefficient~$\alpha$ and heat capacity at constant pressure $C_{\rm p}$ versus reduced temperature $\hat{\tau}_{\rm X}$ are obtained from the calculation of these properties in terms of the parameter $y$ together with $\hat{\tau}_{\rm X}(y)$ with $y$ being an implicit parameter.

\begin{widetext}
\bse
\bea
\kappa_{{\rm T}g}p_{\rm c} &=& \frac{4 e^{6 y} \sinh^2(y) [2 y^2 -1 + \cosh(2y) - 2 y \sinh(2y)]^2}{27 [1 + e^{2y} (y-1) + y]^2 [e^{6 y} + e^{2 y} (7 - 4 y) - 2 y -3 + e^{4 y} (-5 + 6 y - 8 y^2)]},\\
\kappa_{{\rm T}l}p_{\rm c} &=& \frac{-e^{-y} \sinh^2(y) [1 - 2 y^2 - \cosh(2 y) + 2 y \sinh(2 y)]^2}{54 [y \cosh(y) - \sinh(y)]^2 \left\{(4 y^2-1 + y) \cosh(y) - (y-1) \cosh(3 y) - 2 y [3 + 2 y + \cosh(2 y)] \sinh(y) + 8 \sinh^3(y)\right\}}\nonumber\\
\\
\Delta\kappa_{\rm T}p_{\rm c} &\equiv& (\kappa_{{\rm T}g}-\kappa_{{\rm T}l})p_{\rm c} = \frac{\Delta\kappa_{\rm T}^{(1)}}{\Delta\kappa_{\rm T}^{(2)}},\label{Eq:DeltaKappa}\\
\Delta\kappa_{\rm T}^{(1)} &=& 4 \sinh^4(y) \{2 y [2 + \cosh(2 y)] - 3 \sinh(2 y)\} [2 y^2 -1 + \cosh(2 y) - 2 y \sinh(2 y)]^2\nonumber\\
\Delta\kappa_{\rm T}^{(2)} &=& 27 [y \cosh(y) - \sinh(y)]^2 \Big\{42 + 12 y^2 + 32 y^4 - (61 + 48 y^2) \cosh(2 y) + (22 + 36 y^2) \cosh(4 y) - 3 \cosh(6 y)\nonumber\\*
&& -\ 128 y^3 \cosh^3(y) \sinh(y) - 16 y [9 \cosh(y) - \cosh(3 y)] \sinh^3(y)\Big\},\nonumber
\eea
\ese
\bse
\bea
\frac{\alpha_g\tau_{\rm c}}{k_{\rm B}} &=& \frac{\alpha_g^{(1)}}{\alpha_g^{(2)}},\\*
\alpha_g^{(1)} &=& 8 \sinh^2(y) [2 y^2 -1 + \cosh(2 y) - 2 y \sinh(2 y)]^2\nonumber\\*
\alpha_g^{(2)} &=& 27 [y \cosh(y) - \sinh(y)] [2 y - \sinh(2 y)] \Big\{[ y (4 y-1)-1] \cosh(y) + (1 + y) \cosh(3 y)\nonumber\\
&& +\ [6 + y (4 y-5)] \sinh(y) - (2 + y) \sinh(3 y)\Big\},\nonumber\\
\frac{\alpha_l\tau_{\rm c}}{k_{\rm B}} &=& \frac{\alpha_l^{(1)}}{\alpha_l^{(2)}},\\*
\alpha_l^{(1)} &=& 8 \sinh^2(y) [2 y^2 -1 + \cosh(2 y) - 2 y \sinh(2 y)]^2,\nonumber\\
\alpha_l^{(2)} &=& 27 [y \cosh(y) - \sinh(y)] \Big\{(y -1 + 4 y^2) \cosh(y) - (y-1) \cosh[3 y]\nonumber \\*
&& -\ 2 y [3 + 2 y + \cosh(2 y)] \sinh(y) + 8 \sinh^3(y)\Big\} [2 y - \sinh(2 y)]\nonumber
\eea
\bea
\frac{\Delta\alpha_l\tau_{\rm c}}{k_{\rm B}} &\equiv& \frac{(\alpha_g-\alpha_l)\tau_{\rm c}}{k_{\rm B}} = \frac{\Delta\alpha^{(1)}}{\Delta\alpha^{(2)}},\label{Eq:DeltaAlpha}\\
\Delta\alpha^{(1)} &=& 64 \sinh^3(y) [1 - 2 y^2 - \cosh(2 y) + 2 y \sinh(2 y)]^2 [2 \cosh(2 y) - y \sinh(2 y) - 2 - 2 y^2],\nonumber\\*
\Delta\alpha^{(2)} &=& 27 [y \cosh(y) - \sinh(y)] \Big\{42 + 12 y^2 + 32 y^4 - (61 + 48 y^2) \cosh(2 y) + (22 + 36 y^2) \cosh(4 y)\nonumber\\*
&& -\ 3 \cosh(6 y) - 128 y^3 \cosh^3(y) \sinh(y) + 16 y [\cosh(3 y)-9 \cosh(y)] \sinh^3(y) \Big\} [2 y -\sinh(2 y)].\nonumber
\eea
\ese
\bse
\label{Eqs:Cplgy}
\bea
\frac{C_{{\rm p}\,g}}{Nk_{\rm B}} &=& \frac{3}{2} + \frac{(e^{2 y}-1)^2 (e^{2 y} - 2 y-1)}{ e^{6 y} + e^{2 y} (7 - 4 y) - 2 y + e^{4 y} (-5 + 6 y - 8 y^2)-3 },\\*
\frac{C_{{\rm p}\,l}}{Nk_{\rm B}} &=& \frac{3}{2} + \frac{(e^{2 y} - 1)^2 [1 + e^{2 y} (2 y - 1)]}{1 + e^{6 y} (2 y-3) + e^{4 y} (4 y + 7) - e^{2 y} (8 y^2 +6 y + 5)}.
\eea
\bea
\frac{\Delta C_{\rm p}}{Nk_{\rm B}} \equiv \frac{C_{{\rm p}\,l}-C_{{\rm p}\,g}}{Nk_{\rm B}} &=& \left(e^{2y}-1\right)^2 \bigg[\frac{1 - e^{2y} + 2 y}{e^{6y} + e^{2y} (7 - 4 y) - 2 y -3 + e^{4y} (-5 + 6 y - 8 y^2)}\label{Eq:DeltaCpThy}\\*
&&\hspace{0.9in} +\ \frac{1 + e^{2y} (-1 + 2 y)}{1 + e^{6y} (-3 + 2 y) + e^{4y} (7 + 4 y) - e^{2y} (5 + 6 y + 8 y^2)}\bigg]. \nonumber
\eea
\ese
\end{widetext}
These quantities are calculated versus $\hat{\tau}_{\rm X}$ using Eq.~(\ref{Eq:tauFromy}) with $y$ as an implicit parameter.

\end{document}